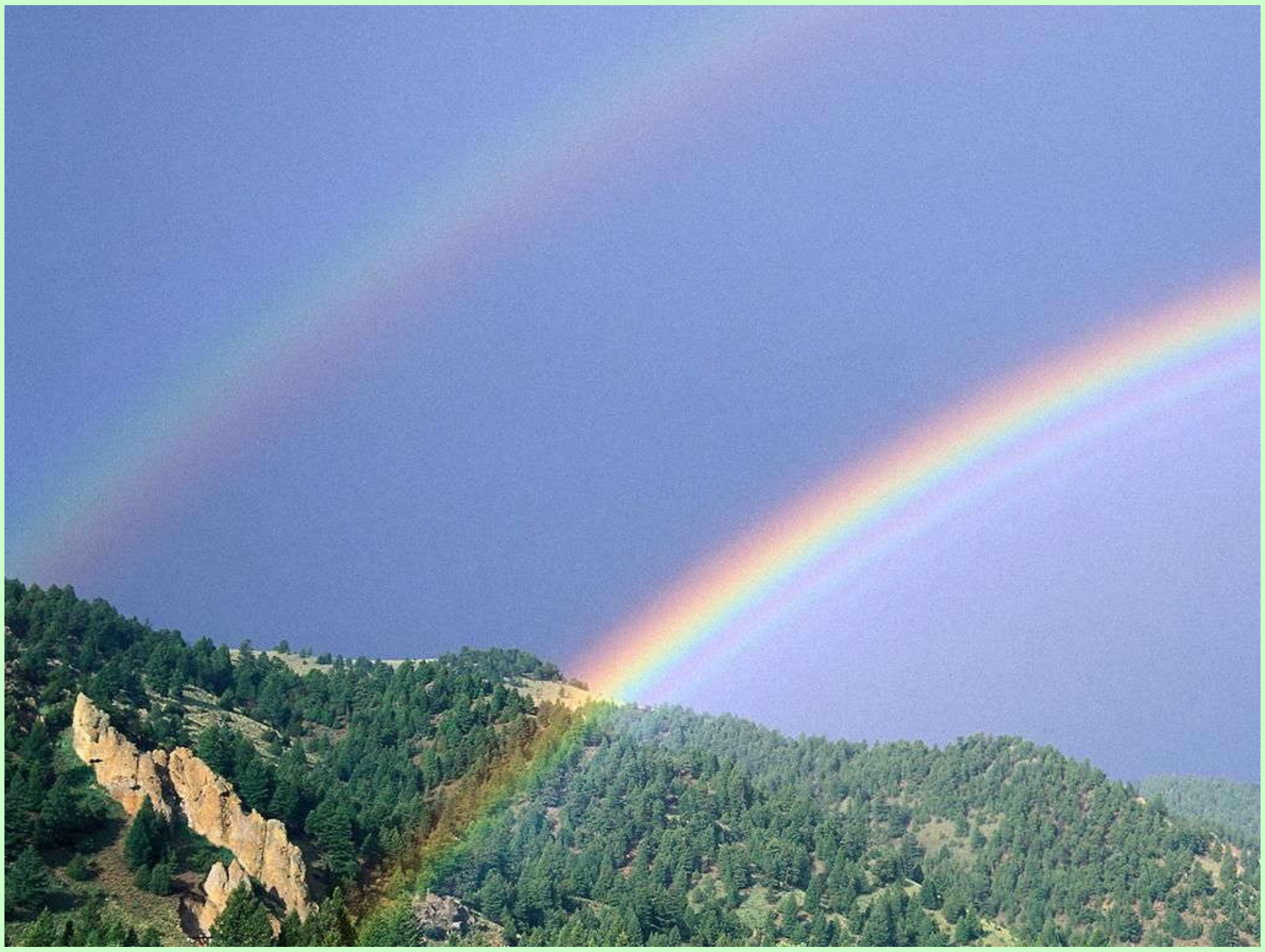

# Model of pathogenesis of psoriasis

## Part 1. Systemic psoriatic process

Edition e4.0

**Mikhail Peslyak**

---

**Part 1 (edition r4.0) and Part 2 (edition r1.3) are in agreement.**

---

**Web: www.psorias.info, E-mail: info@psorias.info**

---

---

**The given book is the authorized translation of the book, published in Russian. Translation has been carried out with active support of Translation agency "LingLab".**
**The author thanks Vladimir Turbin for the help with the proofreading.**
**Periodic reprinting at occurrence of new materials or detection of serious errors is occured. Further, updated texts (Russian and English) of the monograph will be regularly placed in the Internet.**
**Details and links on www.psorias.info.**

## Abstract

**Review and analytical study of results of experimental and theoretical works on etiology and pathogenesis of psoriatic disease was conducted. Psoriasis is dermal implication of systemic psoriatic process (SPP). New SPP model explaining results of clinical and laboratory experiments was formulated. According to the model (hereinafter Y-model) there are two main factors: hyperpermeability of small intestine for bacterial products and colonization of its walls by Gram+ bacteria (incl. psoriagenic bacteria PsB) and Gram(-) TLR4-active bacteria. Inside SPP there is a vicious cycle which is supported by disturbance of production and-or circulation of bile acids.**

**SPP central subprocess is PAMP-nemia, namely chronic kPAMP-load on blood phagocytes (neutrophiles, monocytes and dendritic cells). The load results in increase of blood kPAMP level. The major key PAMP (kPAMP) are LPS and PG (incl. PG-Y – peptidoglycan of psoriagenic bacteria). Chronically increased kPAMP-load possibly provides tolerization of some neutrophils Neu, monocytes Mo and dendritic cells DC in blood flow.**

**The chemostatus of tolerized blood Neu-T in process of their aging changes similarly to chemostatus nonactivated Neu and, hence, they carry endocytosed content from blood flow into the bone marrow. Chemostatuses of tolerized Mo-T and DC-T are similar to nonactivated ones. So they don't bring endocytosed content to lymph nodes or spleen and remain in blood.**

**Tolerized phagocytes degrade endocytosed fragments of bacterial products containing kPAMP slowly and incompletely, Tolerized phagocytes appeared to be (PG-Y)-carriers are named by R-phagocytes and are designated as Neu-R, Mo-R and DC-R.**

**SPP severity predetermines possibility of psoriasis initialization and maintenance because Mo-R and DC-R along with normal Mo and DC participate in homeostatic and inflammatory renewal of pool of dermal macrophages and DC of non-resident origin. Mo-R and DC-R enter derma and they can transform to mature maDC-Y (in particular – to TipDC), presenting Y-antigen to specific TL-Y. Local processes in derma and epidermis will be described in details in the second part of the monograph.**

**The given book is the authorized translation of the book, published in Russian (ISBN 978-5-905504-01-3; edition r4.0).**

## Keywords

**Systemic psoriatic process, psoriasis, psoriatic arthritis, intestinal permeability, intestine microflora, bile acids, psoriagenic bacteria, bacterial products, peptidoglycan, muramyl dipeptide, lipopolysaccharide, endotoxin, PAMP-nemia, neutrophils, monocytes, dendritic cells, chemostatus, tolerization, tolerized phagocytes, R-phagocytes, kPAMP-carriage, (PG-Y)-carriage.**

## Content

---

## *List of figures*

*«Tank with kerosene was strapped to fireman's shoulders…»*
Ray Bradbury, «Fahrenheit 451»

# Introduction

Epidermis self-renewal is regular process. New cells are born in basal layer. They mature, vary, migrate outside and form external horny layer. Then they die away and exfoliate. Standard duration of epidermis cell life (renewal period) for areas of skin with average thickness is 20-25 days. Psoriasis accelerates self-renewal. Cells live 4-10 days (Baker 2000, Iizuka 2004, Weinstein 1985). Cells migrating outside have no time to differentiate and they aren't quite functional. Psoriatic plaques have red shade. They are tender, they are covered by white flakes due to intensive lost of cells and they are much thicker.

Psoriasis isn't contagious. There are various types of psoriasis: vulgaris or plaque (L40.0), flexural or inverse (L40.83-4), erythrodermic (L40.85), pustular (L40.1-3, L40.82), guttate (L40.4). Codes of diseases are given according to ICD-10. Chronic plaque psoriasis (CPs) is the most frequent type (more than 80% of total number of cases). Up to 15% of psoriatics also suffer from psoriatic arthritis (L40.5). Psoriasis strikes about 2-3% of population (120-180 million people). New diagnosis of psoriasis gets 4-6 million people every year. Disease appears after birth or in extreme old age. Psoriasis is a chronic disease so there are periods of aggravation and remission. Sometimes there is no cause for period change and sometimes aggravation can be decreased as a result of treatment. Serious psoriasis can result in disability. Psoriasis course is similar in men and women. Afro-Americans, Indians, Chineses and Japaneses suffer from psoriasis less frequently and Eskimos don't suffer from psoriasis at all (Piruzian 2006, Giardina 2004).

Psoriasis is registered in "Online Mendelian Inheritance in Man" at number OMIM*177900. Psoriasis is disease with hereditary predisposition: concordance of uniovular twins is 70%. If one parent suffers from psoriasis children are diagnosed the disease in 15-25% of cases; if both parents suffer from psoriasis children are diagnosed the disease in more than 40-60% of cases. The interrelation of allele HLA-Cw*0602 (chromosome 6p21) and psoriasis of the first type which is characterized by early beginning is proved (Weisenseel 2002). This allele is found in more than 60% of psoriatics (not more than 15% of healthy people). Locuses of other chromosomes have weaker interrelations (Piruzian 2006, Giardina 2004). Psoriasis can't begin only in presence of genetical deflections. External exposure is necessary for beginning and maintenance of psoriasis. Infections, skin traumas, stresses, reaction to medications, climatic changes and other causes can provoke onset of psoriasis or its aggravation (Molochkov 2007, Bos 2005, Fry 2007b, Gudjonsson 2004).

Accelerated proliferation of keratinocytes is likely to be caused by erroneous actions of mechanisms of skin antibacterial protection. Influence of beta-hemolitic streptococci (first of all causing tonsillar infections) on initialization and aggravations of psoriasis is avowed (Molochkov 2007, Khairutdinov 2007, Baker 2006a, Baker 2003, Baker 2000, Fry 2007b, Gudjonsson 2004). There is no uniform point of view on etiology and pathogenesis of psoriasis. Researches offer various models (Baltabaev 2005, Korotkii & Peslyak 2005, Khardikova 2000, Baker 2000, Baker 2006b, Bos 2005, Danilenko 2008, Gudjonsson 2004, Gyurcsovics 2003, Iizuka 2004, Lowes 2007, Majewski 2003, Nestle 2005b, Nickoloff 2000, Sabat 2007).

---

Note. Appendix 11 contains the list of all essential changes and additions (in comparison with the pregoing edition) and the reference to particular places. In the text the significant new or revised fragments are marked with a vertical line on the right. New works included into the bibliography are marked the same way.

**Objective**

Present work is review and analytical study of interrelation of disfunction of small intestine, hepatobiliary system, intestinal microflora and blood phagocytes (neutrophiles, monocytes and dendritic cells). The aim of the study was substantiation of new model of psoriasis pathogenesis (further Y-model). Model of pathogenesis of psoriasis was formulated in previous work (Korotkii & Peslyak 2005). It was based on colonization of intestine by beta-hemolitic streptococci (BS) associated with hyperpermeability of intestinal walls for BS-proteins. Appreciable part of new results confirms this model. However results of new studies of psoriasis (Baker 2006a, Baker 2006b, Boyman 2007, Boyman 2004, Clark 2006, De Jongh 2005, Fry 2007b, Gudjonsson 2004, Lande 2007, Majewski 2003, Nestle 2005a, Nestle 2005b, Ritchlin 2007, Sabat 2007, Gumayunova 2009a, Nesterov 2009) and mechanisms of cellular immune protection (Bachmann 2006, Blander 2006, Blander 2007, Buckley 2006, Fasano 2005, Medvedev 2006, Merad 2007, Myhre 2006, Noverr 2004, Sozzani 2005, Toivanen 2003) demanded specification, development and more detailed formulation of model.

**Methods**

Publications with subject of examination of GIT and hepatobiliary system, investigation of intestine and biliary microflora or antiendotoxic and antistreptococcal immunity in psoriatics were searched and analyzed. Attention was also paid to publications with subject of prepsoriatic skin investigation and evaluation of events initiating the beginning of psoriatic plaque. Publications in Medline and Embase, and Russian publications were searched in Central Scientific Medical Library and in Scientific Electronic Library. Bibliography includes about 25% from total number of analyzed publications. Review and/or analysis study results are given in description of subprocesses. Real or prospective mutual influence of subprocesses is given at the end of description of dependent subprocess.

**Abbreviations and terms**

There are two tables in appendices: "Abbreviations" (Appendix 1) and "New terms" (Appendix 2). There are links on Wikipedia or references on works where these terms are well described in the tables.

**Novelty and hypotheses**

There are references where the subprocess is proved and/or described in details or hypothesis is summarized (Appendix 3).

Hypothesises about fractionation of blood phagocytes under chronic PAMP-load (Appendix 10) and about chemostatus of tolerized phagocytes (Appendix 4) are essentially new. Suggested systemic psoriatic process SPP is essentially new also.

# Model of pathogenesis. Systemic process.

### *Review of models of pathogenesis*

The majority of studies on psoriasis state that principal cause of the disease is in skin itself. There is far less number of authors, who don't restrict themselves with investigation of only local processes, but conduct search and find of evidences, proving that psoriasis is local sign of systemic psoriatic process SPP (Baltabaev 2005, Nesterov 2009, Stenina 2004, Baker 2006b, Gyurcsovics 2003, Ojetti 2006, Ritchlin 2007, Zaba 2009b). In work (Scarpa 2006) suggested largely simplified model of traffic of activated T-lymphocytes moving from gut-associated lymphoid tissue to skin and/or joints. It results in initialization and maintenance of psoriasis and/or psoriatic arthritis (it was suggested in 1999 for the first time). There were no assumptions on the mechanism of process in publications of the authors. On the contrary the authors of work thought that systemic process is caused only by disorder of production and circulation of bile acids (Baltabaev 2005). The authors are undoubtedly right, but it isn't the only one and main disorder which causes systemic process. It is discussed in details considering new model.

In the work (Nesterov 2009) a complex model of pathogenesis of the chronic dermatoses based on trigger role of blastocystosis (BLC) was offered. The blastocystosis can entail serious dysbiotic deviations in microflora of intestine and disturbance of its barrier function which occurs because of both blastocystosis and disbacteriosis of parietal microflora (Glebova 2007, Tan 2008). As a result too many toxic products of vital activity of blastocystes and microflora (it is not specified, which exactly) get to blood flow and support chronic endointoxication. The chronic endointoxication in its turn breaks the work of immune system and the balance of the oxidizers-antioxidants system. There is hyperactivation of processes of free-radical oxidation and decrease of antioxidatic activity. The author assumes that disturbance of work of these systems appears to be sufficient for the initialization and support of chronic dermatoses, including psoriasis.

Professor Korotkii NG and I suggested model of SPP pathogenesis in 2005 for the first time (Fig. 1). The model supposed that products of beta-streptococci moved to skin directly from blood. The products move to blood from intestine with hyperpermeability (Korotkii & Peslyak 2005).

However B.Baker et al. proved presence of BSPG (beta-streptococcal peptidoglycan) in psoriatic plaques only **inside** dermal and epidermal monocytes Mo and possibly in dendritic cells (Baker 2006a). On the other hand, other scientists proved that attraction of monocytes, their activation and secretion of TNF-alpha was defining link of escalation of psoriatic plaques (Clark 2006).

In his book-atlas Lionel Fry surveyed the role of BS in the initialization and support of guttate and chronic psoriasis and analysed the ways of BSPG getting to skin from the streptococci localized in tonsils (Fry 2005). A year later Barbara Baker, Anne Powles and Lionel Fry offered the model of pathogenesis (further BF-model) based on the new role of peptidoglycan (further PG) (Baker 2006b) (Fig. 2). Process of initialization of temporary guttate psoriasis is simulated on right part of figure. Streptococci temporary situated in tonsils produce toxins-superantigens. They activate TL of tonsils or skin lymph nodes. PG-specific TL are selected due to contacts with PG+Mo (transformed to PG+MoDC). Other TL become anergy or apoptotic.

Similar sequence of events can be observed in chronic psoriasis (left part of figure) if streptococci and/or streptococcal antigens stay in tonsils and/or intestine for a long time. Plaques appear after PG+Mo and PG-specific TL enter in derma. Autoantigen (e.g. keratin) has aggravating effect. BF-model doesn't give answers to the next two questions:

**1. Why do PG+Mo appear in skin though PG was endocytosed by Mo in other place of organism?**

**2. Why do PG+Mo become PG+MoDC and present PG?**

The first question is connected to disorder of traffic of immune cells. Let's answer this question. The answer to the second question will be given while discussing local processes (Part 2).

In the works (Zaba 2009a, Zaba 2009b) proved that dermal TipDC are main factors of psoriasis maintenance. TipDC are mature dendritic cells maDC which actively secrete iNOS and TNF-alpha and present unknown antigen Y.

Baker proposes that Y are parts of BSPG interpeptide bridge (Baker 2006a).

The authors of work (Zaba 2009a) showed that 90% of TipDC are CD11c+BDCA-1(-)DC of non-resident origin while 10% of TipDC are CD11c+BDCA-1+DC of resident origin.

Blood monocytes Mo and/or dendritic cells DC were proved to be precursors of larger part BDCA-1(-)TipDC. However two questions remain; which Mo and-or DC fraction is it; and why does this fraction act in such a way in case of psoriasis? TipDC is present in visible healthy prepsoriatic skin and in healthy people skin, but its quantity is rather less than in psoriatic derma. It means that TipDC-precursors moving from blood and their transformation to TipDC take place in homeostasis also.

Part PG+ dermal cells expresses CD68 (Baker 2006a). However increased expression CD68 is available at dermal inflammatory BDCA-1(-)DC in comparison with BDCA-1(+)DC (Zaba 2010). I.e. some from CD68+PG+ cells can be BDCA-1(-)DC.

Results received by Baker and Zaba correlate with each other. It is possible to assume that TipDC-precursors are PG+Mo.

BF-model has been developed recently in work (Valdimarsson 2009). Authors recognize role BSPG, but as well as earlier (Gudjonsson 2004), assume that, the basic antigens are parts of beta-streptococcal M-protein (BSMP). As the area of chronic localization of BS only tonsils are admitted. The essential role in their model is played by cross-reactivity between BSMP and autoantigens - peptides of epidermal keratins K14, K16 and K17.

In 2005-7 works of Garaeva ZS et al. were published. The authors showed that majority of psoriatics had high blood LPS level and also increased LPS-load on blood Mo and DC (Garaeva 2005, Garaeva 2007).

(See Appendix 8. PAMP-nemia in psoriasis. Study results.)

TLR2 and TLR4 are membranous PAMP ligands, while NOD1 and NOD2 are intracellular PAMP ligands (Fig. 3). As fragments of peptidoglycan PG (BLP (bacterial lipoprotein), LTA (lipoteichoic acids), MDP (muramyl dipeptide), DAP (diaminopimelic acid)) are also active PAMP so blood Mo and DC in SPP have combined (LPS and PG) PAMP-load giving synergic effect (Myhre 2006, Traub 2006). Term "load" hereafter means linkage, endocytosis and/or contact with PAMP. Abovementioned facts and recently published studies on action and transformation of blood Mo and DC (Auffray 2009, Buckley 2006, Serbina 2008) forced to reconsider model 2005 and to assume that unknown antigen move to skin mainly inside involved Mo and/or DC (Fig. 7).

Hyperpermeability of small intestine for bacterial products (subprocess SP1) and growth of bacterial populations (including beta-streptococci) on its walls (subprocess SP2) are main factors (as before) (Korotkii & Peslyak 2005). Let’s describe systemic psoriatic process SPP.

## Systemic psoriatic process SPP.
## Increased kPAMP-carriage of tolerized phagocytes.
## Increased (PG-Y)-carriage of R-phagocytes.

The process was partly studied earlier. However its name and influence on psoriasis development are suggested for the first time. Systemic psoriatic process SPP (Fig. 6, Fig. 7) was partly (SP1, SP2) suggested in (Korotkii & Peslyak 2005). The process involves GIT, hepatobiliary, vascular and immune systems, elimination organs; F - fragments of bacterial products containing PAMP (including kPAMP) chronically enter the blood as part of this process. Therefore PAMP-load (contact, linkage, endocytosis) on blood phagocytes (including Mo and DC) becomes constant, so tolerized phagocytes appear.

Term "phagocytes" hereinafter is used not only for neutrophils and monocytes (macrophages), but also for dendritic cells. Immature dendritic cells are also professional phagocytes (Nagl 2002, Savina 2007).

Phagocytes (neutrophils Neu, monocytes Mo, dendritic cells DC) exposed to chronic PAMP-load (contact, linkage, endocytosis) can be tolerized (Buckley 2006, Cavaillon 2006, Cavaillon 2008).

For SPP defining role play two properties of tolerized phagocytes (Neu-T, Mo-T, DC-T):

Property 1. Despite contact, linkage with PAMP and endocytosis of PAMP (as a part of endocytosed F-content) their chemostatuses are similar to nonactivated;

Property 2. As tolerization occurs under chronic kPAMP-load then there is kPAMP in endocytosed F.
Its degradation occurs slowly and not completely, i.e. kPAMP-carriage takes place;.

Tolerized phagocytes appeared to be (PG-Y)-carriers are named by R-phagocytes and are designated as Neu-R, Mo-R and DC-R. All R-phagocytes possess properties 1, 2 and 3.

Property 3. In endocytosed F-content is PG-Y. Its degradation occurs slowly and not completely, i.e. takes place (PG-Y)-carriage.

Property 1 - are assumed (App. 4), and properties 2 and 3 - are real (App. 2).

Mo-T and DC-T (incl. Mo-R and DC-R) migrate under the influence of chemokines as nonactivated Mo and DC (including during renewal of pool of tissue Mo and DC) because their chemostatuses are similar to nonactivated ones. Mo-T and DC-T (incl. Mo-R and DC-R) preserve earlier endocytosed F-content some time after entering the tissue and participate in one of two scripts, in particular, depending on level of cytokines-deprogrammers (Adib-Conquy 2002, Cavaillon 2008, Randow 1997):

1. Low level of cytokines-deprogrammers: Mo-T and DC-T (incl. Mo-R and DC-R) mainly preserve tolerance to F-content, gradually degrading it.

2. Increased level of cytokines-deprogrammers: Mo-R and DC-R quickly lose tolerance to F-content. They are activated, they mature and they are transformed: Mo-T in MF-T or MoDC-T (incl. Mo-R in MF-R or MoDC-R). Thus only DC-R and MoDC-R can mature and be transformed into maDC-Y (Fig. 6, local processes).

Until full degradation of PG-Y (inside of Mo-R and DC-R) can enter the second script during the first one (including maDC-Y formation).

Features of certain tolerized phagocyte:

- depth of tolerization which is defined by previous and present PAMP-load;
- volume and assortment of kPAMP (within F), containing in it;

Features of certain R-phagocyte:

- depth of tolerization which is defined by previous and present PAMP-load;
- volume and assortment of kPAMP (within F), containing in it;
- volume of PG-Y (within F), containing in it;

Volume of total (PG-Y)-carriage of all blood R-phagocytes is the important factor for SPP. This index depends on part of R-phagocytes among total number of blood phagocytes and from volumes of PG-Y containing in single R-phagocytes. Low part of R-phagocytes can be combined with high volumes of PG-Y containing in single R-phagocytes and, on the contrary, reprogrammed state of all blood phagocytes can be combined with small volumes of PG-Y containing in single R-phagocytes (SP7).

SPP severity is proportional to total (PG-Y)-carriage of Mo-R and DC-R in blood flow.
SPP severity also depends on total kPAMP-carriage of tolerized phagocytes in blood flow.

(See Appendix 4. Substantiation of hypothesis about chemostatus of tolerized phagocytes.)

(See Appendix 10. Fractionation of blood monocytes under chronic PAMP-load.)

Chronically increased blood level LPS is called endotoxemia and chronically increased LPS-load on phagocytes is the main sign of endointoxication. Let's name set of these deflections (in case of influence of several PAMP) "PAMP-nemia" on the analogy of term "endotoxinemia" (Iakovlev 2003), and set of PAMP causing these deflections - key PAMP, i.e. kPAMP. Major kPAMP in psoriasis are PG (including PG-Y) and LPS (see SP4).

Significant percent of tolerized Mo-T and DC-T appear among blood Mo and DC due to chronic kPAMP-load (SP8). Mo-T and DC-T can respond to homeostatic and proinflammatory chemokines and participate in homeostatic and intensive inflammatory renewal of pool of tissue MF and DC due to chemostatuses similar to nonactivated ones.

Initialization and maintenance of psoriasis with R-phagocyte provide carriage of PG-Y – peptidoglycan from psoriagenic bacteria. Normal skin becomes prepsoriatic skin due to Mo-R and DC-R participation in homeostatic (inflammatory) renewal of pool of dermal MF and DC. It is the main cause of initialization and maintenance of psoriatic plaque in local inflammation.

Source of PG-Y entering the blood can be not only GIT microflora, but also temporary local infections, for example tonsillar infection (SP6). Genetical deflection can be the cause of absence of respond to Mo and DC on PG-content in rare cases. So episodic and low level of (PG-Y)-load can be enough for initialization and maintenance of psoriasis (SP7).

All subprocesses forming process SPP are on scheme (Fig. 7).

Subprocesses SP2, SP4 and SP8 make SPP-basis. Internal dependencies of SPP-basis and separation on two components are on scheme (Fig. 8)

### *Subprocess SP1. Hyperpermeability of intestinal walls for F-content*

Subprocess is known; its influence on psoriasis was investigated.

F-content constantly moves to blood through intestinal wall. It is proved by blood tests of healthy people (Osipov 2001, Iakovlev 2003). F-content consists of macromolecules so small volumes move through inter-enterocytes contacts under control of barrier function. As a result of genetical predisposition or gastroenterologic diseases barrier function of inter-enterocytes transmission is interrupted and volume of F-content increases (Parfenov 2000, Parfenov 1999, Fasano 2005). Long-term therapy of glucocorticoids (hydrocortizone, betamethasone, etc.) or cytostatics (methotrexate, cyclophosphamide, etc.) also can promote the process. Additionally permeability for F-content can increase as a result of disturbed mechanism of transcytosis around Peyer's patches and lymphoid follicles through associated epithelium (Male 2006).

Ovalbumin (OVA) test was used to evaluate the level of intestinal permeability in children with psoriasis (Parfenov 1999, Rudkovskaya 2003, Stenina 2004). Standard blood OVA-level before OVA-load (chicken egg proteins) is close to zero and it shouldn't exceed 1 ng/ml after 3 hours of OVA-load. Initial average OVA-level of 30 children was 1,13 ng/ml and average OVA-level after OVA-load was 15,5 ng/ml (maximum - 104 ng/ml). Average OVA-level in children with advanced psoriasis was 35,4 ng/ml, in children with stable psoriasis - 5,1 ng/ml.

OVA-permeability depends on disease duration in children with advanced psoriasis. It sharply increases for first four months and then doesn't essentially vary. OVA-permeability decreased from 43,2 ng/ml to 23,1 ng/ml during treatment in patients with subacute psoriasis. There was no obvious correlation between OVA-permeability and psoriasis severity (index PASI).

In work (Hendel 1984) in vitro the accelerated proliferation of enteric enterocytes has been shown. In research there were 5 psoriatics (index of proliferation LI=20,26,29,36) and 5 patients from the control (LI = 13,17,23,26) at incubation time (0,5; 1; 2; 3 hours). That is probably caused by removing LPS to intestine from blood (Yimin 2000). Such acceleration defines incomplete differentiation of enterocytes at renewal of epithelium and, as consequence, disturbance of trans-enterocyte permeability.

100 patients with only psoriasis were examined. Decrease of acid production in stomach in stimulated phase was observed in 63% of patients (Khardikova 2005). The more psoriasis duration the more was deterioration of D-xylose and fats adsorption in small intestine. Decrease of trans-enterocyte adsorption of monosaccharides (D-xylose, Mannitolum), as a rule, correlates with increased inter-enterocyte permeability for macromolecules, including F-content. 60% (33/55) of psoriatics and only 3% (2/65) of control group patients had low D-xylose absorption in other study (Ojetti 2006). 2 patients with low D-xylose absorption suffered from celiac disease and 7 patients suffered from SIBO.

The subject of works was also relations between malabsorption syndrome (SM) and psoriasis (Harkov 2008, Harkov 2006, Harkov 2005). SM grade can be measured in grams of D-xylose excreted with urine within 5 hours after oral taking. SM was diagnosed in 83 psoriatics and 20 persons of control group. It was rather lower in psoriatics (average value SM=1,0) in comparison with standard (average value SM=1,8). They found inverse relationship between SM and severity (PASI) and type of psoriasis: vulgar (SM=1,2, PASI=14), exudative or arthropathic (SM=1,0; PASI=18), erythrodermic (SM=0,8; PASI=39). Also they found that the lower SM, the longer is disease duration. These results are represented in the dissertation (Shiryaeva 2007) in more details and with a larger group of psoriatics (103 patients).

Phytalbumin gluten is part of many graminoids. If the patient has predisposition (16% psoriatics) gluten has an adverse effect on enterocytes. Villuses atrophy and permeability of intestinal walls increases (Parfenov 1999, Abenavoli 2007). In average, level of IgA to tissue transglutaminase and gliadine in case of psoriasis is increased. Level of antibodies to gliadine was 14,8 (67 psoriatics) vs. 5,7 (85 – control group). This is a sign of latent celiac disease (Damasiewicz-Bodzek 2008). It is likely to be the result of beneficial influence of gluten-free diet on psoriasis course in patients involved by this sign (Wolters 2005).

SP1 depends on SP2. Some intestine bacteria are capable to influence function of inter-enterocyte transmission and transmission through LPS-influence and excrete toxins (Chin 2006, Fasano 2004, O'Hara 2008). So increased inter-enterocyte permeability can be a direct result of composition of enteric microflora, especially in SIBO (Husebye 2005, Ojetti 2006), including Gram(-)TLR4-active microflora.

SP1 depends on SP3. Inter-enterocyte transmission depends on quantitative and qualitative composition of bile entering the intestine (Assimakopoulos 2007). Chronic insufficiency of its entering interrupts barrier function of inter-enterocyte contacts and increases inter-enterocyte permeability, including for F-content (Khardikova 2000).

SP1 depends on SP4. Significant rising of blood LPS level interrupts intestine barrier function (O'Dwyer 1988).

SP1 depends on SP5.1

### *Subprocess SP2. Growth of populations of Gram(-) TLR4-active and Gram+ bacteria on small intestine mucosa.*

Subprocess is known; its influence on psoriasis was partly investigated.

Subprocess, as a rule, takes place within the limits of SIBO (small intestine bacterial overgrowth) (Bondarenko 2006, Maev 2007, Bures 2010, Husebye 2005, Sullivan 2003). However, subprocess SP2 may also not be accompanied with SIBO. Probably, the growth of parietal microflora frames conditions for SIBO.

Detailed researches of transient enteric microflora at psoriatics have been made for the first time by authors of the following works (Gumayunova 2009a, Nesterov 2009). Their results have confirmed the presence of serious dysbiotic deviations at BLC+ psoriatics (with blastocystosis) and at BLC(-) psoriatics (without blastocystosis). These results are given in detail in App. 6.

(See Appendix 6. Small intestine microflora in SIBO (without psoriasis) and in psoriasis.)

The information about blastocystosis and its role in the change of colic microflora can be found in the following works (Glebova 2007, Tan 2008).

Within the limits of subprocess SP2 special subprocess SP2.1 takes place (Fig. 8)

#### *SP2.1. Growth of populations of psoriagenic PsB*

The results of work (Gumayunova 2009a), forced to reconsider the exclusiveness of BS (beta-streptococci) offered in previous editions of this book, and to determine a more exact term, namely PsB - psoriagenic bacteria. For the definition of this term we will enter the following symbols:

**IB-Y** - interpeptide bridges of peptidoglycan Str.pyogenes: L-Ala(2-3) or L-Ala-L-Ser (Waksman 2005, p. 107-9). Bacteria Str.pyogenes are skin pathogens causing streptoderma, cellulitis, erypsipelas etc. (more than 100 million cases of skin forms of diseases annually) (Carapetis 2005). IB-Y - is marked in red color (Tab. 1)

**Y-antigen** - part(s) of interpeptide bridge IB-Y.

**PG-Y** - peptidoglycan of type A3alpha (Goldman 2008, p. 396) containing interpeptide bridges of type IB-Y (L-Ala (2-3) and-or L-Ala-L-Ser), but probably some others, too. The muropeptides formed at PG-Y degradation, contain entirely and/or fragmentary interpeptide bridges IB-Y. A good analysis of the set of such muropeptides is given in work (Billot-Klein 1996).

**PsB** - psoriagenic bacteria - Gram+ bacteria with peptidoglycan of type PG-Y (in Tab. 1 are marked in red color). All BS - the beta-streptococci having peptidoglycan PG-Y, also are PsB. The more share of interpeptide bridges of type IB-Y is in PG-Y, the stronger psoriagenity of PsB is. Only such PsB can show their psoriagenity, which are able to form stable colonies in such place of human body from which products of bacterial disintegration constantly get to blood flow. Tonsils and mucous small intestines are unfortunately the most suitable to this place.

**nPsB** – non-psoriagenic bacteria = Gram+ bacteria, or having PG type distinct from A3alpha, or having PG type A3alpha, but without interpeptide bridges of type IB-Y.

The left part of Tab. 1 contains the grouped information on Gram+ microflora from Tab. 3 (App. 6), added with statistics on carriage of particular bacterial species from other works. The right part of this table contains the general statistics Gram+ SSTI (skin and soft tissue infections) among all hospitalized patients from the USA, France, Germany, Italy and Spain for 2001 (Jones 2003). It is possible to assume that the role of particular bacterial species at SSTI among the population as a whole is similar.

**Tabl.1. Psoriagenic bacteria - priming and enteric carriage**

| Gram+ microflora | Enteric carriage (Gumayunova 2009a): 121 psoriatics | | 43 control healthy | | types of interpeptide bridges (IB) in peptidoglycan | SSTI shares = % of priming among Gram+ bacteria (Jones 2003): Enterococcus spp. | | Staphylococcus spp. | | Streptococcus spp. | | |
|---|---|---|---|---|---|---|---|---|---|---|---|---|
| | | | | | | 24 | 8 | 41 | 22 | 4 | 1 | 1 |
| | | | | | | E.faecalis | E.faecium | S. aureus | CoNS | VGS | Str.agalactiae | Str.pyogenes |
| | % carriers | lg CFU/ml | % carriers | lg CFU/ml | | L-Ala(2-3) | D-Asp (D-Asn) | Gly(5-6) | Gly(5-6); L-Ala - Gly(4-5); Gly(2-3) - L-Ser(1-2) etc. | L-Ala(1-3) | L-Ala(2) and L-Ala - L-Ser | L-Ala(2-3) and L-Ala - L-Ser |
| Bifidobacterium spp.: including | 93 | 5,3 | 40 | 2,41 | | | | | | | | |
| B.animalis subgr. Lactis | nd | | 23 [1] | | L-Ala(3) | # | | | | # | # | # |
| B.catenulatum | | | 92 [2] | | L-Ala(2) | # | | | | # | # | # |
| B.gallicum | | | nd [3] | | L-Ala - L-Ser | # | | | | # | # | # |
| B.pseudolongum | | | 4 [4] | | L-Ala(2-3) | # | | | | # | # | # |
| Bifidobacterium sp. nPsB | | | | | distinct from IB-Y | | | | | | | |
| Enterococcus spp.: including | 65 | 5,28 | 0 | 0 | | | | | | | | |
| E.faecalis | >40 [6] | nd | 48 [5] | | L-Ala(2-3) | # | | | | # | # | # |
| E.faecium | 25 | nd | 41 [5] | | D-Asp (D-Asn) | | | | | | | |
| Enterococcus sp. nPsB | nd | nd | | | distinct from IB-Y | | | | | | | |
| VGS | 30 | 5,74 | 0 | 0 | L-Ala(1-3) | # | | | | # | # | # |
| Str.pyogenes | 9 | 4,81 | 0 | 0 | L-Ala(2-3) or L-Ala - L-Ser | # | | | | # | # | # |
| Gram+ bacteria with IB distinct from IB-Y | | | | | | | | | | | | |
| Lactobacillus spp. | 84 | 4,66 | 19 | 2,54 | D-Asp (basically) | | | | | | | |
| S.epidermidis | 62 | 5,54 | 40 | 2,70 | Gly(5-6); L-Ala - Gly(4-5); Gly(2-3) - L-Ser(1-2) etc. | | | | | | | |
| S.aureus | 15 | 3,24 | 0 | 0 | Gly(5-6) | | | | | | | |
| Clostridium spp. | 20 | 5,2 | 0 | 0 | mDpm (direct) (basically) | | | | | | | |

## Notes to Tab. 1.

nd - there is no data;

The information on interpeptide bridges bases mostly on (Schleifer 1972), but also was specified and completed from later works.

VGS = Viridans group streptococci, found both at SSTI and in intestine, are basically Str.mutans, Str.salivarius, Str.sanguinis, Str.mitis, Str.anginosus (Goldman 2008, p. 299; Pokrovskii 2006, p. 24). They have the peptidoglycans containing IB of type L-Ala(1-3) (but also others) (Schleifer 1972, Tab. 20 and 22; Kilpper-Balz 1985, Tab. 1).

CoNS = Coagulase-negative staphylococcus, found at SSTI, are basically S.epidermidis, S.lugdunensis, S.haemolyticus, S.saprophyticus, S.sciuri (Crossley 2009, p. 385).

# - the initialization and support of psoriasis is possible.

In the absence of the data in work (Gumayunova 2009a) on the percentage of carriage of bacterial species in small intestines, the data on carriage in excrements of healthy adults is presented:

[1] - to this specie belong 23% of the strains which have been found in the samples taken from 30 healthy adults (Turroni 2009);

[2] - it is found in 44 of 48 (92%) samples at healthy (Matsuki 1999);

[2, 3] - (Dworkin 2006a, p. 327);

[4] - to this specie belong 4% of the strains which have been found on mucous large intestine at 30 healthy adults (Turroni 2009);

[5] - Enterococcus spp. in samples of adults (29 persons) from (Noble 1978). The quantity of E.faecalis as a rule exceeded 10-100 times the quantity E.faecium.

[6] - Estimated calculation was performed. Enterococcus spp. are found at 65% of psoriatics (Gumayunova 2009a), and E.faecium - at 25% of psoriatics (Gumayunova 2009b, Gumayunova 2009c), but the exact data on E.faecalis is absent. As as a rule there is simultaneous carriage of both E.faecalis and E.faecium, while E.faecalis itself is apparently found at more than 40% of psoriatics (65-25=40).

**If in the term prior to the occurrence of psoriasis:**

**firstly, there occurs Gram+ priming of dermal immune system by one of species of psoriagenic bacteria (during SSTI, Tab.1 columns) and, as consequence, the pool of mTL-Y is formed in priming places and also in derma and epidermis, and**

**secondly, on the mucous of small intestines one or several species of psoriagenic bacteria form significant colonies (lines of Tab. 1),**

**there appears a possibility of initialization and support of psoriasis (corresponding cells of Tab. 1 are noted # ).**

**Whether this possibility becomes reality or not - depends on the intensity of the SPP psoriatic process as a whole (**Fig. 7**) and on the genetic predisposition. Particular places of plaques are defined by local processes. How exactly? Assumptions will be made in (Part 2).**

Other species of streptococci also have peptidoglycan PG-Y: Str.pneumonia, Str.thermophilis, Str.equi, Str.dysgalactiae, Str.equisimilis, Str.zooepidemicus (Schleifer 1972, Tab. 20 and 22), however they are practically not found in GIT.

There is a probability to find in GIT the representatives of order Lactobacillales (Weissella spp., Leuconostoc spp.), part of which contain peptidoglycan PG-Y (Dworkin 2006b, p. 274-5) or representatives of family Micrococcaceae (Micrococcus spp., Arthrobacter spp., Kocuria spp., Rothia spp., Stomatococcus spp.), part of which are also containing peptidoglycan PG-Y (Dworkin 2006a, p. 947, 965, 979).

Tab. 1 contains the list of cores PsB, but in each specific case of psoriasis it is necessary to define not only E.faecalis, VGS and Str.pyogenes and the listed species Bifidobacterium spp., but also to carry out search of other significant colonies of PsB in small intestines.

It is necessary to notice that in work (Gumayunova 2009a) transient microflora was investigated, therefore it is only possible to make an indirect conclusion on the condition of parietal microflora. I.e. the absence of significant colonies of particular species in transient microflora most likely means the absence of its significant colonies also in parietal one and, on the contrary, detection of significant colonies of particular species in transient microflora means the probability of presence of its significant colonies also as a part of the parietal. But the absolute value (CFU/ml) may differ essentially.

On the other hand there are indirect and/or direct evidences of PsBP presence in skin and/or bloods in the absence of local PsB-infections in psoriasis and psoriatic arthritis (Molochkov 2007, Baker 2006a, Baker 2003, Baker 2000, Berthelot 2003, Cai 2009, El-Rachkidy 2007, Munz 2010, Rantakokko 1997, Sabat 2007, Weisenseel 2002, Weisenseel 2005). However the authors didn't think that intestine microflora is potential source of PsBP, so they didn't investigate it. Long-term PsB-infection with subsequent long persistence and/or deposition PsBP for example, in tonsillar tissue (Gudjonsson 2004), or directly in skin, was usually supposed. The authors of work (Korotkii & Peslyak 2005) were the first, who supposed that psoriagenic BS can be a part of intestine microbiocenosis.

***In addition about SP2.***

Enteric Gram(-) bacteria are E.coli, Bacteroides spp., Proteus spp., Acinetobacter spp., Klebsiella spp., Moraxella spp. (Tab. 3), and also Pseudomonada aeruginosa, Enterobacter spp., etc. The role of particular Gram(-) bacteria in creation of chronic LPS-load (SP4) defines their activity in relation to human phagocytic receptor TLR4. TLR4-activity, in particular, is defined by structure of lipid A (fragment LPS), and namely by phosphatic groups, length and number of acylic chains. So, for example, LPS of E.coli (six acylic chains in lipid) have the greatest TLR4-activity, and LPS of Bacteroides spp. and Pseudomonada aeruginosa (five acylic chains in lipid) activates TLR4 substantially less. Also the interaction of particular LPS with LBP, CD14 (membranous and soluble) and with coreceptor MD2, necessary for effective TLR4-activation (Munford 2008, Trent 2006, Wang 2010). The growth of colonies of Gram(-) TLR4-active bacteria on small intestine mucous defines income growth of F-content with TLR4-active LPS in the systemic blood flow.

Some influence of Gram(-) on psoriasis was proved in works (Garaeva 2005, Garaeva 2007, Gyurcsovics 2003).

Apart from bacteria entering small intestine together with nutrition (more often transit), there are two more constant sources of enteric populations: microflora of upper respiratory tracts moving in the case of interruption of stomach-acid barrier and microflora from distal parts moving in the case of interrupted peristalsis (Husebye 2005). Growth of enteric populations confines by: antibacterial action of bile acids (SP3); acidity of gastric juice (Khardikova 2005, Gyurcsovics 2003); peristalsis providing fast moving of microorganisms to distal parts (Husebye 2005); release of immunoglobulins; enzymatic activity; state of intestine epithelium and slime excreted by goblet cells (it contains inhibitors of microorganism growth).

Indirect evidence of PsB presence in upper parts of small intestine is their pathogenic presence in gallbladder microflora at its various diseases. In unsterile bile Streptococcus spp. are present in 15% of cases, E.coli – in 20%, Enterococcus spp. – in 18% (Savitskaia 2003, De Waele 2003, Matin 1989, Roa 1999).

Interesting results were received in study (Saxena 2005). The authors supposed that chronic subclinical streptococcal infection (obscure localization) is responsible for chronic psoriasis. 30 patients with moderate-to-serious psoriasis were examined. The majority of patients have been suffering from psoriasis for more than 5 years (21/30). They were treated with various methods without essential recover and they had frequent recurrences. Initial ASLO titer more than 200 IU/ml was found in 15 patients; positive reaction to C-reactive protein in culture of smear from fauces: (2 - GAS; 6 - Str. viridians) was found in 7 patients. Studies lasted two years. The patients were administered benzathine penicillin (1,2 mln U) i/m every fortnight for the first 24 weeks. During 25-48 weeks they were administered benzathine penicillin (1,2 mln U) every month. Marked recover was observed after 12 weeks (average initial PASI – 32,7; after 12 weeks – 19,1; after 24 weeks – 8,7; after 36 weeks – 3,5; after 48 weeks – 1,5). Patients were followed up for 2 years and all of them had remission.

The same group of researchers (Saxena 2010) has studied the effect of long oral reception of azithromycin. 30 of 50 patients (with psoriasis severity from moderated to serious) received azithromycin during 24 courses (two weeks each course). For 4 days - oral dose of 500 mg unitary, then 10 days break, i.e. totally 48 weeks. The other 20 patients received tablet of vitamin C. Appreciable recovery by PASI index has been detected from the 12th week at the majority of the patients receiving azithromycin. In the end of the 48th week, 18 patients (60%) had excellent recovery, 6 patients (20%) – good, and 4 patients – moderate. PASI 75 (decrease of not less than 75% from the initial level) was 80% (i.e. it was observed at 24 patients). No essential changes have been found in the control group.

SP1 and SP2 should be considered together because their combination does influence SP4. In particular, SP4 can be observed at significant SP1 level and insignificant SP2 level and inversely.

Possible influence of changed intestine microflora on immune system at various diseases (microflora hypothesis) is analyzed in works (Huffnagle 2008, Round 2010). Subprocess SP2 in SPP plays a main role, that corresponds to the given hypothesis.

SP2 depends on SP3. Bile has bactericidal properties to many non-commensal enteric bacteria and ability to inactivate F-content or to degrade it to non-toxic fragments thereby reducing level of entering the blood (Gunn 2000, Gyurcsovics 2003). Decrease of production or quality of bile and/or irregularity of its secretion result in decrease of bile bactericidal properties. It promotes bacterial growth in small intestine (Begley 2005, Hofmann 2006).

SP2 depends on SP5.1.

SP2 depends on SP6. PsB (streptococci, enterococci and bifidobacteria) are facultative nonpathogenic inhabitants of small intestine mucosa (Bouhnik 1999, Ciampolini 1996, Conte 2006, Hayashi 2005, Sullivan 2003, Wang 2005). Some of them move from oral cavity and fauces mucosa (where they are commensals) to upper parts of small intestine. They can also move in case of gingivitis (Gubina 2004) or tonsillar infections (Gudjonsson 2003).

### *Subprocess SP3. Disturbance of production and/or circulation of bile acids (BA).*

Subprocess is well-known. It was investigated in patients with psoriasis Baltabaev 2005, Matusevich 2000, Gyurcsovics 2003). Subprocess SP3 is essential part of vicious cycle (letter A) because it directly influences SP1 and SP2.

Disturbance of enterohepatic circulation can be a result of weakening of hepatic function of extraction and conjugation of BA from portal blood. So some part of BA constantly enters peripheral blood. BA pool is reduced, if liver possibilities on BA generation are limited. So the level of BA in bile is low. High blood BA level can be toxic for tissues and cause rising of permeability of membranes and local inflammation. Cholanic acid derivatives can interrupt integrity of blood vessel walls, increase their permeability and dilate lumens of vessels of derma papillary layer (i.e. to influence rate of entering tissues by phagocytes).

(See Appendix 7. BA in psoriasis. Study results.)

SP3 depends on SP5.2. Chronic diseases or congenital defects of hepatobiliary organs result in depression of BA production. Obstructive jaundice or gallbladder excision completely stops their income to intestine. Chronic overload of liver by F-content recycling also reduces BA production.

### *Subprocess SP4. PAMP-nemia. Increased kPAMP-load on blood phagocytes. Increased kPAMP level in blood. The major kPAMP are PG and LPS.*

Subprocess is well-known for LPS (including in psoriasis). But there were only few studies on PG and none on psoriasis (Fig. 8). Note that in work (Valdimarsson 2009) the aim of current investigations was announced: to define both: PG-level and frequency of PG+Mo in the blood of psoriatics.

Previously endotoxins were considered as any products of bacterial degradation (as against exotoxins - toxic secreted products of bacterium vital activity). But now term "endotoxin" means LPS, term "endotoxinemia" means increased blood LPS level and term "endotoxicosis" means increased LPS-load on phagocytes (Tabolin 2003, Chesnokova 2006, Iakovlev 2003). LAL-test is used for measuring blood free LPS level. Standard values are in range from 0 to 1 Eu/ml (0,1 Eu/ml or 10 pg/ml on average). It depends on rate of LPS entry from intestine into portal blood and quality of LPS-elimination made by hepatobiliary system (up to 95% of LPS are destroyed by system of hepatic macrophages before they enter systemic blood flow and excreted with bile). Additionally it depends on porto-caval shunts (portal blood entry in systemic blood flow, i.e. not through liver) and activity of antiendotoxic immunity (Iakovlev 2003).

(See Appendix 8. PAMP-nemia in psoriasis. Study results.)

There are data which allow supposing that obligate kPAMP in psoriasis is PG (including PG-Y) (Baker 2006a, Baker 2006b). PAMP-nemia and endotoxinemia have the same cause. The cause is combined action of SP1 and SP2. Besides, overload and/or disturbance of detoxication systems (SP5) influence its level. Thus, detoxication systems do not have enough time to eliminate excess volume of F-content entering the blood, or they are weakened because of diseases and they don't stand usual load. It is important that kPAMP-load increases to such extent, that it results in reprogramming of significant percent of blood phagocytes. Chronically increase results in their tolerance to kPAMP-load (Buckley 2006, Medvedev 2006). Actually PAMP-receptors (TLR4 in LPS-load, TLR2 and NOD2 in PG-load) of tolerized phagocytes are temporarily blocked (Hedl 2007, Myhre 2006, Nakatani 2002).

(See Appendix 9. Definition of concept of PAMP-load on phagocytes)

PAMP-load on blood phagocytes increases while blood PAMP-level raises slowly during initial stage of PAMP-nemia. PAMP-load can be critical for psoriasis possibility at this stage. PAMP-consumption (phagocyte-dependent and phagocyte-independent) can't stand PAMP-income at the second stage. Blood PAMP-level becomes even higher and PAMP-level can be critical for psoriatic arthritis possibility.

Endotoxinemia can accompany any Gram(-) infections and become the cause of many systemic diseases (Tabolin 2003, Iakovlev 2003), but not psoriasis. LPS-load on phagocytes influences volume and F-content composition (SP8) and, as a result, influences the volume and types of cytokines secreted after possible loss of tolerance (deprogramming) from R-phagocytes in tissues, including because of synergy between LPS and PG (Myhre 2006, Traub 2006). Although TLR4 is membranous receptor, but linkage with LPS and subsequent endocytosis make complex TLR4-LPS able to enter endosomes and cell (Coll 2010, Husebye 2006). However LPS-load itself can't provide local psoriatic process, so LPS has aggravating effect in Y-model of pathogenesis.

Severity of PAMP-nemia also can be estimated on shares of fractions of tolerized Mo-T and DC-T. Than more these shares are - the more severity of PAMP-nemia is (App. 10).

Within the limits of subprocess SP4 special subprocess SP4.1 takes place (Fig. 8)

### *SP4.1. (PG-Y)-nemia.*

(PG-Y)-nemia is called raised (PG-Y)-load on blood phagocytes in combination to increased level PG-Y in blood. All populations of Gram(+) and Gram(-) bacteria (not only PsB) of zones of hyperpermeability of intestine mucosa, define total PG-income into blood and total PG-load on phagocytes. But only growth of PsB populations (SP2.1) is at the bottom of (PG-Y)-nemia.

Existence of fractions of tolerized Mo-T and DC-T is necessary, but not sufficient condition of occurrence of subfractions Mo-R and DC-R (SP8.1). Only essential level of chronic (PG-Y)-load can entail the occurrence of such subfractions, and also provide necessary level of total (PG-Y)-carriage of Mo-R and DC-R for possible initialization and support of psoriasis.

SP4 depends on SP1 and SP2. If there are no local and/or systemic bacterial infections the main source of F-content entering the blood is the microflora of small intestine mucosa.

SP4 depends on SP5. Normal state of detoxication systems inhibits PAMP-nemia increase and, on the contrary, their diseases and/or overload promote its increase. Complex therapy of psoriasis always includes investigation of hepatobiliary organs, kidneys and urinary tract and their treatment if it is required.

SP4.1 depends on SP2.1. (PG-Y)-nemia severity is proportional to growth of PsB-populations on small intestine mucous.

SP4.1 depends on SP6. Tonsillar PsB-infection (as well as any other local PsB-infection) results in temporary (PG-Y)-nemia: temporary increase of (PG-Y)-level and (PG-Y)-load.

### *Subprocess SP5. Overload and/or disorders of detoxication systems.*

Subprocess is well-known (Sharapova 1993, Molochkov 2007). Detoxication systems role in psoriasis was described earlier (Korotkii & Peslyak 2005); also see Appendix 8. Taking into account various influence of the components of the given subprocess on other subprocesses, we will allocate disorders in intestine (SP5.1) and in hepatobiliary system (SP5.2):

#### *SP5.1. Intestine*

Researches show that functional and structural disorders in intestine aggravate psoriasis, and the combined treatment aimed at normalization of its functioning lead to good and long-term results (Garaeva 2005, Harkov 2008, Shagova 2009, Pietrzak 2009, Pagano 2009). Certainly any helminthiasis aggravates the current of psoriasis, in particular, opisthorchiasis (Matusevich 2000, Khardikova 2005, Kuranova 2009) or parasitic disease, in particular, blastocystosis (Gumayunova 2009a, Nesterov 2009). Successful treatment of helminthiases and/or parasitic diseases and correction of dysbiotic deviations lead to much more successful and stable results in psoriasis treatment.

#### *SP5.2. Hepatobiliary system*

Chronic endotoxinemia in psoriasis (Garaeva 2005) can result in dysfunction of liver. Dysfunction severity depends on its level, duration and concomitant diseases (Matusevich 2000). The organic pathology of biliary tract and/or its functional disorders aggravate psoriasis, and cholestasia degree correlates with PASI (Gyurcsovics 2003, Ibliyaminova 2009).

It is well known that liver diseases aggravate psoriasis and make its treatment more complicated. For example, symptoms of non-alcoholic fatty liver disease (NAFLD) have been found at 47% of psoriatics (of 130), while in the control group of healthy - only at 28% (of 260). Psoriatics with NALFD symptoms have higher PASI (14,2 against 9,6), than psoriatics without NALFD symptoms (Gisondi 2009). The review of works with similar results has been published recently (Wenk 2010). It should be mentioned that implication of NALFD can correlate with the disturbance of circulation and transportation of bile acids (Trauner 2010).

**In addition about detoxicating**

Percent of LPS linkage in liver decreases from 90-95% (in healthy patients) to 24% in patients with obstructive jaundice. Experimental study on influence of endotoxinemia on liver and kidneys of experimental rats is described in monograph (Mishnyov 2003). Increscent irreversible changes in structure and function of these organs caused by increase of endotoxin injection dose (LPS E. coli) were studied.

The authors of work (Yimin 2000) showed that enterocytes and colonocytes are important factors of LPS excretion from organism. The process was simulated and studied in details on experimental rats. Rats were intravenously injected high dose of LPS. Previously excretion of LPS absorbed by Kupffer cells of liver along with bile to intestine was studied. Authors of the same work also studied the mechanism of LPS excretion (free or absorbed by macrophages) to intestine by translocation through own plate of mucosa (lamina propria) and basolateral basis directly to enterocytes (colonocytes). Enterocytes (colonocytes) moved from crypts to the top of villus after 4-5 days. They are exfoliated, taking collected LPS away to intestine lumen. Translocation of LPS injected in intestine to the blood was investigated. The results showed that translocation isn’t made through enterocytes or inter-enterocyte space.

<u>SP5 depends on SP4.</u> PAMP-nemia provides constant load on all detoxication systems.

### *Subprocess SP6. Tonsillar PsB-infection*

Subprocess is well-known; it was repeatedly investigated for BS in psoriasis. Work (Prinz 2009) represents the modern detailed review of the facts and assumptions of the influence of streptococcal infection on psoriasis.

Tonsillar PsB-infection (as well as other local PsB-infection) provides temporary, but significant PsBP (including PG-Y) entry in blood. The detailed analysis of these events for BS-infection may be found at (Baker 2006b) (Fig. 2).

> 208 patients with plaque psoriasis were observed for a year. Researches took swabs from fauces in the beginning of follow-up and during each aggravation of psoriasis or throat inflammation (61 cases). 20,7% of patients (percent corresponds to average value) were asymptomatic carriers of BS with M-protein of A, C and G groups. BS provoked throat inflammation with subsequent aggravation of psoriasis in 9% of cases if inflammation lasted not less than 4 days. BS didn't cause throat inflammation and didn't provoke psoriasis aggravation in 20% of cases (Gudjonsson 2003).
>
> Questionnaire of 74 psoriatics who underwent tonsillectomy has shown that within 4.5 years after this operation, one third of psoriatics had full purification of skin, and another third of psoriatics had significant remission (Nyfors 1975). Three patients recovered from psoriasis after tonsillectomy. Identical TL-receptors of TL in tonsils and psoriatic plaques was previously shown (Diluvio 2006). The authors of review work (Wilson 2003) carried out analysis of influence of antibiotics and tonsillectomy on psoriasis course. They concluded that despite variety of positive results additional control studies are needed.
>
> Interesting results have been received in work (Wang 2009). It has been proved that L-forms of bacteria (i.e. bacteria at which there cellular wall is partly or completely absent) are defined in smear from fauces at 74% (92 of 124) psoriatics (65 - with guttate, 59 - with chronic), only at 24% (19 of 81) patients only with adenoid disease and at 6% (5 of 79) healthy of control group. L-forms Gram+ and Gram(-) bacteria have been found, at 79% (73 of 92) psoriatics dominant species were Gram(-) bacilli (Chryseomonas luteola, Burkholderia cepacia, Enterobacter cloacae etc.), and at 21% (19 of 92) psoriatics Gram+ coccuses, mainly Str.pyogenes, were dominant species, After treatment by antibiotics only 9% of psoriatics still had carriage of L-forms of bacteria. These authors have not published the data about the correlation of the condition of psoriasis with carriage of L-forms of bacteria yet, neither about condition of patients before and after treatment .

About 30% of patients with primary guttate psoriasis recover spontaneously. However it regenerates to chronic plaque psoriasis at once or after remission in 70% of cases (Baker 2000, Williams 1976). Probably tonsillar PsB-infection causing primary guttate psoriasis, also becomes a source of stable PsB-populations in upper parts of small intestine (SP2). It can result in development of chronic psoriasis. Chronic psoriasis aggravation during tonsillar PsB-infections is caused by because significant additional PsBP (including PG-Y) entry in blood.

The authors of work (Fry 2007a) discussed preventive streptococcal vaccination for risk groups (genetic or family signs) if CPs hasn't yet begun.

### *Subprocess SP7. Deviation in intracellular signal path "from MDP recognition through NOD2-ligand to chemostatus change" (<1%).*

MDP is PG fragment formed after its intracellular degradation. MDP is the main ligand of intracellular receptor NOD2 (Girardin 2003). "Wrong" blood phagocytes appear when intracellular signal path of bone marrow stem MoDP (and blood Mo and DC) is weakened or interrupted: from contact between MDP and NOD2 to cell chemostatus change (Fig. 11). DNA changes cause congenital or acquired deviation (as a result of mutagenesis).

"Wrong" phagocyte without chronically increased PG-load acts as tolerized phagocyte because endocytosis small volume of PG doesn't change its chemostatus. If MoDP has such deviation psoriasis can develop at low (PG-Y)-load as formed in blood flow PG-Y(+)Mo and PG-Y(+)DC can keep nonactivated chemostatuses and participate in subprocess LP1.1.

Deviation disappears after allogenic transplantation of bone marrow when MoDP (stem precursors DC and Mo) are replaced. But transplantation is made for more serious indications than psoriasis. Blood level of "wrong" phagocytes of recipient decreases and blood level of normal phagocytes of donor increases.

Course of serious disease which was the cause of transplantation is relieved and psoriasis subsides forever or until the moment when kPAMP-load (SP4) and/or (PG-Y)-load (SP4.1) will increase and SP8 will begin.

The authors of work (Kanamori 2002) reported 8 patients with long and full remission of psoriasis after allogenic bone marrow transplantation. They also reported that psoriasis or psoriatic arthritis developed in recipients after similar transplantation from donors with these diseases. The authors of work (He 2005) described 3 similar patients with serious psoriasis and leukemia. The authors consider that remission of these diseases was due to elimination of genetic deviations, including chromosomal instability and defect of DNA reparation of hemopoietic cells.

50000 bone marrow transplantations are made in the world every year; about 30% of them are allogenic (data of 2002). Prevalence of psoriasis among recipients is 2-3% (not low than in population). If each allogenic bone marrow transplantation result in remission of psoriasis the number of such cases would be not less than 300-450 in a year. It wouldn't remain unnoticed, but such cases are single (less than 1%). Therefore genetic deviations of cells of bone marrow origin aren't the cause of psoriasis in overwhelming number of cases.

NOD2-mutations provoke susceptibility to Crohn's disease (Kullberg 2008). NOD2-mutations can be the cause of increased risk (more than 10%) of psoriasis development in patients with Crohn's disease (Piruzian 2009, Christophers 2006).

In work (Chatzikyriakidou 2010) statistically significant differences (P=0,003) for intracellular protein IRAK-1 in rs3027898 polymorphism between patients with psoriatic arthritis (29 people) and the control group of healthy (66 people) have been found for the first time.

In the following works (Gudjonsson 2009, Nair 2009) are described recently found some significant differences for protein A20 (=TNFAIP3) and binding it protein TNIP1 in polymorphisms rs610604 (OR=1,19, p = $9*10^{-12}$) and rs17728338 (OR = 1.59, p =$1*10^{-20}$) between psoriatics (more than 5000) and control group of healthy people (more than 5000). Normally, interaction of these proteins blocks signal paths TRAF6, RIP1 and RIP2 for prevention of the superfluous inflammatory response (Coll 2010) (Fig. 11). The found out differences may promote increase in share of blood Mo-R and DC-R at the same level of PAMP-load that raises probability of occurrence and support of psoriasis.

### *Subprocess SP8. Growth of tolerized fractions Mo-T and DC-T. Increased kPAMP-carriage.*

This subprocess (incl. SP8.1) is the final stage of systemic process SPP and it is formulated here for the first time. Chronically increased kPAMP-load provides tolerization (reprogramming) of some Mo and DC. They stop to recognize F-content (fragments of bacterial products containing PAMP, including kPAMP) as pathogenic material and become F-carriers.

Mo-T and DC-T are abbreviations of tolerized Mo and DC.

**Assumption 1: Tolerized Mo-T and DC-T continue to participate in renewal of pool of tissue MF and DC (LP1.1), degrading F-content slowly and incompletely.**

Why can it happen? Nonactivated Mo and DC under temporary kPAMP-load activate, secrete more TNF-alpha and their chemostatuses temporarily undergoes significant changes (Buckley 2006, Møller 2003). However activated Mo and DC tolerize and almost stop secreting of TNF-alpha under chronically increased PAMP-load. So, may be chemostatuses of tolerized Mo and DC became similar to nonactivated?

**Assumption 2: Chemostatuses of Mo-T and DC-T are similar to chemostatuses of nonactivated Mo and DC.**

Possible intracellular signal paths from PAMP-receptors to TNF-alpha secretion and CCR7-expression are given in Appendix:

(See Appendix 4. Substantiation of hypothesis about chemostatus of tolerized phagocytes.)

Transformation of certain nonactivated Mo in tolerized Mo-T (and, in particular, in Mo-R) depends on time of stay in systemic blood flow under chronic PAMP-load. At first the small number of occurrings with F-content converts nonactivated Mo into activated, and then the subsequent occurrings convert activated Mo into tolerized Mo-T (App. 10).

(See Appendix 10. Fractionation of blood monocytes under chronic PAMP-load)

The more certain Mo (or DC) meets F-content, the more of formed intracellular protein IRAK-M occures. IRAK-M temporarily blocks intracellular signal paths going from concrete PAMP-receptors (TLR4 at LPS-load, TLR2, NOD1 and NOD2 at PG-load) (Buckley 2006).

The share of fractions tolerized Mo-T and DC-T in blood flow is proportional to total kPAMP-load on Mo and DC. This kPAMP-load takes place in blood flow (SP4) and, possibly, in bone marrow (SP10).

Within the limits of subprocess SP8 special subprocess SP8.1 takes place (Fig. 8) .

#### *SP8.1. Growth of subfractions Mo-R and DC-R. Increased (PG-Y)-carriage.*

Subprocess SP8.1 occurs only if SP4.1 operates, i.e. when as a part of kPAMP-load on blood phagocytes (SP4) is (PG-Y)-load.

Mo-R is PG-Y(+)Mo-T. Similarly DC-R is PG-Y(+)DC-T. Thereby Mo-R and DC-R are subfractions of fractions of tolerized Mo-T and DC-T. Mo-R and DC-R can be also carriers of others PAMP (besides PG-Y). Total (PG-Y)-cariage of Mo-R and DC-R in blood flow is proportional to:

(a) shares of fractions tolerized Mo-T and DC-T and

(b) total (PG-Y)-load on Mo and DC in blood flow (SP4.1) and, possibly, in bone marrow (SP10).

SPP severity is proportional to total (PG-Y)-carriage of Mo-R and DC-R in blood flow.

Total (PG-Y)-carriage should be normed on total amount of blood of the patient. Such norming will allow to compare SPP severity at psoriatics of various weight. I.e. SPP severity is proportional to total (PG-Y)-carriage of Mo-R and DC-R in one ml of blood. Y-antigen is part of the interpeptide bridges IB-Y necessarily containing in PG-Y.

Therefore SPP severity also is proportional to total Y-carriage of Mo-R and DC-R in one ml of blood.

***In addition about SP8.***

According to part of PG+Mo among total quantity (Baker 2006a) not more than 5-10% of blood monocytes are Mo-R in psoriasis. Blood CD14+CD16+Mo are more likely to be reprogrammed. Possible sequence of events leading to tolerization (reprogramming) and formation of CD14+CD16+Mo-R is given at scheme.

It is supposed that as a part of kPAMP-load is (PG-Y)-load (Fig. 9)

Let S2 = {CCR1, CCR2, CCR5, CXCR4} - assortment of the chemokine receptors CD14+CD16+Mo which is responsible for attraction in tissue.

Then S2+ = {CCR1+, CCR2+, CCR5+, CXCR4+} - at inactive or at tolerized chemostatus, and S2(-) = {CCR1(-), CCR2(-), CCR5(-), CXCR4(-)} - at active chemostatus.

Normal CCR7(low)S2+Mo has chances to accept kPAMP-load many times (1) because low CCR7 expression ambiguously influences its action. It can bring F-content to lymph node (2) or it can degrade F-content, remaining in blood and return to initial chemostatus (3). However the level of intracellular blocking protein IRAK-M will slightly increase.

If subsequent kPAMP-load takes place in the near future each cycle of transformations (1,3) will increase IRAK-M level. (IRAK-M level gradually decreases in the absence of kPAMP-load). As soon as IRAK-M level becomes blocking, transformation (4) to F+CCR7(-)S2+Mo-R will take place. Such Mo-R is mainly attracted to non-lymphatic tissues (5).

Monocytes CD14+CD16+Mo has lower CCR7 expression than CD14++Mo does, but high potential of phagocytosis remains (Auffray 2009, Serbina 2008). They well express CCR2, CCR5, CXCR4 and CX3CR1 (Ancuta 2003, Serbina 2008), most actively secretes TNF-alpha and iNOS and also is more capable (in comparison with CD14++Mo and CD14(-)CD16+Mo) to be quickly transformed (without division) to MoDC (Ziegler-Heitbrock 2007). CD14+CD16+Mo-R seem to be precursors of BDCA-1(-)TipDC (90% of total number of TipDC (Zaba 2009a)).

This assumption correlates with data that many inflammatory BDCA-1(-)DC in psoriatic derma coexpress monocytic markers CD14 and CD16. These markers can remain during fast transformation CD14+CD16+Mo into BDCA-1(-)DC after their attraction into derma from blood flow (Johnson-Huang 2012).

The results of transcriptome of skin BDCA-1(-) DC research have shown the increased level of CCR2, CD14, CD64 that quite corresponds to such assumption (Zaba 2010). The essential expression of CD14 available at BDCA-1(-)DC, makes it improbable that they originate from blood DC which practically do not express this receptor (Piccioli 2007). On the other hand assumption of their origination from CD14++Mo, which practically lose FPRL1 receptor at the transformation in MoDC, is improbable (Yang 2001).

Blood CCR2+Mo are precursors of TipDC during intracellular Gram+L.monocytogenes-infection (Serbina 2008, Serbina 2003). Mo (and their derivatives MoDC) are activated only through intracellular receptor NOD2. Similar activation is likely to arise during transformation PG+Mo-R in MF-R and in MoDC-R.

Transformation (activation or even tolerization) of Mo and DC can begin during their hemopoiesis in bone marrow (SP10). Senescent neutrophils Neu-R coming back to bone marrow to perish (SP9) (Martin 2003). Neu-R can bring in bone marrow F-content endocytosed in blood. F-content is situated in extracellular space after their apoptosis (Fig. 12) (Tacke 2006). Reprogramming of bone marrow phagocytes finishes in 1-3 hours after LPS-load (Fitting 2004). Details are given in Appendix.

(See Appendix 5. Relay variant. Transformation of Mo and DC by means of apoptotic Neu in bone marrow.)

If SP7 operates, all Mo and DC can be considered reprogrammed. Endocytosis PG-Y (within F-content) R-phagocytes increase (PG-Y)-carriage level. (PG-Y)-load should be increased in case of SP4 and can be episodical in case of SP7 to make total (PG-Y)-carriage significant. Sufficient rate of (PG-Y)-entering inside Mo-R and DC-R into derma will be supported for initialization and maintenance of psoriasis only under these conditions.

When Mo-R and DC-R enter tissues their action also essentially differs from action of normal Mo and DC. They are «delayed-action mines» because they contain non-degraded PG-Y (Fig. 6).

Tolerance of R-phagocytes to the major kPAMP (e.g. LPS) can cause tolerance to minor kPAMP (e.g. flagellin) if intracellular signal path of minor kPAMP completely coincides with one of such ways of the major kPAMP (De Vos 2009). I.e. endocytosed material can contain not only major kPAMP, but also minor kPAMP and chemostatuses of tolerized phagocytes became similar to nonactivated also (Hedl 2007, Medvedev 2006).

Blood tolerized Mo-T and DC-T chronically contact and endocytose LPS (free or bound). Their chemostatuses became similar to nonactivated and they don't bring endocytosed content to lymph nodes and/or spleen. Thereof level of T-independent humoral immune response to LPS decreases (Tabolin 2003).

Lowered expression HLA-DR at blood monocytes is a sign of immune supression. It happens because of their reprogramming, for example during CARS (Cavaillon 2008, chapter 15). Expression HLA-DR also is lowered and at psoriatic Mo (it is essential at CD14++ Mo and slightly less at CD16+Mo) (Zheng 1997). The expression is restored by an incubation of monocytes with IFN-gamma.

And as IFN-gamma is cytokine-deprogrammer these results confirm reprogramming of psoriatic Mo. Hypersensitivity of peripheral blood Mo to low LPS concentration results in intensive secretion of TNF-alpha, IL-1beta and IL-6 in psoriatics (in comparison with healthy people) (Mizutani 1997).

Blood psoriatic Mo spontaneously secreted 2-4-fold levels of IL-1alpha, IL-1beta, IL-6 and IL-8 (Teranishi 1995). They reached maximum after 48 hours (Okubo 1998). Blood psoriatic Mo spontaneously secreted the following quantities of cytokines (after 20 hours in supernatant): in patients with serious psoriasis - TNF-alpha (150 pg/ml) and IL-6 (184 pg/ml), in patients with moderated psoriasis - TNF-alpha (63 pg/ml) and IL-6 (30 pg/ml) against in healthy people - TNF-alpha (37 pg/ml) and IL-6 (24 pg/ml) (Bevelacqua 2006).

Blood Mo have similar status in atopic dermatitis when 2-10-fold levels of IL-6, IL-10, TNF-alpha are spontaneously secreted (Mandron 2008). PAMP-load (especially under PG-load) always results (Hadley 2005) in increase of part of blood CD14+ Mo (in psoriatics - 90-99% against in healthy people - 85%) (Teranishi 1995, Zaba 2009a).

The facts listed above mean that blood Mo at psoriasis basically are activated. However it does not exclude existence of other fractions of monocytes, as, in particular tolerized fraction (including subfraction Mo-R). (App.10)

Treatment for psoriasis with MDP-immunomodulators - licopid (Korotkii 2001) and GMDP (Williamson 1998) - gives positive results. They activate monocytes and neutrophils, increase their englobing and destroying properties, microbicidal function (Pinegin 1998, Karsonova 2007). Since MDP is PG-fragment kPAMP-load on blood Mo and DC significantly increases after MDP-immunomodulator taking (half-life - 4,3 hours) in comparison with previous chronic PG-load. It activates not only usual Mo and DC, but part of already tolerized Mo-T and/or Mo being tolerized (Hedl 2007). Reprogramming stops for a while and/or completely and all Mo and DC degrade F-content faster. Increase of blood level of cytokines-deprogrammers IFN-gamma, IL-12, GM-CSF, also caused by MDP, prevents tolerization (Pinegin 1998). As a result part of reprogrammed Mo-R and DC-R and F-content volume decreases. According Y-model this process causes psoriasis remission.

HIV (human immunodeficiency virus) is known to increase risk of development of serious and extended psoriasis (Fry 2007b, Leal 2008). TLR2-receptor activation increases CCR5-expression in HIV+Mo more, than in normal Mo (Heggelund 2004). Taking into account increased CCL5 level in psoriatic skin (De Groot 2007, Raychaudhuri 1999) in comparison with prepsoriatic (where it is also increased) it is possible to assume that increased PG-load on HIV+Mo results in more active attraction of HIV+Mo to prepsoriatic and/or psoriatic skin (including Mo-R). On the other hand HIV-infection significantly changes proportions of monocyte fractions, in particular 40% of monocyte pool is CD14(low)CD16(high) fraction (while in healthy people it doesn't exceed 15-20%) (Thieblemont 1995). Abovementioned assumption that CD14+CD16+Mo are the most probable precursors of Mo-R makes clear increased risk of psoriasis initialization and severity in HIV+ patients. Increase of fraction of potential precursors CD14+CD16+Mo raises fraction CD14+CD16+Mo-R .

In overwhelming majority of cases subprocess SP7 is not present. For total (PG-Y)-carriage becoming significant, (PG-Y)-load should be chronically increased (share of Mo-T and DC-T is insignificant). However, if SP7 operates (and shares of Mo-T and DC-T are conditionally possible to assume as equal to 100%), it is weak or incidental (PG-Y)-load enough for support of SP8.1.

Rate of (PG-Y)-entry (inside Mo-R and DC-R) into derma is the main factor of psoriatic plaque maintenance.

SP8 depends on SP4. The share of fractions Mo-T and DC-T depends on kPAMP-load level in blood flow.

SP8.1 depends from SP4. The share of subfractions Mo-R and DC-R depends on kPAMP-load level in blood flow.

SP8.1 depends from SP4.1. The share of subfractions Mo-R and DC-R and their level (PG-Y)-carriage depend on level of (PG-Y)-load in blood flow. Really share of fractions PG-Y(+)Mo and PG-Y(+)DC and their level of (PG-Y)-carriage obviously depend of SP4.1 (Fig. 8). And as subfractions Mo-R and DC-R are crossing of tolerized fractions and fractions PG-Y(+)Mo and PG-Y(+)DC, so their share depends on the size of each of crossed fractions.

SP8 depends on SP7 (instead of dependence on SP4). Deviation grade in functioning of intracellular signal path influences level of «tolerization» of blood Mo and DC.

SP8 depends from SP10. Truly, if there exists relay variant (App. 5). The share of fractions Mo-T and DC-T depends on level of bone marrow kPAMP-load. Total (PG-Y)-carriage of Mo-R and DC-R depends on level of bone marrow (PG-Y)-load.

## Local subprocess LP1.1. Attraction of Mo and DC, Mo-T and DC-T (incl. Mo-R and DC-R) from blood.

Homeostatic or inflammatory renewal of pool of dermal macrophages MF and dendritic cells of non-resident origin.

Moving to derma, only Mo-R and DC-R can transform to mature maDC-Y, presenting unknown Y-antigen to specific TL-Y (Fig. 6, derma). According (Baker 2006a) this antigen is a part of interpeptide bridge (IB) of BSPG. According (Fry 2005, Valdimarsson 2009) this antigen is a part of BSMP.

Here it is supposed that Y is part of interpeptide bridge IB-Y of psoriagenic bacteria PsB (in particular some of BS). These transformations depend on type of derma: psoriatic or prepsoriatic (Clark 2006, Zaba 2009b).

In renewal of pool of dermal macrophages MF and dendritic cells DC of non-resident origin are participating all Mo-T and DC-T. Having arrived in psoriatic derma Mo-T and DC-T appear under the influence of cytokines-deprogrammers, so they lose tolerance and secrete proinflammatory cytokines, chemokines and AMP and, thereby, promote aggravation of psoriatic inflammation. Thus Mo-T will be transformed in MoDC-T or in MF-T (Fig. 6, derma).

Activity of such secretion is defined by quantity and assortment of kPAMP (except PG-Y) which carriers they (and cells derived from them) are.

At the description of subprocess SP8.1 it is offered to estimate SPP severity through total (PG-Y)-carriage of Mo-R and DC-R in one ml of blood. To consider influence of secretion of Mo-T and DC-T, caused by others kPAMP (except PG-Y), it is necessary to apply raising factor. This factor will be proportional to total kPAMP-carriage (except PG-Y) of Mo-T and DC-T in one ml of blood. And summation should be lead with correction multipliers for each of kPAMP, depending on their activity.

Such correction of SPP severity considers all kPAMP (and not just PG-Y) that corresponds to direct correlation (R=0,46) between degree of expression SIBO and PASI (App. 6). To pick up correction multipliers it will be possible only after carrying out of researches of fractions of tolerized Mo-T and DC-T at psoriatics.

Subprocess LP1.1 is a component of complex of the local processes which are taking place in prepsoriatic and in psoriatic skin and is in detail stated in (Part 2).

**Systemic psoriatic process SPP defined by set of interconnected subprocesses (Fig. 7):**

- **Hyperpermeability of intestinal walls for F-content (SP1)**
- **Specific small intestine disbacteriosis (SP2)**
- **Disorder of production and/or circulation of bile acids (BA) (SP3).**
- **Overload and/or disorder of detoxication systems (SP5)**
- **Chronic PAMP-nemia (SP4)**
-

**SPP severity is proportional to total (PG-Y)-carriage of Mo-R and DC-R in blood flow (SP8.1).**

**Each of these subprocesses can be caused or strengthened by genetic (Gudjonsson 2009) and/or functional deviations.**

# Discussion

Systemic psoriatic process SPP is a dynamic interaction of all listed subprocesses. The chronically increased PAMP-load (SP4) lays constant impact on blood immune system and includes chronic proinflammatory and antiinflammatory (in particular SP8) mechanisms. The result of it is the dynamic equilibrium condition of blood immune system. It is possible to draw an analogy with the interaction between SIRS (systemic inflammatory response syndrome) and CARS (compensatory anti-inflammatory response syndrome) which take place for more serious reasons than SP4. Tolerization (reprogramming) monocytes and dendritic cells (SP8) is limited by the compartment of systemic blood flow (and probably bone marrow) and corresponds to the definition of CARS (Adib-Conquy 2009, Cavaillon 2008).

SPP affects all organs because tolerized phagocytes participate in homeostatic and/or inflammatory renewal of pool of any tissue phagocytes. However problems can begin if F-content brought by R-phagocytes contains enough antigenic material wrongly accepted by local immune system as proof of presence of pathogenic bacteria. Antigenic material in psoriasis is PG-Y (in F-content) accepted by local immune system as proof of S.pyogenes presence in skin.

The SPP-basis is made by three obligatory subprocesses SP2, SP4 and SP8. Without any of them SPP will be incomplete. The SPP-basis can be parted on two components: tolerization of phagocytes and (PG-Y)-carriage of phagocytes (Fig. 8).

Tolerization of phagocytes assumes growth in small intestine of bacterial populations without PsB (SP2 without SP2.1). As a result PAMP-nemia without participation of PG-Y (SP4 without SP4.1) takes place. Growth of fractions Mo-T and DC-T in blood flow will be a consequence of joint LPS-load and PG-load on blood phagocytes (SP8 without SP8.1). And the main contribution is done, as a rule, by LPS-load.

(PG-Y)-carriage of phagocytes assumes growth on small intestine mucous of PsB populations only (SP2.1). Thus arises (PG-Y)-nemia (SP4.1) and in blood flow there are fractions PG-Y(+)Mo and PG-Y(+)DC. But (PG-Y)-load it is not enough for occurrence of fractions Mo-T and DC-T as activating (tolerizing) abilities of PG are essentially lower, than such LPS abilities. Occurrence of tolerized fractions only because of (PG-Y)-load is theoretically possible at essential growth of PsB-populations only (in the absence of any others). In Fig. 8 this variant is represented as questionable.

Each of these two components alone, is not enough for occurrence of subfractions Mo-R and DC-R in blood flow (SP8.1). Each of components separately is incomplete SPP-basis (pre-SPP) and can precede development SPP in the concrete patient.

Only their joint action (tolerization and (PG-Y)-carriage of phagocytes) provides growth of subfractions Mo-R and DC-R (SP8.1) – the final subprocess of SPP. Without subprocess SP8.1 systemic psoriatic process SPP is incomplete and cannot entail initialization and support of psoriasis.

If SPP completely operates, local conditions are necessary for occurrence and support of any psoriatic plaque. In particular for giving Y-antigen (in PG-Y) time to be presented before its complete degradation. So Mo-R should be transformed to MoDC-R and DC-R, MoDC-R should be transformed to maDC-Y before full degradation of PG-Y. For detail information see (Part 2).

**Basic hypotheses of this work (App. 3):**

- **Systemic psoriatic process SPP as the main factor of psoriasis initialization and maintenance**
- **Psoriagenic bacteria PsB as a key factor (SP2.1)**
- **(PG-Y)-nemia as key factor (SP4.1)**
- **Growth of tolerized fractions Mo-T and DC-T under chronic PAMP-load (SP8, App.10)**
- **Increased (PG-Y)-carriage of blood Mo-R and DC-R (SP8.1)**

10-15% of patients with psoriasis have psoriatic arthritis (Molochkov 2007, Berthelot 2003, Gladman 2006, Ibrahim 2009). There is no doubt now that LPS (Stoll 2004, Wiedermann 1999) and blood MF under PG-load (Erridge 2008, Nijhuis 2006) have aggravating effect on atherosclerosis pathogenesis. Probably new effect of tolerized phagocytes (in particular R-phagocytes) on pathogenesis of other diseases will explain why patients with psoriasis are at high risk of their development.

The results of statistical study conducted in 2008 are given at (Fig. 10). The study included data of 16851 psoriatics and 48681 control group patients without psoriasis (Cohen 2008). 13% of psoriatics suffered from diabetes, against 7,3% in control group. 27,5% of psoriatics suffered from arterial hypertension, against 14,4% of control group. 8,4% of psoriatics suffered from obesity, against 3,6% of control group. 14,2% of psoriatics suffered from coronary heart disease, against 7,1% of control group. Similar study (46095 psoriatics) also showed increased comorbidity of diabetes (OR=1,27; 95%CI 1,1-1,48) and atherosclerosis (OR=1,28; 95%CI 1,04-1,59) (Shapiro 2007). The results are of high statistical reliability $p < 0.001$ because of large scale sample group.

The detailed analysis of conditions of comorbidity for cases of obesity and for cardiovascular diseases is executed in (Davidovici 2010).

Further tests for psoriatics and control group patients will check abovementioned SPP model. Apart from well-known LAL-test, allowing to determine endotoxinemia level (Iakovlev 2003) and its modification realized in (Garaeva 2005), it is necessary to use SLP (silkworm larvae plasma) - test for determination of blood peptidoglycan level (Kobayashi 2000).

Phagocyte-dependent PAMP-load can be evaluated by blood neutrophils condition because they take up essential part of this PAMP-load. Intensive LPS entry to blood is known to increase LPS+Neu count up to 100% (standard count doesn't exceed 10%) (Chesnokova 2006). Not only percent of LPS+Neu should be determined, but also volume of LPS they brought.

Test for determination of level of intracellular protein IRAK-M (Hedl 2007, van't Veer 2007) in psoriatic blood Mo, DC and Neu and control healthy group of patients would determine share of Mo-R, DC-R and Neu-R fractions among normal Mo, DC and Neu. According to SPP model IRAK-M level in Mo-R, DC-R and Neu-R should be increased. Recently increased level IRAK-M has been found out in Neu during a sepsis (Reddy 2010).

Only comparative qualitative and quantitative monitoring of intestine parietal microflora can prove SIBO effect and to determine specific types of psoriagenic and arthritogenic bacteria (SP2). Respiratory tests (hydrogen with glucose; hydrogen with lactulose; 14S-D-xylose or 13C-D-xylose) give possibility for indirect evaluation. Bacteriological analysis of aspirate and scrape of small intestine is direct method (Gumayunova 2009b, Maev 2007). Introduction of perspective techniques (Osipov 2001, Ott 2004, Suau 1999) can provide effective monitoring and, as consequence, it will promote direct correction of intestine parietal microflora.

I hope that the publication will stimulate new cooperative studies of dermatologists, rheumatologists, gastroenterologists and microbiologists and will give possibility to solve a puzzle of psoriatic disease. I believe that future treatment for psoriatic disease will not be aimed at cosmetic and/or anti-inflammatory correction of local symptoms only, but at elimination and/or alleviation of underlying causes, i.e. at treatment for intestine dysfunctions in most cases (SP1, SP2). Efficiency of such treatment will depend mainly on patient's wish and possibility of lifestyle and diet control. So remission will be long-term (or life-long!). It can be supported by regular or periodic drug administration (bacteriophages, pre - and probiotics).

The proposed systemic process SPP is a component of Y-model of pathogenesis of psoriasis. Description of Y-model for prepsoriatic and psoriatic skin will be the content of the subsequent publication (Part 2).

**The author declares no conflict of interest.**

# Appendices

### Appendix 1. Abbreviations

| | | Abbr. | Term and comments | Notes |
|---|---|---|---|---|
| | | BA | Bile acids (BA) | Baltabaev 2005, Gunn 2000 |
| | | BLC | Blastocystosis | Glebova 2007, Tan 2008 |
| | | BLP | Bacterial lipoprotein | Buckley 2006 |
| * | | BS | Beta-hemolytic streptococci | Pokrovskii 2006 |
| | | BSMP | M-protein of beta-hemolitic streptococci | Fry 2005, Valdimarsson 2009 |
| * | | BSPG | Peptidoglycan of beta-hemolitic streptococci | |
| * | | CPs | Chronic psoriasis | |
| | | DAP | Diaminopimelic acid - component of Gram(-) PG, ligand NOD1 | |
| | #2 | DC | Dendritic cells | Bachmann 2006, Male 2006 |
| * | #2 | DC-T | Tolerized DC | App.2 |
| * | #2 | DC-R | Reprogrammed (tolerized) and repleted by PG-Y dendritic cells are subfraction of tolerized fraction DC-T. | App.2 |
| * | | F | Fragments of bacterial products with PAMP (including kPAMP) | |
| | | FPRL1 | Formyl peptide receptor-like 1 | Yang 2001 |
| | | GAS | Group A streptococci | Pokrovskii 2006 |
| | | GIT | Gastrointestinal Tract | |
| | | GP | Guttate psoriasis (temporary) | Baker 2000 |
| | | IB | Interpeptide bridges bind stem peptides attached to glycan chains of peptidoglycan | Baker 2006a, Schleifer 1972, Dworkin 2006b, Vollmer 2008 |
| * | | IB-Y | Interpeptide bridges from peptidoglycan of Str.pyogenes: L-Ala(2-3) or L-Ala-L-Ser | |
| | | iNOS | Inducible nitric oxide synthase | |
| | #1 | LPS | Lipopolysaccharide (endotoxin) | Iakovlev 2003, Husebye 2006, Kitchens 2005 |
| | | maDC | Mature DC | Bachmann 2006, Male 2006 |
| * | #2 | maDC-Y | Mature DC derived from DC-R or from MoDC-R, which present Y-antigen | |
| * | #2 | mTL-Y | Y-specific mTL | |
| | | MDP | Muramyl dipeptide - component of Gram+ and Gram(-) PG, ligand NOD2 | Girardin 2003, Traub 2006, Windheim 2007 |

**Abbreviations (continuation)**

| | | Abbr. | Term and comments | Notes |
|---|---|---|---|---|
| | #2 | Mo | Monocytes | |
| * | #2 | Mo-T | Tolerized Mo | App.2 |
| * | #2 | Mo-R | Reprogrammed (tolerized) and repleted by PG-Y monocytes are subfraction of tolerized fraction Mo-T. | App.2 |
| | | MoDC | DC, derived from Mo | Nagl 2002, Randolph 2008a |
| * | #2 | MoDC-T | DC-T, derived from Mo-T | |
| * | #2 | MoDC-R | DC-R, derived from Mo-R | |
| * | #2 | MoDP | CD34+ cells-precursors of monocytes and immature dendritic cells of bone marrow | |
| | #2 | MF | Macrophages, derived from Mo | |
| * | #2 | MF-T | Macrophages, derived from Mo-T | |
| * | #2 | MF-R | Macrophages, derived from Mo-R | |
| | #2 | Neu | Neutrophils | |
| * | #2 | Neu-T | Tolerized Neu | App.2 |
| * | #2 | Neu-R | Reprogrammed (tolerized) and repleted by PG-Y neutrophiles are subfraction of tolerized fraction Neu-T. | App.2 |
| | | NOD1 | Intracellular receptor - ligand to DAP | Fig.3; |
| | | NOD2 | Intracellular receptor - ligand to MDP | Fig.3; Girardin 2003 |
| | | PAMP | Pathogen-associated molecular patterns | Fig.3; Havkin 2008, McInturff 2005, Strober 2006 |
| | | PsA | Psoriatic arthritis | Berthelot 2003, Gladman 2006, Ibrahim 2009, Ritchlin 2007 |
| * | #1 | PsB | Psoriagenic bacteria = Gram+ bacteria with peptidoglycan PG-Y. | |
| * | #1 | PsBP | Products of vital activity and-or disintegration of psoriagenic bacteria | |
| | #1 | PG | Peptidoglycan (in particular PG-Y) | Myhre 2006, Schleifer 1972, Vollmer 2008 |
| * | #1 | PG-Y | Peptidoglycan A3alpha with interpeptide bridges IB-Y (but can contain and others also) | |
| | | PRR | Pattern-recognition receptors | |
| | | SIBO | Small intestine bacterial overgrowth | Bondarenko 2006, Maev 2007, Bures 2010, Husebye 2005 Sullivan 2003 |
| * | | SPP | Systemic psoriatic process | |
| * | | SP | Subprocess | |
| | | TipDC | Mature dendritic cells actively secreting TNF-alpha and iNOS (in particular maDC-Y) | Auffray 2009, Serbina 2008, Serbina 2003, Zaba 2009a |

**Abbreviations (continuation)**

| | | Abbr. | Term and comments | Notes |
|---|---|---|---|---|
| | | TL | T-lymphocytes | |
| | | TLR | Toll-like receptors | Fig.3; Blander 2006, McInturff 2005, Strober 2006 |
| | | TLR2 | Membranous receptor - ligand to PG-fragments LTA, BLP | |
| | | TLR4 | Membranous receptor - ligand to LPS | |
| * | | TL-Y | Y-specific TL | |
| | | Y-model | Model of pathogenesis of psoriasis proposed in this monograph | |
| * | #1 | Y | Y-antigen = part(s) of interpeptide bridge IB-Y | Baker 2006a |
| * | | YM | YM-antigen = part(s) of BSMP | Valdimarsson 2009 |

* - new abbreviation; images: #1 - Fig. 4,  #2 - Fig. 5.
Links are given for good articles in Wikipedia.

**Appendix 2. New and specified terms**

| English | Description | Notes |
|---|---|---|
| Chemostatus | Range of chemokine receptors expressed by cell (CCR1, CCR2, CCR3 etc.) including the amount of each. Chemostatus of cell defines it's behavior in reply to homeostatic and/or inflammatory chemokines. | Bachmann 2006, Male 2006, Sozzani 2005 |
| PAMP-load | Phagocyte-dependent consumption (linkage, endocytosis) of PAMP and contact with PAMP. | Fig. 13 |
| Nonactivated phagocytes | Phagocytes under weak PAMP-load are not activated. Their chemostatuses remain nonactivated. | |
| Activated phagocytes | Nonactivated phagocytes under temporarily increased PAMP-load are activated. Particularly, the secretion of TNF-alpha raises and their chemostatuses change significantly. | Buckley 2006, Moller 2003 |
| Tolerized phagocytes | The activated phagocytes under chronically increased PAMP-load are tolerized (reprogrammed).<br>Tolerized phagocytes are designated as Neu-T, Mo-T and DC-T. It is supposed that tolerized phagocytes (except known properties) have:<br><br>**Property 1. Despite contact, linkage with PAMP and endocytosis of PAMP (as a part of endocytosed F-content) their chemostatuses are similar to nonactivated (App.4);**<br><br>**Property 2. As tolerization occurs under chronic kPAMP-load then there is kPAMP in endocytosed F-content. Its degradation occurs slowly and not completely, i.e. kPAMP-carriage takes place;**<br><br>Properties 1 and 2 seem to depend on the blocking intracellular protein IRAK-M. Level of this protein in tolerized phagocytes becomes more than critical as a result of chronically increased PAMP-load.<br>Property 2 depends on parity between rate of endocytosis F and intracellular proteolytic activity.<br><br>In process of aging chemostatus of tolerized Neu changes similarly to chemostatus of nonactivated Neu.<br>Those and others start to express actively CXCR4 that provides their homing into bone marrow (SP9). | Biswas 2009, Buckley 2006, Hedl 2007, Medvedev 2006, Nakatani 2002, Savina 2007, van't Veer 2007 |
| R-phagocytes = Reprogrammed (tolerized) and repleted by PG-Y phagocytes | R-phagocytes are tolerized phagocytes which also are (PG-Y)-carriers. They are designated as Neu-R, Mo-R and DC-R and have additional:<br><br>**Property 3. There is PG-Y in endocytosed F-content. Its degradation occurs slowly and not completely, i.e. takes place (PG-Y)-carriage.**<br><br>Property 3 (as well as Property 2) depends on parity between rate of endocytosis F and intracellular proteolytic activity.<br>As endocytosis and conservation of PG-Y (as a part of F-content) by phagocytes occurs in a random way, also subfraction R-phagocytes in tolerized fraction is formed in a random way. | |

**New and specified terms (continuation)**

| English | Description | Notes |
|---|---|---|
| kPAMP = key PAMP | Not all PAMP but only key PAMP (kPAMP) provide tolerization of phagocytes. Tolerized phagocytes were likely to contact with kPAMP than with other PAMP and/or endocytosed F-content contained more kPAMP than other PAMP. | |
| Major kPAMP | All kPAMP can be conditionally divided to major and minor PAMP. Only major kPAMP-load can help to achieve critical level of IRAK-M in tolerized phagocytes because major kPAMP form the greater part of chronic load. The major kPAMP for systemic psoriatic process are LPS (TLR4-ligand) and PG (PG-fragments BLP, LTA - TLR2-ligands, MDP - NOD2-ligand and DAP - NOD1-ligand). | |
| Minor kPAMP | Minor kPAMP also positively influence IRAK-M level in phagocytes. However they bring essentially lesser contribution, their load can be neither chronic nor increased. Besides minor kPAMP-load is not necessary for tolerized phagocyte formation though the load enlarges depth of their reprogramming.<br>Minor kPAMP bring contribution because their intracellular signal paths completely coincide with paths used by major kPAMP.<br>Minor kPAMP for systemic psoriatic process seem to be flagellin (TLR5-ligand), CpG DNA (TLR9-ligand) and others. | |
| kPAMP-load | Phagocyte-dependent consumption (linkage, endocytosis) of kPAMP and contact with kPAMP. (Appendix 9). | |
| PAMP-nemia | Chronic increasing of kPAMP-load on blood phagocytes resulting in<br>a) formation of essential share of tolerized phagocytes;<br>b) kPAMP-level increasing in blood.<br>c) increased kPAMP-carriage of tolerized phagocytes | |
| Cytokines-deprogrammers | Normal phagocytes aren't tolerized (reprogrammed) and tolerized phagocytes (incl. R-phagocytes) are quickly lose tolerance (deprogrammed) in the presence of such cytokines.<br>IFN-gamma, GM-CSF and IL-12 are known. IL-12 stimulates non-monocytic cells to produce IFN-gamma. | Adib-Conquy 2002, Cavaillon 2008, Randow 1997 |

**Appendix 2a. Properties of tolerized phagocytes and R-phagocytes**

| | Tolerized phagocytes | R-phagocytes (subfractions of tolerized fractions) |
|---|---|---|
| 1. Chemostatuses are similar to nonactivated | + | + |
| 2. kPAMP-carriage | + | + |
| 3. (PG-Y)-carriage | | + |

**Appendix 3. (Sub)processes, references and hypotheses**

| | | References with facts | | |
|---|---|---|---|---|
| **NN** | **Process or subprocess** | **With psoriasis** | **Without psoriasis** | **Hypotheses and comments** |
| **SPP** | Systemic psoriatic process SPP. Increased kPAMP-carriage of tolerized phagocytes. Increased (PG-Y)-carriage of R-phagocytes. | Unknown<br>SPP-H2: Baker 2006b, Valdimarsson 2009 | SPP-H1: Bachmann 2006, Buckley 2006, Cavaillon 2008, Juffermans 2002, Krappmann 2004, Medvedev 2006, Parker 2004, Randolph 2008a, Windheim 2007 | SPP-H1. Fractions of tolerized phagocytes exist at psoriasis;<br><br>SPP-H2. (PG-Y)-carriage of R-phagocytes is necessary for psoriasis initialization and maintenance. |
| **SP1** | Hyperpermeability of intestinal walls for F-content | Stenina 2004, Khardikova 2000, Harkov 2008, Shiryaeva 2007, Ojetti 2006 | Parfenov 2000, Assimakopoulos 2007, Fasano 2005 | Permeability for D-xylose, lactulose, mannitolum etc. were investigated by many scientists. SP1-H1. These data correlate with F-content permeability. |
| **SP2** | Growth of populations of Gram+ and Gram(-)TLR4-active bacteria on small intestine mucosa. | Garaeva 2005, Ojetti 2006, Gumayunova 2009a, Gumayunova 2009b, Gumayunova 2009c, Nesterov 2009 | Bouhnik 1999, Bures 2010, Ciampolini 1996, Conte 2006, Hayashi 2005, Husebye 2005, Ott 2004, Sullivan 2003, Wang 2005, Zilberstein 2007 | SP2-H1. PsB-populations define psoriasis possibility.<br><br>SP2-H2. Y-antigen define psoriagenity of PsB. |
| **SP2.1** | Growth of populations of psoriagenic PsB | | | |
| **SP3** | Disturbance of production and/or circulation of bile acids (BA). | Baltabaev 2005, Matusevich 2000, Gyurcsovics 2003, Itoh 2007 | * | SP3-H1. This subprocess worsens psoriasis |
| **SP4** | PAMP-nemia. Increased kPAMP-load on blood phagocytes. Increased kPAMP level in blood. The major kPAMP are PG and LPS. | LPS –<br>Garaeva 2005, Garaeva 2007 | LPS –<br>Tabolin 2003, Chesnokova 2006, Iakovlev 2003, Biswas 2009<br>PG –<br>Buckley 2006, Myhre 2006, Nakatani 2002 | SP4-H1. (PG-Y)-nemia is present and it plays essential role in psoriasis |
| **SP4.1** | (PG-Y)-nemia | | | |

**(Sub)processes, references and hypotheses (continuation)**

| | | References with facts | | |
|---|---|---|---|---|
| **NN** | **Process or subprocess** | **With psoriasis** | **Without psoriasis** | **Hypotheses and comments** |
| **SP5** | Overload and/or disorders of detoxication systems | Garaeva 2005, Gumayunova 2009a, Kuranova 2009, Nesterov 2009, Matusevich 2000, Molochkov 2007, Khardikova 2005, Harkov 2008, Shagova 2009, Sharapova 1993, Gyurcsovics 2003, Wenk 2010 | Mishnyov 2003, Yimin 2000 | Reprogramming blood Mo-R and DC-R and their increased kPAMP-carriage (certainly with PG-Y). |
| **SP5.1** | Intestine | | | |
| **SP5.2** | Hepatobiliary system | | | |
| **SP6** | Tonsillar PsB-infection | BS: Baker 2000, Prinz 2009, Valdimarsson 2009, Wang 2009 | * | |
| **SP7** | Deviation in intracellular signal path "from MDP recognition through NOD2-ligand to chemostatus change" (<1%). | Christophers 2006, He 2005, Kanamori 2002, Chatzikyriakidou 2010 | Kullberg 2008 | SP7-H1. Possibility of such deviation |
| **SP8** | Growth of tolerized fractions Mo-T and DC-T. Increased kPAMP-carriage. | Baker 2006a Bevelacqua 2006, Fuentes 2010, Mizutani 1997, Okubo 1998, Zaba 2009a, Zaba 2010 | Biswas 2009, Nijhuis 2006, Turnis 2010 | SP8-H1. Defining role of the subprocess in psoriasis maintenance.<br><br>SP8-H2. SPP severity is proportional to total (PG-Y)-carriage of Mo-R and DC-R in blood flow. |
| **SP8.1** | Growth of fractions Mo-R and DC-R. Increased (PG-Y)-carriage. | | | |

**(Sub)processes, references and hypotheses (continuation)**

| | | References with facts | | |
|---|---|---|---|---|
| **Additional subprocesses (Appendix 5)** | | | | |
| **SP9** | Increased kPAMP-carriage of Neu-T. Increased (PG-Y)-carriage of Neu-R. Return of senescent Neu from blood to bone marrow and their apoptosis. | Bos 2005, Sabat 2007, Numerof 2006, Toichi 2000 | Rankin 2010 LPS – Tabolin 2003, Chesnokova 2006, Medvedev 2006; PG – Clarke 2010, Ekman 2010, Hoijer 1997, Male 2006, Martin 2003, Tacke 2006 | SP9-H1. Neu transform to Neu-T under PG-load; SP9-H2. Senescent Neu-T behave as senescent Neu – they go to die in bone marrow; SP9-H3. Role of the subprocess in psoriasis maintenance. |
| **SP10** | Transformation of Mo and DC during hemopoiesis in bone marrow. | Altmeyer 1983, Zhang 2010a, Zhang 2010b | Mo – Bartz 2006, Fitting 2004, Tacke 2006 | SP10-H1. DC are activated under relay kPAMP-load, that promotes their tolerization after exit into blood flow; SP10-H2. Role of this subprocess in psoriasis maintenance. |

*- references are available; SPn-Hm = hypothesis Hm, proposed in subprocess SPn.

**Appendix 4. Substantiation of hypothesis about chemostatus of tolerized phagocytes.**

Functional F-carriage of blood Mo and DC is standard. It arises occasionally at low changeable PAMP-load due to single endocytosis of F-content. Episodic PAMP-load activate blood Mo (DC) and changes their chemostatuses. Expression of proinflammatory and homeostatic chemokine receptors (CCR1, CCR2, CCR5, CXCR4) is reduced to prevent endocytosed F-content (probably pathogenic) translocation made by phagocytes from blood to tissues.

DC is transformed to licensed (semimature) liDC which move to spleen and/or lymph nodes under the influence of chemokines CCL19 and CCL21 (CCR7 - ligands), preserving endocytosed F-content. Then they finally mature, simultaneously degrading and presenting F-content brought from blood. So they activate and support humoral immune response (Bachmann 2006, Randolph 2008b, Sozzani 2005). Blood CD16+ Mo seem to behave similarly, supporting the T-independent humoral response (Randolph 2008a). However blood CD14++Mo reduce expression of all chemokine receptors (getting reduced chemostatus) for time necessary for degradation of endocytosed content under the influence of episodic PAMP-load. Then they restore expression of chemokine receptors (coming back to normal chemostatus), remaining in blood all the time: CCR1, CCR2, CCR5 (Sica 1997), CXCR4 and CCR5 (Juffermans 2002); CCR2 (Møller 2003); CCR1 and CCR2 (Parker 2004).

Work (Biswas 2009) is the state-of-the-art review of the publications devoted to condition of ET - endotoxic tolerance. Intracellular signal paths are in details illustrated, the short review of influence of ET on expression of chemokine receptors at monocytes/macrophages is made, and series of important open questions is formulated. Assuming the existence of R-phagocytes (as a part of tolerized fraction) we can foresee answers (certainly incomplete) to some of them:

| The question (Biswas 2009) | The answer |
|---|---|
| It would be worth investigating whether downregulation of other monocyte/macrophage chemokine receptors (except those mentioned in this work) occurs during ET and thereby it has a role in mediating immunosuppression? | Yes it does, to be exact, the level of expression CCR7 raises insufficiently (though it should rise essentially). It occurs during the chronically increased PAMP-load (in particular at ET). Traffic F+DC and F+CD16+Mo into spleen and/or into lymph nodes is thus reduced and, as consequence, the level of the humoral answer decreases. That is a characteristic sign of immunosuppression. It is confirmed by recent researches (Turnis 2010). |
| Besides tuning-down of inflammatory responses and modulating phagocytosis, does ET have a regulatory role over other aspects of monocyte/macrophage functions? | Yes, the chronically increased PAMP-load (as well as at ET) modulates chemotactic characteristics of blood monocytes/macrophages. It makes their chemostatus almost unchangeable. |
| What similar phenotype have endotoxin-tolerized monocytes/macrophages at different human diseases? | For example the same as at R-monocytes (see definition in App. 2.) |
| What other immune cells and non-immune cells may also be affected by ET? | During the chronically increased PAMP-load (as well as at ET) the condition of other blood phagocytes also changes. In this monograph the author analyses its influence on neutrophils and dendritic cells. |

Mo and DC are tolerized under the chronic kPAMP-load to subsequent influence of kPAMP (Buckley 2006, Medvedev 2006). In particular they stopping TNF-alpha secretion.

Changes of chemostatuses of PAMP-tolerized Mo and DC were not investigated until recently. Recently in work (Turnis 2010) the chemotactic behavior of mouse bone marrow DC after LPS-load of various duration was studied. Expression CCR7 (and other chemokine receptors) was studied, and also the ability of DC to migrate in lymph nodes. It is shown that the longer LPS-load lasts, in lesser degree DC expresses CCR7, and the worse their migration goes. The opposite interrelation between the level of intracellular blocking protein IRAK-M and the level of expression CCR7 is proved. Therefore it is possible to assume that signal paths providing PAMP-load influence on TNF-alpha secretion partially coincide with signal paths providing influence on expression of CCR7 in DC (Fig. 11) (Krappmann 2004, Windheim 2007).

One of the mechanisms of tolerization is increasing of level of blocking intracellular proteins IRAK-M, Tollip, SOCS-1. Chronic LPS-load (TLR4-ligand) or PG-load (in the form of fragments BLP and/or LTA) on phagocytes leads to increase of IRAK-M. IRAK-M can block link IRAK-1, IRAK-4 (Biswas 2009, Buckley 2006, Coll 2010, Geisel 2007, Myhre 2006, van't Veer 2007). In particular low LPS-load only raises IRAK-M level while higher LPS-load also reduces IRAK-1 level (van't Veer 2007). IRAK-M production is similarly increased due to chronic load of PG-fragments MDP or DAP (NOD2 and NOD1 ligands respectively) on phagocytes. However the exact mechanism of blocking of RIP2 link (and/or subsequent links) is unknown (Hedl 2007). It is known that at chronic MDP-load tolerization occurs because of level decrease of phosphorylated TAK1, but how it is reached, is not known (Canto 2009). The fragment of intracellular signal paths of monocyte from kPAMP-load to CCR7 and CD163 expression and TNF-alpha secretion is schematically presented (Fig. 11).

The activated MF and DC located in psoriatic derma have increased expression of CD163 receptor which appeared to be characteristic for the definition of these cells among the others (Fuentes 2010, Zaba 2010). It is possible to assume that such increased expression occurs for the same reason, as the active secretion of TNF-alpha. I.e. due to the action of cytokines-deprogrammers on dermal MF-T, MoDC-T and DC-T (incl. MF-R, MoDC-R and DC-R) and the subsequent loss of their tolerance to F-content (Fig. 6, derma). Receptor CD163 belongs to the family of scavengers and promotes binding and endocytosis of bacteria. Blood monocytes express it in reply to the single PAMP-load, thus its essential part is disconnected from cellular membrane (shed by cell) to pass into the soluble form sCD163. After a while after shedding, its expression is restored in much larger volume than it was before the PAMP-load (Fabriek 2009, Nijhuis 2006, Weaver 2007). Intracellular signal paths defining the expression of CD163 after the PAMP-load partly coincide with the similar ways defining secretion of TNF-alpha (Adcock 2004). It is possible to assume that blood Mo-R and DC-R being reprogrammed keep the expression of CD163 at the usual level though they have F-content. But if they appear in the environment of cytokines - deprogrammers (i.e. in psoriatic derma), they begin the active expression of this receptor.

Chronic kPAMP-load tolerizes Mo and DC and leads to decrease of secretion of proinflammatory cytokines (first of all TNF-alpha), simultaneously reduces changes of chemokine receptors which can be observed at single PAMP-load. Under chronic kPAMP-load chemostatuses of Mo and DC became similar to nonactivated.

Transformation of certain nonactivated phagocyte into tolerized one (and, in particular, into R-phagocyte) depends on time of stay in the systemic blood flow under the chronic PAMP-load. The small number of occurrings with F-content converts nonactivated phagocyte into activated, and large number - converts the activated phagocyte into tolerized one (for monocytes - see App.10).

R-phagocytes remain those if predominant volume of PAMP-load on them is kPAMP-load (including obligatory (PG-Y)-load). Proportion of blood tolerized phagocytes among all phagocytes depends on level of chronic kPAMP-load. And share of Mo-R and DC-R among Mo and DC in blood flow also depends on level of chronic (PG-Y)-load.

Mo-T and DC-T along with nonactivated DC and Mo participate in homeostatic and/or inflammatory traffic due to their chemostatuses similar to nonactivated. As a result Mo-T and DC-T (incl. Mo-R and DC-R) move to tissues. It is very dangerous if there is local inflammation. Mo-T (Mo-R) can transform into MF-T (MF-R) and aggravate any inflammation. Being (PG-Y)-carriers Mo-R and DC-R can transform into maDC-Y providing complete maintenance of psoriatic inflammation.

Phagocytes containing F-content can be found in tissues and/or blood of patients with different diseases: rheumatoid and psoriatic arthritis, atherosclerosis and CNS diseases (Laman 2002, Myhre 2006, Toivanen 2003), including in psoriatics (Baker 2006a, Okubo 2002, Weisenseel 2005).

**Appendix 5. Relay variant. Transformation of Mo and DC by means of apoptotic Neu in bone marrow.**

The subject of the appendix is the possible role of neutrophils in Mo and DC activation and reprogramming (full or partial) in hemopoiesis. Subprocesses SP9 and SP10 as an addition to subprocess SP8, especially in DC reprogramming are offered.

Chronic kPAMP-load in blood is enough for tolerized Neu-T and Mo-T formation because, as a rule, endocytosis of F-content doesn't promote these phagocytes to leave blood. However blood DC change of chemostatus at once and express CCR7 under impact of kPAMP after linkage and endocytosis of F (including kPAMP) (Bachmann 2006). Then F+DC respond to chemokines CCL19 and CCL21 and move to the nearest lymph node or spleen (Sozzani 2005) to provide the humoral immune response. I.e. the certain DC may have no time to undergo chronic kPAMP impact in blood, which is necessary for its tolerization because, as a rule, the DC leaves blood after the first kPAMP impact.

***Subprocess SP9. Increased kPAMP-carriage of Neu-T. Increased (PG-Y)-carriage of Neu-R. Return of senescent Neu from blood to bone marrow and their apoptosis.***

Subprocess components are partly known, but they haven't been investigated in psoriatics.

LPS+Neu percent can be increased up to 100% during endotoxinemia (in norm no more than 10%) (Tabolin 2003, Chesnokova 2006), depending on LPS-load. On blood Neu lays the largest part of work on PG utilization and degradation, and for that purpose they actively produce lysozyme and PRGP2 (Hoijer 1997). Neu status in PG-nemia hasn't been investigated, but it is possible (similarly to LPS) that PG+Neu percent also significantly increases.

In work (Ekman 2010) human neutrophils were investigated ex vivo. It is found out that NOD1 is not practically expressed at Neu, and, as consequence, they are not activated after the influence of DAP. At the same time NOD2 is well expressed at Neu, and consequently Neu are activated after the influence of MDP.

Recently in experiments on mice it has been proved that fragments of PG get from intestine bacteria (containing labeled DAP) to systemic blood flow. Then Neu endocytose these fragments, and subsequently senescent Neu bring them into bone marrow (Clarke 2010).

Chronic LPS-load provides Neu tolerization (reprogramming) (Medvedev 2006), in particular it decreases their response to LPS. It is possible that chronic kPAMP-load (LPS- or PG-load) also results in Neu tolerization (reprogramming), in particular it decreases level of degradation endocytosed PG.

Significant percent of tolerized (reprogrammed) neutrophils Neu-T containing F-content (including kPAMP) occur among blood Neu due to increased kPAMP-load. Neu are formed from cells-precursors in bone marrow and then move to blood to perform "fast reaction" function both in blood and in tissues where they migrate in the case of inflammation (Male 2006). Migrated Neu are lost in tissues. Remained Neu senesce in blood, express CXCR4 and above all they come back to die in bone marrow (Martin 2003, Rankin 2010). Senescent Neu can bring material, endocytosed in blood, to bone marrow. This material appears in extracellular space after Neu apoptosis (Tacke 2006) (Fig. 12)

It is supposed that in process of aging the chemostatus of tolerized Neu-T changes like chemostatus of nonactivated Neu. Both of them start to express actively CXCR4 that provides their homing into bone marrow. Whether CXCR4 express activated kPAMP+Neu in process of aging it is not known. Tolerized Neu-T appeared (PG-Y)-carriers are designated Neu-R. Neu-R make subfraction of fraction tolerized Neu-T.

Neu role in psoriasis is generalized in reviews (Bos 2005, Sabat 2007). Experimental decrease of Neu level in psoriatic skin isn't beneficial for psoriasis course (Numerof 2006). So skin Neu aren't necessary for support of local psoriatic plaque.

On the other hand, it is known a special case of neutropenia (caused by medications), leading to full disappearance of psoriasis, which plaques started to reappear again after the renewal of the Neu level in blood (Toichi 2000).

In work (Nesterov 2009) phagocytic activity of Neu at psoriatics (183 people, PASI >= 18) and in control group of healthy (52) people has been defined. Results: phagocytic indicator (42,2 % against 56,7 %), phagocytic number (6,8 against 5,9) and indicator of completeness of phagocytosis (38,6% against 42,6%). Function of killing at psoriatic Neu also has appeared weakened. It correlates with changes of phagocytic activity of Neu under LPS influence (Grebowska 2008).

SP9 depends on SP4. Level of chronic kPAMP-load defines share of fraction of tolerized Neu-T in blood flow and level of their kPAMP-carriage. Level of chronic (PG-Y)-load defines share of subfraction Neu-R in blood flow and level of their (PG-Y)-carriage.

### *Subprocess SP10. Mo and DC transformation during hemopoiesis in bone marrow.*

Part of DC and Mo can already leave bone marrow being already activated or tolerized (and, in particular, reprogrammed DC-R and Mo-R). These transformations can occur during their development in bone marrow when DC and Mo are trained, endocytosing apoptotic bodies of various cells and including dying Neu-R (Fig. 12). Activation of DC and Mo occurs, if the relay PAMP-load (transmitted through apoptotic bodies Neu-R) in the period before the exit of Mo and DC from bone marrow, takes place incidentally. Transformation of activated DC in DC-R, and activated Mo into Mo-R, can occur, if the relay PAMP-load in the same period appears constant and essential (Bartz 2006, Tacke 2006).

Mo and DC are formed from MoDP - mutual CD34+ cells-precursors in bone marrow (Merad 2007). DC are formed longer than Mo. Blood Mo can turn to MoDC. For example, it takes place in renewal of pool of epidermal dendritic cells LC (Ginhoux 2006).

Reprogramming of bone marrow phagocytes takes place in 1-3 hours after essential LPS-load (Fitting 2004) which simultaneously increases formed Mo number due to reduction of formed DC number (Bartz 2006).

Activated or tolerized Mo and DC leave bone marrow containing some non-degraded F-content (endocytosed together with apoptotic bodies Neu-R), including kPAMP. After entering the blood they are again under kPAMP-load. They continue contact, linkage and endocytosis of F-content and support and/or strengthen their own activated or tolerized status.

It is known that in bone marrow of psoriatics the activity of monocytopoiesis is raised and the phagocytic hyperplasia takes place (Altmeyer 1983, Zhang 2010a). At psoriasis peripheric blood and bone marrow stromal cells secrete cytokines abnormally and express receptors (Zhang 2010a, Zhang 2010b). It breaks the normal functioning of the bone marrow hemopoietic microenvironment. It shall be mentioned that the chronic LPS-load can be one of the reasons of bone marrow hyperplasia (Hirahata 1987). It is possible to assume that the given subprocess is component of such disturbances.

SP10 depends on SP9. Without formation of Neu-T, their aging and apoptosis in bone marrow it is impossible relay kPAMP-load on Mo and DC during hemopoiesis. This load activates Mo and DC already in the bone marrow that accelerates their activation and subsequent tolerization after exit into blood flow.

Without formation of Neu-R, their aging and apoptosis in bone marrow is impossible relay (PG-Y)-load on Mo and DC during hemopoiesis. This load provides occurrence of PG-Y(+)Mo and PG-Y(+)DC already in bone marrow that promotes increasing level of their (PG-Y)-carriage after exit into blood flow (Fig. 16).

**Appendix 6. Small intestine microflora in SIBO (without psoriasis) and in psoriasis.**

In norm total bacterial count of duodenum and upper parts of small intestine does not exceed $10^4$-$10^5$ CFU/ml. There are lactobacillus, bifidobacteria, bacteroides, enterococci, streptococci, yeast-like fungi, etc. Their count in ileum is $10^8$ CFU/ml of chymus. About 10% of obligate microflora is Gram(-) microflora (Ott 2004).

The research of small intestines microflora is a difficult procedure. Materials for such research are received by capture of aspirate during the GIT intestinoscopy. Usually this procedure is prescribed at GIT diseases for the purpose of diagnosis specification (Matsulevich 2007). It is seldom carried out in the absence of diseases, or at the diseases which etiology traditionally isn't bound with GIT diseases.

The report of the results of the researches of transient microflora of the small intestines described in two works is shown in Tab. 2. In work (Bouhnik 1999) the proximal department of jejunum at 63 patients having various diseases of GIT was surveyed. At 55 from them SIBO was found (total bacterial count more than $10^5$ CFU/ml). The microflora of various departments of small intestines at the group of healthy volunteers was studied in work (Zilberstein 2007). The data presented in Tab. 2 allows to make a comparison with the results received recently (Gumayunova 2009a) for psoriatics (* - in the first column) and healthy volunteers (Tab. 3.).

Density of parietal microflora of small intestine (especially in distal part) is comparable with density of parietal microflora of large intestine - $10^{11}$ CFU/ml (Parfenov 2001). F - fragments of bacterial products of parietal microflora first of all penetrate through intestinal walls and enter the blood. Therefore PAMP blood level and kPAMP-load on phagocytes depend on count and types of parietal bacteria.

The author of dissertation (Garaeva 2005) uses original method of bacteria count evaluation in small intestine mucosa by means of proteolytic activity of coprofiltrates. Lymphoid system of small intestine mucosa (Peyer's patches) responds to bacterial antigens of intestinal microbiocenosis. Antibodies (sIgA, IgG) are transferred to intestine through mucosa epithelium by means of endocytosis (Male 2006). Bacteria actively produce proteinases, which are capable to break antibodies (Shenderov 1998). The quantity of bacterial proteinases in excrements corresponds to volume of bacterial colonies of intestine mucosa. Psoriatics had 4–5 times higher than normal level of thiolic proteinases before treatment. 3–4 times higher than normal level of thiolic proteinases was found after standard method of psoriasis treatment and less than 1,5 times higher than normal level was found after complex treatment. It was the feature of complex method efficiency. Therefore the method provided reduction of volume of bacterial colonization of small intestine and, as a result, longer remission of psoriasis. Thereby they proved influence of volume of bacterial colonization of small intestine mucosa on psoriasis.

**Tabl.2. Transient microflora of small intestine at patients with SIBO (without psoriasis) and at healthy person**

| Microflora | Gumayunova (Tabl.3) | SIBO - 55 pers. (Bouhnik 1999) | | Healthy (Zilberstein 2007) | | | | | |
|---|---|---|---|---|---|---|---|---|---|
| | | JEJUNUM proximal + JIT diseases | | JEJUNUM proximal (22 pers.) | | JEJUNUM distal (22 pers.) | | ILEUM proximal (20 pers.) | |
| | | % of carr. | lg CFU/ml | % of carr. | lg CFU/ml | % of carr. | lg CFU/ml | % of carr. | lg CFU/ml |
| Acinetobacter spp | * | 9% | 8 | nd | | nd | | nd | |
| Bacillus spp | | nd | | - | - | - | - | - | - |
| Bacteroides spp | * | 29% | 6,9 | nd | | nd | | nd | |
| Bacteroides spp (pig) | | nd | | 9% | 2 | 9% | 3 | - | - |
| Bacteroides spp (npg) | | nd | | 36% | 3,5 | 45% | 3 | 40% | 4,5 |
| Bifidobacterium spp | * | nd | | 9% | 3 | - | - | - | - |
| Candida | * | nd | | 18% | 3 | - | - | 10% | 1 |
| Clostridium spp | * | 25% | 5,5 | 18% | 2,5 | 18% | 2,5 | 18% | 2,5 |
| Clostridium ramosum | | nd | | - | - | - | - | - | - |
| Clostridium spp (gel-) | | nd | | 18% | 2,5 | 27% | 5 | 20% | 3,5 |
| Clostridium spp (gel+) | | nd | | - | - | - | - | - | - |
| Corynebacterium spp | | nd | | 9% | 2 | 18% | 5,5 | 30% | 3 |
| E.coli | * | 69% | 7,2 | 18% | 5,5 | 36% | 6 | 30% | 3 |
| Enterobacter spp | | 7% | 7,3 | 27% | 5 | 27% | 7 | 50% | 7 |
| Enterococcus spp | * | - | - | 18% | 4,5 | 18% | 6 | 20% | 5 |
| Enterococcus faecalis | | - | - | - | - | - | - | - | - |
| Eubacterium spp | | nd | | - | - | - | - | - | - |
| Fusobacterium spp | | 13% | 4,8 | 18% | 2 | 18% | 2 | 10% | 3 |
| Klebsiella spp | * | 20% | 7,1 | 18% | 3,5 | 36% | 7 | 10% | 9 |
| Lactobacillus spp | * | 75% | 6,1 | 45% | 3 | 27% | 4 | 20% | 4,5 |
| Leptotrichia spp | | nd | | 9% | 1 | - | - | - | - |
| Micrococcus spp | | 22% | 6 | nd | | nd | | nd | |
| Moraxella spp | * | nd | | nd | | nd | | nd | |
| Neisseria spp | | 16% | 6,5 | nd | | nd | | nd | |
| Peptococcus spp | | nd | | 9% | 3 | 27% | 3 | 30% | 4 |
| Peptostreptococcus spp | | 13% | 6,1 | - | - | - | - | 10% | 3 |
| Propionibacterium spp | | nd | | 9% | nd | 36% | 2 | - | - |
| Proteus spp | * | 11% | 6,1 | 45% | 4 | - | - | 20% | 6 |
| Pseudomonas spp | | nd | | - | - | - | - | - | - |
| Rodothorula spp | | nd | | - | - | 18% | 4 | 20% | 4,5 |
| Staphylococcus spp | | 25% | 6,2 | nd | | nd | | nd | |
| Staphylococcus spp (coag+) | * | nd | | 18% | nd | 45% | 2 | 50% | 3 |
| Staphylococcus sp (coag-) = CNS | * | nd | | - | - | - | - | - | - |
| Streptococcus spp | | 71% | 6,2 | nd | | nd | | nd | |
| Str.viridans group = SVG | * | nd | | 27% | 2 | 9% | 4 | 10% | 5 |
| Torulopsis spp | | nd | | - | - | - | - | - | - |
| Veillonella spp | | 25% | 5,3 | 55% | 4 | 64% | 4 | 50% | 4 |
| Total bacterial count (TBC) | | | 7,6 (8,1) | | nd | | nd | | nd |

Notes to Table 2: nd - no data; pig = pigmented; npg = nonpigmented; gel+ = gelatinase positive; gel- = gelatinase negative; coag- = coagulase negative; coag+ - coagulase positive;

Positive dynamics of psoriasis in children as a result of treatment with method of interval normobaric hypoxia proved that psoriasis associated with aerobic microflora (streptococci, enterobacteria) (Rudkovskaya 2003, Stenina 2004).

Researches of microflora of large intestine is carried out by research of microflora of faeces - so the transient microflora is defined, or by research of the materials taken with mucous at colonoscopy - so the parietal microflora (Eckburg 2005) is defined. Such researches allow to define dysbiotic deviations in large intestine, but cannot reflect the condition of microflora of small intestines qualitatively or quantitatively. It is proved by the special comparative researches of the enteric and colic microflora which review of those is presented in (Huffnagle 2008). Researches show that dysbiotic deviations in large intestine as a rule precede and may be one of principal causes of dysbiotic deviations in small intestines.

In several dissertations the comparative analysis of fecal microflora in particular has been carried out. In work (Falova 2004) 152 psoriatics and 80 healthy people were surveyed. There have been established dysbiotic changes of microflora of the large intestine mostly expressed at exacerbation of widespread psoriasis and psoriasis followed with arthropathy. These changes were aggravated with the increase in duration of disease. In work (Shagova 2009) 140 psoriatics and 30 healthy people were surveyed. It has been established that the maximum disturbances of colic microflora are observed at 75% of psoriatics with moderate and at 97% of psoriatics with serious psoriasis. Less appreciable dysbiotic shifts are found at 47 % of psoriatics with the limited psoriasis. Severity of disturbances of microflora correlates with increase in index PASI (to 26 and more), duration of disease over 5 years and presence of psoriatic arthropathy.

In work (Nesterov 2009) the intestine microflora at patients with various dermatoses (psoriasis, eczema, atopic dermatitis and planus) and their BLC-carrier states was investigated. In particular 193 psoriatics have been surveyed (PASI>18, average PASI = 37,9), among them has appeared 146 (79,8 %) BLC+ psoriatics. The expressed dysbiotic changes in large intestine have been found out in majority BLC(-) and BLC+ patients in comparison with healthy people. These deviations appeared to be most essential at BLC+ patients. Investigation of enteric microflora at patients with chronic dermatoses has been carried, in 69,6 % BLC+ patients is found out SIBO (the specific data on psoriatics not present).

Than in the work (Gumayunova 2009b) was represent the first investigation of transient microflora of small intestine. They examined 80 psoriatics and 20 healthy persons of control group. 70% of psoriatics were diagnosed SIBO (more than $10^5$ CFU/ml). 21% of psoriatics had SIBO $10^9$-$10^{11}$ CFU/ml, and 35% of psoriatics had SIBO $10^7$-$10^8$ CFU/ml. All patients with SIBO had Gram(-) E.coli, Bacteroides and Gram+ Clostridium. So colon microflora supposed to be migrating to small intestine and cause increasing of LPS and PG level entering the blood (SP4). 25% of psoriatics had Enterococcus faecium, 10% psoriatics had Klebsiella pneumonia and 5% psoriatics had Proteus vulgaris. There were no SIBO and pathogenic flora in control group. Evaluation of biopsy materials from distal duodenum of all psoriatics with SIBO showed morphological signs of chronic inflammation. Correlation analysis shown direct relationship between SIBO grade and PASI value (correlation coefficient R=0,46) and between SIBO grade and disease duration (R=0,43). Similar results are presented in (Gumayunova 2009c) where the quantity of the surveyed psoriatics is 100 persons.

In the dissertation the results of previous researches are generalized (Gumayunova 2009a). 121 psoriatics were surveyed: 52 people with moderate psoriasis (PASI in range 20-30) and 69 people with severe psoriasis (PASI more than 30). At all patients the psoriasis was in progressing stage. 43 healthy persons have been included into the control group.

Let's notice that psoriatics with PASI >= 20 make approximately 12% of the whole contingent of psoriatics. This estimation can be received from the statistical data presented for 2260 psoriatics (Grashin 2009). I.e. the results received in works (Gumayunova 2009a, Nesterov 2009) characterize BLC-carriage and condition of intestine microflora concerning a small subgroup (about 12%) of the whole contingent of psoriatics. However it is these patients who have psoriasis in the most severe form (moderate and serious degree), and many of them also have psoriatic arthritis.

At treatment of BLC+ patients with psoriatic arthritis under the combined scheme which (in addition to the standard scheme) included Tinidazolum, Intestibacteriophage, Enterosgel, Linex and Hilak-forte, appreciable improvement has come at 46,8% of patients. At treatment under the standard scheme - only at 36,1% (Nesterov 2009).

Summary results of researches of transient microflora of proximal department of small intestines (Gumayunova 2009a) are presented in Tab. 3. Level SIBO more than $10^5$ CFU/ml (TBC> 5) was found at 95 (78,5%) psoriatics. TBC (total bacterial count) for psoriatics has made average $3x10^6$ CFU/ml that is much more than in the control group - $1,1x10^3$ CFU/ml. The correlation between SIBO level and type, severity and duration of psoriasis disease has been found.

At 93% of psoriatics Bifidobacterium spp. was found - on average $2x10^5$ CFU/ml (in the control group at 40%, on average 250 CFU/ml). At 84% of psoriatics Lactobacillus spp. was found, on average $4,6x10^4$ CFU/ml; (in the control group at 19%, on average 350 CFU/ml). At 65% of psoriatics Enterococcus spp. was found - on average $2x10^5$ CFU/ml (in the control group not found at all). At part of psoriatics Str.pyogenes (9%) and Str.viridans (30%) were found (in the control group not found).

Maximum (100 to 10000 times) excess took place at BLC+ psoriatics, at BLC(-) psoriatics the excess over the control group was also essential (10 to 100 times).

Various researchers tried to estimate influence of Helicobacter pylori (HP) on the condition of psoriasis as well as on the permeability of intestine. The authors of work (Pavlenok 2001) conducted comparative study of two groups of psoriatics (38 and 12 patients) who had or had no Helicobacter pylori (HP). 55% of HP+ psoriatics suffered from intensive itch, while the HP(-) psoriatics didn't. Psoriatic lesions of nail plates were observed in 47% of HP+ psoriatics against 17% of HP(-) psoriatics. 21% of HP+ psoriatics complained of periodic mild pains in joints. Average PASI value for HP+ psoriatics was 20% more. In work (Qayoom 2003) 50 psoriatics and 50 healthy people were surveyed. Among the psoriatics there were 40% HP+, while among the healthy people 5% only. In work (Fukuda 2001) 33 patients were surveyed: 15 HP+ and 18 HP(-). It was shown that the average permeability of intestine defined by sucrose test at HP+ patients is 345 mg/g against 59 mg/g at HP(-), i.e. is increased more than 5 times. Others refer to this work trying to explain the remission of psoriasis at HP+ patient who received complex treatment, including anti-Helicobacter therapy (Pietrzak 2009).

**Tabl.3. Transient microflora of proximal small intestine at psoriatics and at control healthy (Gumayunova 2009a)**

| | Psoriatics (121 pers.) | | | Control healthy (43 pers.) | | |
|---|---|---|---|---|---|---|
| **Microflora** | **carrier** | **% of carrier** | **lg CFU/ml** | **carrier** | **% of carrier** | **lg CFU/ml** |
| Bifidobacterium spp. | 112 | 93% | 5,3 | 17 | 40% | 2,41 |
| Lactobacillus spp. | 102 | 84% | 4,66 | 8 | 19% | 2,54 |
| Bacteroides spp. | 20 | 17% | 3,3 | 5 | 12% | 2,86 |
| E.coli typical | 81 | 67% | 5,04 | 11 | 26% | 2,94 |
| E.coli lactose-negative | 4 | 3% | 3,62 | 0 | | |
| E.coli hemolytic | 18 | 15% | 3,6 | 0 | | |
| Enterococcus spp. | 79 | 65% | 5,28 | 0 | | |
| Str.viridans | 36 | 30% | 5,74 | 0 | | |
| S.aureus | 18 | 15% | 3,24 | 0 | | |
| Str.pyogenes | 11 | 9% | 4,81 | 0 | | |
| S.epidermidis | 75 | 62% | 5,54 | 17 | 40% | 2,70 |
| Candida | 45 | 37% | 4,76 | 10 | 23% | 2,43 |
| Acinetobacter spp. | 7 | 6% | 3,56 | 4 | 9% | 2,40 |
| Proteus spp. | 24 | 20% | 4,1 | 7 | 16% | 2,14 |
| Clostridium spp. | 24 | 20% | 5,2 | 0 | | |
| Klebsiella spp. | 17 | 14% | 3,13 | 0 | | |
| Moraxella spp. | 63 | 52% | 4,45 | 0 | | |
| Total bacterial count | | | 6,49 | | | 3,05 |

The note: The analysis and comparison of the data resulted in various tables (Gumayunova 2009a) allows to assume that the average level of bacterial carriage (lg CFU/ml) was defined by calculation of the arithmetic average of logarithmic value of carriers only, and to define the average TBC – the arithmetic average of logarithmic value of all patients was calculated. A similar way of averaging is also used at (Bouhnik 1999). This way of averaging of the absolute value (CFU/ml) means the calculation of their geometrical average instead of their arithmetic average. The average value received in such way is underestimated, and the error (arithmetic average minus geometrical average) is proportional to the disorder of averaged absolute value.

**Appendix 7. BA in psoriasis. Study results.**

The authors of monograph (Matusevich 2000, chapter 3) summarized results of evaluation of functional condition of hepatobiliary system in psoriatics. 213 psoriatics were examined. Diseases of GIT, liver and bile ducts were diagnosed in 88 patients (58 patients suffered from chronic cholecystitis-cholangitis). Bile of 67 psoriatics was received with duodenal intubation and analyzed for composition. 1st group (28 persons) included psoriatics with hepatobiliary system disorder; 2nd group (39 persons) included psoriatics without such disorder. Both groups (in comparison with control group of 15 persons) had significant reduction of bile acid level in bile. Results of 2nd group are most interesting: portion B: 11,9 against 27,4 g/l, portion C: 3,8 against 6,0 g/l. The level of 3-OH- and 2-OH-acids and bilirubin was reduced similarly. Cholatocholesterol coefficient also was decreased for portion B: 3,14 against 7,47.

Total blood level of BA and their fractions was determined with thin-layer chromatography (Tab.4). The level was increased 50 times higher than normal and more. There was a correlation between reduction of bile BA level and increasing of BA blood level. The results show disturbance of enterohepatic circulation of BA, cholestasis and diffuse damage of liver parenchyma in psoriatics. Hepatobiliary system disorders haven't been diagnosed in these psoriatics before now.

Similar results are summarized in work (Baltabaev 2005). 50 psoriatics were examined. Blood BA-fractions – deoxycholic acid (DC), cholic acid (C), glycochenodeoxycholic acid along with glycodeoxycholic acid (GCDC + GDC) and glycocholic acid (GC) - were determined before and after hepatotropic therapy. High BA blood level (10-20 times higher than standard) correlated with disease severity and stability (Tab.5). Free secondary DC-BA in blood is a sign of disorder of its enterohepatic circulation because the main site of its formation under the influence of microflora is small intestine. Increase of primary C-BA concentration, synthesized by hepatic cells is a sign of conjugation disorder and/or disorder of patency of bile ducts of the functional or organic genesis promoting cholestasis.

There was apparently no GC-BA in blood of patients with complications (erythrosis, arthropathy) before and after treatment. It is likely to characterize more serious disorders of BA pool formation, than those found in patients with chronic psoriasis only. Decrease of blood fractions of free BA, psoriasis remission, itch alleviation - were found after complex hepatotropic therapy.

Similar results were received for studies of bile lithogenicity (Kuranova 2009). Sample of average values for control group (15 persons) and psoriatics (45 patients) is given in Tab.6. Levels of bile acids and phospholipids in bile were significantly decreased and lithogenic indexes were 2-4 times increased in psoriatics.

Authors of work (Rudkovskaya 2003) examined 30 children with psoriasis. Dyskinesia of bile ducts was the most frequent concomitant disease of GIT (72%).

Authors of work (Gyurcsovics 2003) supposed that deficiency of bile acids (BA) raised LPS volume moving from intestine to blood and aggravated psoriasis. Test group included 551 psoriatics (average PASI = 19,1). All patients were administered oral dehydrocholic acid. 434 psoriatics (78,8%) were asymptomatic and 117 psoriatics had significant improvement (average PASI = 2,7) after treatment. Control group included 249 psoriatics. They were administered standard therapy. Only 62 psoriatics (24,9%) were asymptomatic after the therapy. 319 of 551 (57,9%) test group psoriatics against 15 of 249 (6%) control group psoriatics were asymptomatic two years later. Dehydrocholic acid (DA) temporary increases BA (including DA) volume. So volume LPS entering blood through intestinal walls decreases. Authors of work explain that the results are associated with BA destructive ability in relation to LPS and don't take into account BA bactericidal action.

Similar results were obtained in psoriatics who took ursodeoxycholic acid every day for 5-12 months (Itoh 2007). Three patients were administered the drug to treat liver disease and the drug improved psoriasis. However the authors don't think that this BA can influence intestine microflora.

**Tab. 4. Average blood BA level (mkg/ml) in psoriatics (in aggravation phase) (Matusevich 2000)**

| Psoriatics | Pathology of hepatobiliary system | Quan-tity | Gene-ral BA | Including: | | | | | |
|---|---|---|---|---|---|---|---|---|---|
| | | | | TC | TCDC +TDC | TLC | GC | GCDC +GDC | C |
| 1st group | yes | 31 | 695,0 | 14,6 | 160,5 | 161,2 | 34,1 | 59,1 | 59,1 |
| 2nd group | no | 38 | 752,5 | 22,2 | 164,4 | 176,1 | 27,8 | 37,2 | 112,9 |

(TC - taurocholic, TCDC - taurochenodeoxycholic, TDC - taurodeoxycholic, TLC - taurolithocholic)

**Tab. 5. Average blood BA level (mkg/ml) in healthy patients, patients with chronic hepatitis and psoriatics before and after hepatotropic therapy (Baltabaev 2005)**

| Patients | Quan -tity | DC | | C | | GCDC + GDC | | GC | |
|---|---|---|---|---|---|---|---|---|---|
| | | Before | After | Before | After | Before | After | Before | After |
| Control group (healthy) | 15 | - | - | 0,57 | - | 0,67 | - | - | - |
| Chronic hepatitis | 15 | 9,9 | - | 6,4 | - | 5,4 | - | 4,3 | - |
| Psoriasis | | | | | | | | | |
| progressing stage | 15 | 9,4 | 5,6 | 9,6 | 6,4 | 12 | 6,3 | 3 | 3,6 |
| stationary stage | 22 | 7,2 | 4 | 7,5 | 4,6 | 7,2 | 4,4 | 5 | 4,3 |
| complicated forms | 13 | 12,2 | 4,6 | 13,2 | 5,4 | 12,4 | 5,1 | - | - |

(DC – deoxycholic, C – cholic, GCDC – glycochenodeoxycholic, GDC – glycodeoxycholic, GC – glycocholic)

**Tab. 6. Biochemical structure and indicators of lithogenicity gallbladder (B) and hepatic (C) bile (Kuranova 2009)**

| Indicators | Control (n=15) | | Psoriatics (n=45) | |
|---|---|---|---|---|
| Portion | B | C | B | C |
| Bile acids (mmol/l) | 21,5 | 7,2 | 8,9 | 4,5 |
| Cholesterol (mmol/l) | 2,3 | 0,78 | 1,6 | 0,53 |
| Bilirubin (mmol/l) | 665,9 | 255,2 | 432,0 | 245,3 |
| Total lipids(g/l) | 7,6 | 2,9 | 7,6 | 3,0 |
| Phospholipids (mmol/l) | 4,8 | 2,4 | 1,8 | 0,9 |
| Cholatocholesterol coefficient | 11,1 | 7,5 | 8,3 | 8,6 |
| Lithogenic index (Rubens) | 0,47 | 0,33 | 2,37 | 1,19 |
| Lithogenic index (Thomas-Hofmann) | 0,82 | 0,57 | 1,74 | 1,15 |

## Appendix 8. PAMP-nemia in psoriasis. Study results.

The author of dissertation (Garaeva 2005) showed that total blood LPS-level (free and bound LPS) correlated with psoriasis severity. Plasma samples were dissolved in purified water in ratio of 1:10 and incubated in water bath 30 minutes to release endotoxin from complexes with LPS-binding proteins. Such modification of LAL-test allows estimating LPS-load on phagocytes better. LPS-load is realized also in the form of complexes with LPS-binding proteins LBP and sCD14 (Kitchens 2005).

Total LPS-level for 16 patients with moderate psoriasis (average PASI = 16,5) was 7,2 Eu/ml. Total LPS-level for 30 patients with serious psoriasis (average PASI = 24) was 35,8 Eu/ml. Total LPS-level for patients with erythrodermic and exudative psoriasis was from 1000 to 2800 Eu/ml. Total LPS-level in control group (112 healthy persons) was about 0,1 Eu/ml. The authors of this work also investigated antiendotoxic immunity. They determined the level of antibodies to glycolipid (structural part of LPS consisting of lipid A and R-core). Humoral antiendotoxic immunity was decreased from 2 to 15 times and, as a rule, it had inverse relationship with blood LPS-level.

Authors of work (Garaeva 2005) offered complex method of treatment including not only standard, but specific therapy according to condition of antiendotoxic immunity and intestinal microflora. The authors received good results: average PASI in patients with serious psoriasis decreased from 24,5 to 1, with exudative psoriasis - from 26 to 3,2 and with stationary psoriasis - from 16,4 to 0. Significant decrease of blood LPS-level was observed before beneficial changes. Average duration of remission after complex method (in comparison with standard method) increased from 6 months to 1,5 years.

However there was no direct dependence between total LPS-level and psoriasis severity in some cases. So the patients with localized form of psoriasis (palmar, plantar or psoriasis on hairy part of head) had low PASI level (average 1,9) and higher LPS-level (average 39 Eu/ml), than patients with stationary psoriasis (7,2 Eu/ml) with much higher PASI (average 16,4). It means that not only total LPS-level influences psoriasis severity. These results are summarized in the work (Garaeva 2007).

Pagano regimen is sometimes associated with temporary aggravation before remission, so-called Herxheimer reaction (Pagano 2008). Psoriasis aggravation is due to mass destruction among populations of psoriagenic bacteria in intestine. Thus we can observe high, but temporary growth of (PG-Y)-load on phagocytes. The condition is transient and, as a rule, there is no need to treat it.

Subprocesses SP4 and SP5 have dynamic interaction.

Normalization of antiendotoxic immunity (in particular level of humoral immune response for LPS) became possible after blood LPS-level was normalized. The process was observed in 38 psoriatics after complex treatment (Garaeva 2005, Garaeva 2007). LPS-level decreased simultaneously with decrease of LPS-load on phagocytes. It resulted in decrease of tolerized phagocytes (incl. R-phagocytes) part and provided long remission.

**Appendix 9. Definition of concept of PAMP-load on phagocytes**

Let's use functions PI(t) - PAMP-income and PC(t) - PAMP-consumption per unit of time to understand correlation between blood PAMP-level and PAMP-load on phagocytes (neutrophils, monocytes, dendritic cells). Blood PAMP-level in specific moment T is defined as

(1) $PL(T) = \int (PI(t) - PC(t))dt + PL(T0)$,

where integral is taken on interval from T0 to T, and PL (T0) - PAMP-level at initial moment T0. PAMP-consumption PC(t) depends on PI(t) and PL(t). The more income and level, the more is consumption. Degradation, linkage, endocytosis and elimination of PAMP at moderate rate of PAMP-income seem to be carried out so that PL(t) doesn't exceed certain admissible level.

PAMP-consumption can be presented as

(2) $PC(t) = PCi(t) + PCd(t)$,

where PCi (t) - phagocyte-independent PAMP-consumption, and PCd (t) - phagocyte-dependent PAMP-consumption (contact and linkage, endocytosis). Phagocyte-independent PAMP-consumption provide degrading enzymes (for example, lysozyme and PGPR2 for PG), proteins and antibodies binding PAMP in complexes, elimination organs etc. PAMP-consumption is completely phagocyte-independent if its PRR-ligandic properties are fully lost without participation of phagocytes.

PAMP-consumption is phagocyte-dependent, if PAMP (as a part of F-content, complexes or fragments) is endocytosed (is bound) by blood phagocytes and PAMP still has PRR-ligandic properties to the moment of endocytosis (binding).

Binding of free PAMP into complexes and/or its degradation by enzymes at conservation of any PRR-ligandic properties can result in phagocyte-dependent consumption.

PAMP-load on phagocytes depends on phagocyte-dependent PAMP-consumption. Part of tolerized phagocytes (of total number of phagocytes), depth of their reprogramming and kPAMP-carriage level depends on duration and intensity of PAMP-load.

PAMP-income and PUMP-consumption can be presented as container with untight (non-hermetical) walls (systemic circulation). Water enters container through a pipe (PI) and it is filtered outside through walls - consumed (PCi, PCd) (Fig. 13).

Rates of PAMP-income and consumption to container correspond to length and number of white arrows. They are 2 times larger on scheme B than on scheme A, however PAMP-level raised less than in 2 times (calculation by formula for such container). The functional interrelation of PAMP-level and PAMP-load in blood seems to be similar (graph C).

The consumption depends on water level in container PC(T)=FUN(PL(T)). This dependence is calculated by formula (3).

(3) $PC(T) = A*PL(T)**(3/2)$,

For example, if rate of income is constant PC (T) = PICO, then equation (1) will be transformed in (4), and then in (5) after differentiation.

(4) $PL(T) = \int (PICO – FUN(PL(T))dt + PL(T0)$,

(5) $PL'(T) = PICO – FUN(PL(T))$

If equation (5) at T -> infinity has solution PLB, it can be found from equation (6) as PL´(T) aims to 0 at stabilization PL(T).

(6) PICO = FUN(PLB), i.e.

(7) PLB = FIN(PICO), where FIN – inverse function for FUN.

For example, if FUN is dependence of type (3), than PLB is calculated by formula (8). In particular it means that if PICO increases in 2 times PLB increases only in 2**(2/3) =1,6 times.

(8) $PLB = (PICO/A)**(2/3)$.

Thus water level fluctuates near PLB.

**Appendix 10. Fractionation of blood monocytes under chronic PAMP-load**

After the exit from the bone marrow monocytes stay in the blood flow before they will be involved into tissues or lymph nodes. Leaving from the blood flow of certain monocyte occurs in a random way. Exponential distribution (Fig. 14) takes place for time of stay of monocytes in the blood flow (9)

(9) F(T) = LAM*EXP(-LAM*T), where LAM = ln(2)/Thalf

Calculations and estimations of different authors for Thalf - time of half-life of monocytes in human blood flow are resulted in Tab. 7.

**Tab. 7. Time of half-life of monocytes in blood flow**

| Thalf | References | Comments |
|---|---|---|
| 8,4 hours | Wintrobe 2008, table and text on p. 254 (with ref. on Meuret 1973, Meuret 1974). Tmid=Thalf/ln(2) =12 hours (Degos 1999, p.65; Wintrobe 2008) | In (Meuret 1973) the quantity of monocytes, in systemic blood flow for 8 people surveyed (middle age 57 years) on the average is estimated as 8,0*E+7/kg, the same value is specified in (Wintrobe 2008). Now it is well-known that averagely it is nearby 2,8*E+7/kg. Marks of own monocytes were carried out by 3H-DPF in vitro, and then their injection was carried out (autohemotrasfusion). Authors note the loss of up to 40% of the marked monocytes within the first 1-2 hours after injection, assuming their apoptosis owing to damages at reception, marks in vitro and injections. Authors consider that main reason of abbreviation of number of the marked monocytes in the blood flow in the next hours, is their attraction in tissues or lymph nodes. They completely neglect possible essential apoptotic loss of the marked monocytes in the blood flow in the next hours. Also have received as underestimated Thalf, as much essential this loss was. |
| 17,3 hours | Abdulkadyrov 2004, p.48 | Average time of stay of monocytes in the blood flow Tmid =12-48 hours is resulted. Calculation Thalf=Tmid*ln(2)=0,5*(12+48)*ln(2)=17,3 hours is executed here. References are given on (Degos 1999) and on (Javorkovsky LI, 1987). |
| 21 hour | Berger 2008, p.7; Hoffbrand 2005, p.99 | In these works average time of stay of monocytes in the blood flow Tmid = 20-40 hours is resulted. Calculation Thalf=Tmid*ln(2)=0,5*(20+40)*ln(2)=21 hour is executed here. Comments and references are not present. |
| 71 hour | Goto 2003, Wintrobe 2008, p. 255 (both with ref. on Whitelaw 1972); Hughes-Jones 2008, p.5; Vorob'ev 2002, p. 31. | (Whitelaw 1972): 7 patients of hospital (middle age of 61 years) without hematological problems are surveyed. Marks of blood monocytes in vivo occurred during their bone marrow division owing to injection of tritiated thymidine. The marked monocytes have appeared in the blood flow every other day, their quantity has reached a maximum in 3 days, and then began to decrease. Owing to injection neutrophils were marked also. As the full cycle of their bone marrow development reaches 4-6 days, so the marked neutrophils have appeared in the blood flow only on 5th day, and their quantity has reached maximum on 8th day. Authors certainly could not consider recently opened effect of return of senescent neutrophils back into the bone marrow with their subsequent apoptosis (Martin 2003, Rankin 2010). This effect (since 5 days) could affect the additional relay marks of bone marrow monocytes and by that on essential increase of Thalf. |

It is easy to notice, what even with modern researchers estimations of Thalf strongly differ. In work (Wintrobe 2008) two extreme value, based on results (Meuret 1973) and (Whitelaw 1972) are resulted. Surprisingly, but from 1972-3 any group of researchers did not execute similar experiments on humans. And, hence, untill now there are no proofs present, which of these values (8,4 hours or 71 hour) are closer to reality.

In later years experiments by definition of Thalf were executed only for the experimental animal (mice, rats, rabbits or monkeys). The comparative review of techniques and results contains in (Goto 2003). In the same work the stay of monocytes in the blood flow for rabbits (marks by BrdU of donor monocytes in vivo and their subsequent injection to recipients) is investigated and are received Thalf(rabbits)=12,7 hours. According to authors their result correlates with received earlier by other researchers: Thalf(mice)=17,4 hours, Thalf(rats) =42 hours and Thalf(human) =71 hour.

On the basis of results (Hasegawa 2009) it is possible to make calculation and to receive Thalf (macaques-rhesus)=63 hours.

Which Thalf (Tab. 7) is closer to reality does not influence a sense of the assumptions stated further about the fractionation of blood monocytes under the chronic PAMP-load. But to accurately formulate these assumptions, it is necessary to take concrete value, for example, average between the extremes.

(10) Thalf = (8,4+71)/2 = 39,7 hours = 1,65 days

The graph of exponential distribution (Fig. 14) is executed at Thalf, defined under the formula (10).

Rate of production of monocytes at serious psoriasis is above the norm in Ka=1,6, and their quantity in blood flow is above the norm in Kb=1,9 (Meuret 1976). Proceeding from it at serious psoriasis

(11) GAM1 = GAM*Ka/Kb, where GAM = 1 - EXP(-LAM) and GAM1 = 1 - EXP(-LAM1)

(12) Thalf1 = ln(2)/LAM1 = 2,04 days

Graphs of exponential distribution (Fig. 15 and Fig. 16) are executed at Thalf1, defined under the formula (12).

The total PAMP-load on certain monocyte, is directly proportional to time of its stay in the blood flow. Staying in the blood flow under chronic PAMP-load the certain monocyte at first is activated, and then if yet has not left blood flow, is tolerized.

It is possible to assume that at chronic PAMP-load

(a) in the systemic blood flow three fractions of monocytes simultaneously coexist: nonactivated, activated and tolerized.

(b) at SPP the activated fraction is the greatest.

Assumptions (a) and (b) allow to compound the facts about activated blood Mo at psoriasis (Bevelacqua 2006, Mizutani 1997, Teranishi 1995) and hypothesis about existence of reprogrammed blood Mo-R (SP8).

At SPP the nonactivated fraction basically consists of the monocytes which have recently left bone marrow. These are monocytes of short-term stay (Fig. 15-A, Fig. 16, yellow zone).

At SPP the activated fraction consists of monocytes of intermediate-term stay and constantly replenishes with monocytes from nonactivated fraction (it is possible and from bone marrow - App. 5, Fig. 16). Nonactivated monocytes during the first events of essential PAMP-load (contacts, linkage, endocytosis) become activated (Fig. 15-A, Fig. 16, pink zone).

The activated monocytes secrete TNF-alpha more strongly, than nonactivated (as it is spontaneous and in reply to the subsequent PAMP-load). The activated monocytes change chemostatus, they reduce the expression of the basic chemokine receptors S which are responsible for their attraction in tissue. S is (CCR1, CCR2, CCR5, CXCR4) for CD14+CD16+Mo (Fig. 9) and (CCR1, CCR2, CXCR3, CXCR4) for CD14++Mo.

On Fig. 15, Fig. 16 and Fig. 17 probable fractionation of blood CD14+CD16+Mo under chronic PAMP-load is represented. Distribution and fractionation of all blood monocytes under chronic PAMP-load is supposed to be similar.

Monocytes with increased expression of CCR7 carry endocytosed F-content into lymph nodes, while in blood flow mainly remain monocytes with lowered expression of CCR7.

In the activated fraction there are monocytes F(+)TNF-alpha(+)S(-)Mo which not completely degraded F-content. But is also TNF-alpha(+)S(-)Mo with the lowered F-content which were activated mainly due to contacts and linkage with PAMP and-or already thoroughly degraded the endocytosed earlier F-content.

At SPP the tolerized fraction consists of monocytes of long-term stay and constantly replenishes with monocytes from the activated fraction. The activated monocytes after the long essential PAMP-load become tolerized. It occurs at achievement IRAK-M of blocking level (Fig. 9). Tolerized monocytes secrete TNF-alpha more weakly, than nonactivated (as it is spontaneous and in reply to the subsequent PAMP-load). Tolerized monocytes change chemostatus, restoring expression of the basic chemokine receptors S which is responsible for attraction into the tissues. Tolerized monocytes essentially reduce expression CCR7. As a result the chemostatus of tolerized monocytes becomes similar to the chemostatus of nonactivated monocytes (Property 1). For this reason, under the influence of chemokines they migrate similarly to nonactivated monocytes, including at renewal of pool of tissue monocytes.

In the tolerized fraction there are monocytes PG-Y(+)F(+)Mo-R which are keeping essential part of PG-Y. But as well there are also PG-Y(-)F(+)Mo-T which were tolerized mainly due to contacts and linkage with PAMP different from PG-Y and-or already degraded endocytosed earlier PG-Y.

All tolerized fraction of monocytes has Properties 1 and 2, then they are CCR7(-)S(+). But only the part of tolerized fraction has simultaneously Properties 1, 2 and 3 and there are R-monocytes PG-Y(+)F(+)Mo-R.

Under the influence of chemokines the part of monocytes constantly leaves all three fractions, going into tissues or lymph nodes.

On (Fig. 15-A) distribution on fractions of blood monocytes is represented. Here it is supposed that bone marrow activation is absent, i.e. all monocytes leave bone marrow while being nonactivated. Conditionally it is supposed that level of chronic PAMP-load is such, that the monocytes which have stayed in the blood flow more of 2,5 days, become activated. Transition from nonactivated condition (yellow zone) to activated (pink zone) occurs gradually. Also conditionally it is supposed that for all monocytes the period of stay in the blood flow from 5 till 8 days is transitive from the activated condition (pink zone) to tolerized (green zone).

The areas of each of three zones correspond to share of each of fractions: nonactivated (~39%), activated (~51%) and tolerized (~10%).

In Fig. 15-B possible graphs of five characteristics (F, IRAK-M, CCR7, S, TNF-alpha) which change in process of stay of monocytes in the blood flow under the chronic PAMP-load are represented.

In Fig. 16 is represented the distribution by the fractions of blood monocytes in the assumption that bone marrow activation takes place (App. 5). In this case the part of monocytes leaves bone marrow already being activated (conditionally 50%). It is supposed that events occur in advance of 1,5 days before (in comparison with Fig. 15-A).

The monocytes which have stayed in the blood flow more, than 1 day, become activated. Transition from nonactivated condition (yellow zone) to activated (pink zone) occurs gradually. It is supposed that for all monocytes the period of stay in the blood flow from 3,5 till 6,5 days is transitive from the activated condition (pink zone) to tolerized (green zone). The areas of each of three zones correspond to share of each of fractions: nonactivated (~8%), activated (~75%) and tolerized (~17%).

I.e., at bone marrow activation the essential increase in shares of the activated and tolerized fractions takes place.

How to check up the reality of assumptions of fractionation of monocytes at SPP? For representative group of psoriatics to execute the flowing cytometry of blood monocytes on several of the characteristics set forth above (Fig. 15-B). As at SPP the main kPAMP are LPS and PG (incl. PG-Y) and it is necessary to take them as representatives of F-content.

For pair of PG-Y and IRAK-M in Fig. 17 expected distribution is represented.

It is necessary to assume that at SPP similar fractionation (Fig. 15) and distribution (Fig. 17) takes place also for the blood dendritic cells.

## Appendix 11. List of essential changes and additions

This appendix is intended for readers who are familiar with the previous edition of this book (e3.2) and allows to get briefly acquainted only with essential changes and additions (Tab. 8). In the text the significant new or revised fragments are marked with a vertical line on the right. The new works included into the bibliography are also marked the same way.

**Tab. 8. Changes and additions**

| What | Where |
|---|---|
| There, where is represented materials concerning phagocytes of various types (for example about Mo and DC) discrepancy is corrected: the word "chemostatus" is used in plural. | In several places. |
| All figures are moved to "Figures" section.<br>Fig. 4 and Fig. 7 have exchanged in places.<br>Fig. 4 and Fig. 5 are corrected and added so that to become identical about fig. 2-1 and fig. 2-2 of Part 2.<br>Fig. 8 is added.<br>Fig. 4, Fig. 7, Fig. 9, Fig. 7, Fig. 12, Fig. 15, Fig. 16, Fig. 17 are changed. | Figures |
| Tolerized phagocytes are designated Neu-T, Mo-T and DC-T.<br>Also derived from Mo-T macrophages and dendritic cells are designated MF-T and MoDC-T.<br>For all tolerized phagocytes special images are offered (Fig. 5) .<br>In figures these images as a rule are used for PG-Y(-) tolerized phagocytes.<br>Concerning chemostatus of senescent Neu-T specification is made. | App. 1 and App 2. |
| Definition of tolerized phagocytes is specified:<br>kPAMP-carriage is recognized by obligatory (Property 2).<br>Definition of R-phagocytes is specified:<br>(PG-Y)-carriage is recognized by obligatory (Property 3).<br>As a result there was actually renaming entities:<br>Before: a) tolerized b) R-phagocytes and c) (PG-Y)+ R-phagocytes;<br>Now: a) and b) tolerized and c) R-phagocytes.<br>It has entailed the most part from changes listed further. | App. 1 and App 2.<br>App 2a. |
| In all fragments where earlier there were materials about R-phagocytes (Mo-R and DC-R) corrections were made:<br>Before: "R-phagocytes",<br>Now: «Tolerized phagocytes (incl. R-phagocytes)».<br>Before: «Mo-R and DC-R»,<br>now: «Mo-T and DC-T (incl. Mo-R and DC-R)», etc. | In several places. |
| PAMP-nemia definition is specified:<br>Before: «….essential share of R-phagocytes …»;<br>Now: «… essential share of tolerized phagocytes …»<br>Similarly, in some other places the term "R-phagocyte" is replaced with «tolerized phagocyte». | App 2.<br>Paragraph |
| Name SPP is changed:<br>Before:<br>«Increased kPAMP-carriage of R-phagocytes.»<br>Now:<br>«Increased kPAMP-carriage of tolerized phagocytes.<br>Increased (PG-Y)-carriage of R-phagocytes».<br>It is made in connection with change of the formulation of subprocess SP8. | SPP |

**Changes and additions (continuation)**

| What | Where |
|---|---|
| In subprocess SP2 is allocated<br>«SP2.1. Growth of populations of psoriagenic PsB.»<br>It is made for more accurate formulation of dependencies. | Subprocess SP2.1.<br>Fig. 8 |
| Subprocess SP4. The name is corrected.:<br>Before:<br>«PAMP-nemia. Increased kPAMP-load on blood phagocytes (Mo and DC). Increased kPAMP level in blood. The major kPAMP are PG (including PG-Y) and LPS.»<br>Now:<br>«PAMP-nemia. Increased kPAMP-load on blood phagocytes. Increased kPAMP level in blood. The major kPAMP are PG and LPS.»<br>In subprocess SP4 is allocated:<br>«SP4.1. (PG-Y)-nemia».<br>It is made for more accurate formulation of dependences and in connection with change of formulation SP8. | Subprocess SP4.<br>SP4.1<br>Fig. 8 |
| Subprocess SP8. The formulation is changed:<br>Before:<br>«Reprogramming blood Mo-R and DC-R and their increased kPAMP-carriage (certainly with PG-Y).»<br>Now:<br>«Growth of tolerized fractions Mo-T and DC-T.<br>Increased kPAMP-carriage.»<br>In subprocess SP8 is allocated:<br>«SP8.1. Growth of subfractions Mo-R and DC-R.<br>Increased (PG-Y)-carriage.» | Subprocess SP8<br>SP8.1. |
| SP2, SP4 and SP8 are obligatory subprocesses SPP.<br>Their set is named as SPP-basis.<br>SPP-basis consists from two component: tolerization of phagocytes and (PG-Y)-carriage of phagocytes, each of them is the precondition for SPP-basis.<br>It is made for short description of SPP.<br>In Fig. 8 dependences of SPP-basis and both components both separately and together are represented. | Discussion<br>Paragraphs<br>Fig.8 |
| Name of LP1.1 is changed.<br>Before:<br>«Attraction of Mo, Mo-R, DC, DC-R from blood.»<br>Now:<br>«Attraction of Mo and DC, Mo-T and DC-T (incl. Mo-R and DC-R) from blood.»<br>It is made in connection with specification of concept of R-phagocytes.<br>The description of role Mo-T and DC-T in psoriatic inflammation is added. | LP1.1<br>Fig. 6 |

**Changes and additions (continuation)**

| What | Where |
|---|---|
| Subprocess SP9. The formulation is changed:<br>Before:<br>«Increased kPAMP-carriage of blood Neu-R.<br>Return of senescent Neu-R from blood to bone marrow and their apoptosis.»<br>Now:<br>«Increased kPAMP-carriage of Neu-T.<br>Increased (PG-Y)-carriage of Neu-R.<br>Return of senescent Neu from blood to bone marrow and their apoptosis.»<br>Changes in Fig. 12 are made. It is made in connection with specification of Neu-R and change of formulation SP9. | SP9 |
| Additional variants of estimation of SPP severity are offered. | SP8.1. Paragraphs. |
| The comment to Fig. 9 is specified. | SP8. Paragraph. |
| Figures to Appendix 10 are corrected.<br>The image of tolerized PG-Y(-)F(+)Mo-T is added. | Fig. 15, Fig. 16, Fig. 17 |
| It was inexact:<br>«SPP severity is proportional to total kPAMP-carriage of Mo-R and DC-R in blood flow.»<br>Now:<br>«SPP severity is proportional to total (PG-Y)-carriage of Mo-R and DC-R in blood flow.» | In several places. |
| In connection with the listed changes the abstract is corrected. | Abstract |

# Figures

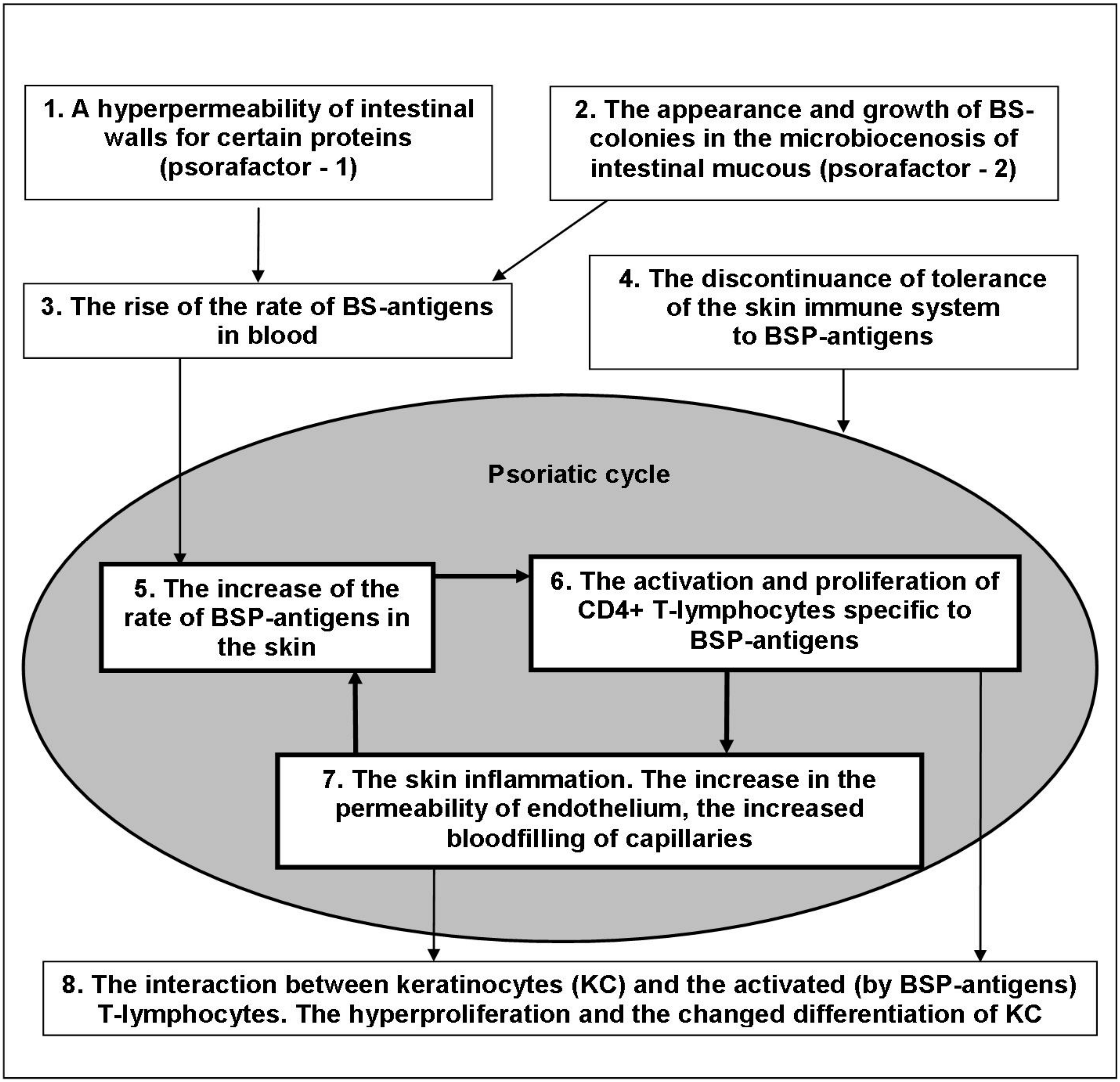

**Fig. 1. Model of pathogenesis of psoriasis (Korotkii & Peslyak 2005).**

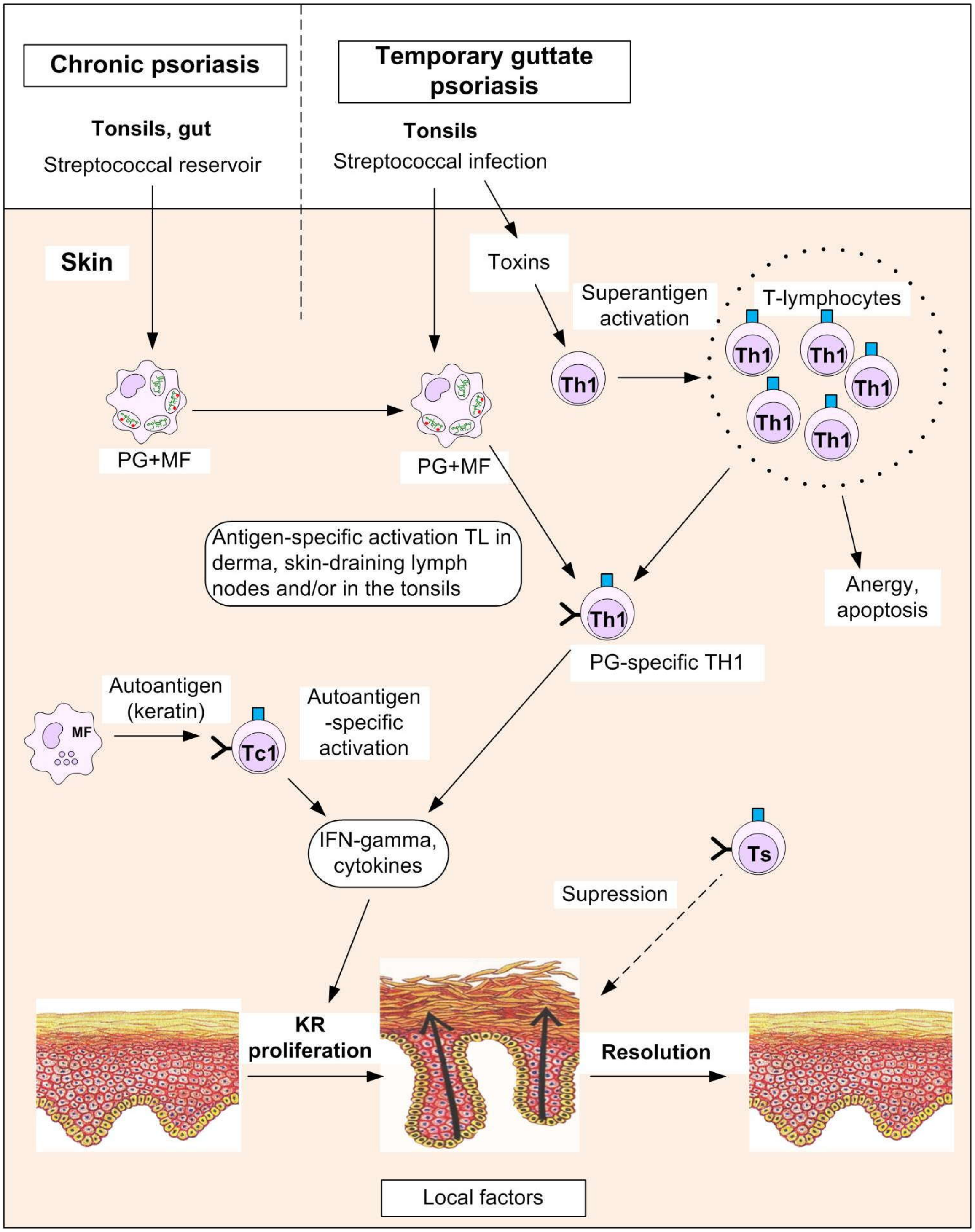

**Fig. 2. PG-induced TL (T-lymphocytes) activation in temporary guttate psoriasis (GP) and chronic psoriasis (CPs).**

In GP, streptococci in the tonsils produce toxins (e.g. SPE-C) that function as superantigens and activate TL(Th1). Superantigens also induce TL expression of the skin-homing receptor, CLA (blue). Most of these activated TL become anergic to further activation, or die (by apoptosis), whereas a subset PG-specific TL are rescued by stimulation by PG+MF (PG-carrying macrophages) have migrated from the tonsils. Cytokines, including IFN-gamma, produced by these PG-specific TL induce keratinocyte (KR) proliferation and skin lesion development. The resolution of skin lesions is mediated by T-suppressor cells (Ts) and local factors by unknown mechanisms. In CPs, streptococci and/or streptococcal antigens persist in the tonsils and/or gut. PG+MF migrate to the skin to activate pathogenic PG-specific Th1. The disease process is maintained by cytokines produced by Tc1 cells (CD8+TL that produce IFN-gamma but not IL-4), activated locally by skin-derived antigen (e.g. keratin), preventing resolution. Mainly based on Fig.2 from (Baker 2006b).

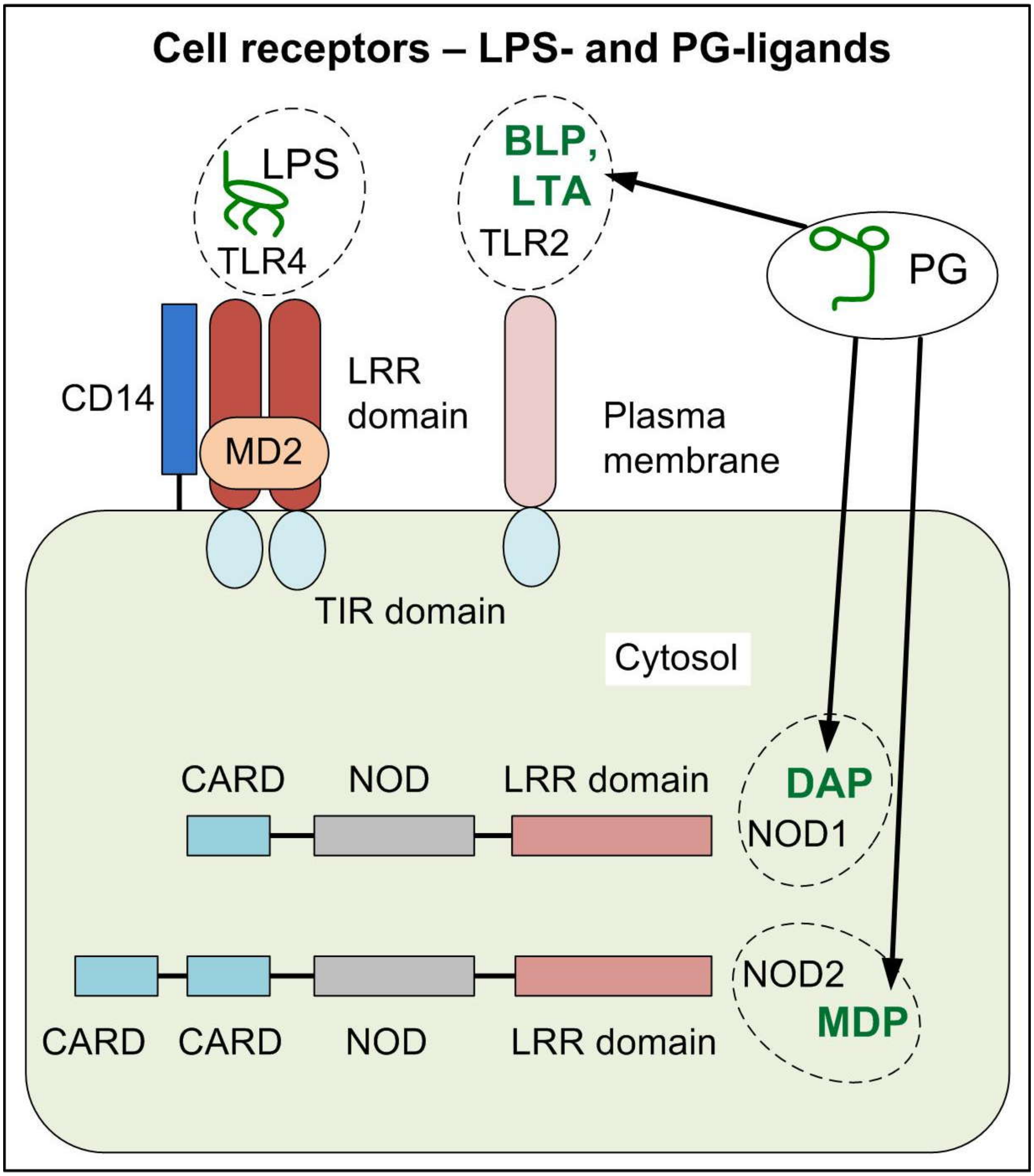

**Fig. 3. Structure and cell's localization of TLR2, TLR4, NOD1 and NOD2.**
TLR4 – LPS-ligand; Receptors for PG(peptidoglycan)-fragments: TLR2 – ligand of BLP and LTA; NOD1 – ligand of DAP; NOD2 – ligand of MDP. Based on (Strober 2006).

| | | | |
|---|---|---|---|
| 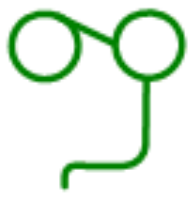 | PG - any peptidoglycan (in particular PG-Y) | 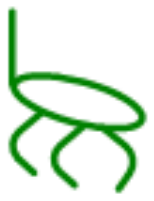 | LPS - lipopolysaccharide, free and bound in complexes with LBP, sCD14, etc. |
| 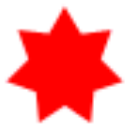 | Y-antigen = part(s) of interpeptide bridge IB-Y | 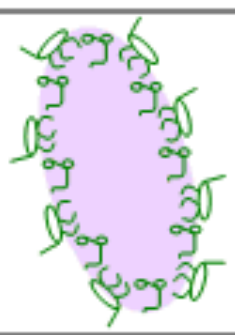 | Gram(-) TLR4-active bacteria |
| 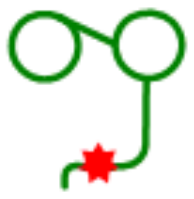 | PG-Y - peptidoglycan A3alpha with interpeptide bridges IB-Y (but can contain and others also) | 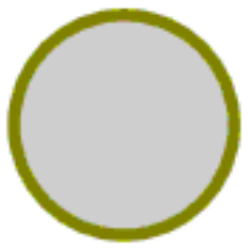 | Gram+ and Gram(-) bacteria - intestine commensals |
| 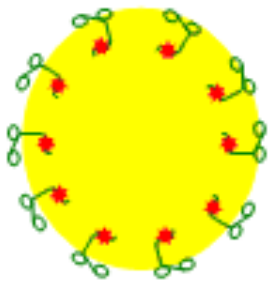 | PsB - psoriagenic bacteria = Gram+ bacteria with peptidoglycan PG-Y. | 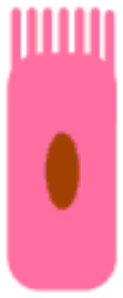 | Enterocytes - epithelial cells covering mucous intestine |
| 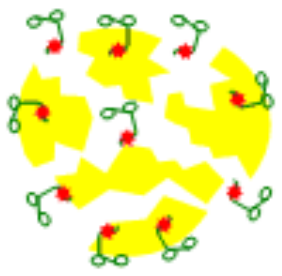 | PsBP - vital activity and/or degradation products of PsB | 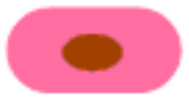 | EC - endothelial cells |
| 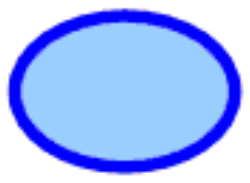 | Bacteria - skin commensals | 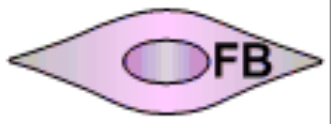  | FB - fibroblasts |
| 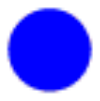 | HPV (Human Papilloma Virus) | 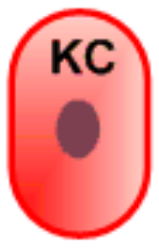  | KC - keratinocytes |
| 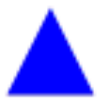 | Z - dominant antigen. At LP2(IN) - antigen of commensals. At LP2(HPV) - virus antigen. | 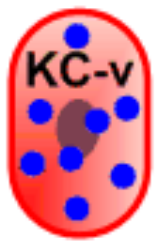  | KC-v = HPV-carring keratinocytes |
| | | 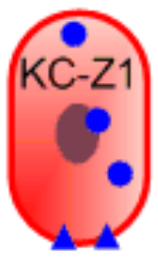  | KC-Z1 = HPV-carring keratinocytes presenting Z1-antigen |

**Fig. 4. Bacteria, bacterial products, viruses and tissue cells (symbols).**
See also Appendix 1.

| | | | |
|---|---|---|---|
| 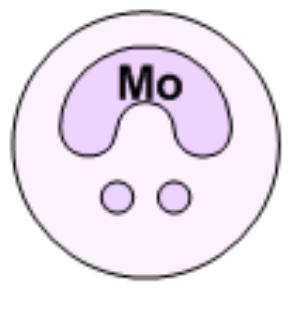 | Mo - monocytes | 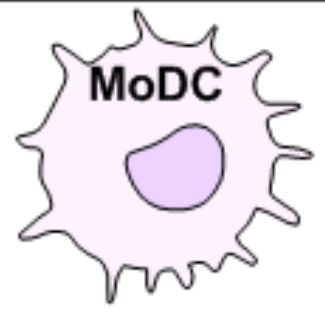 | MoDC - dendritic cells, derived from Mo |
| 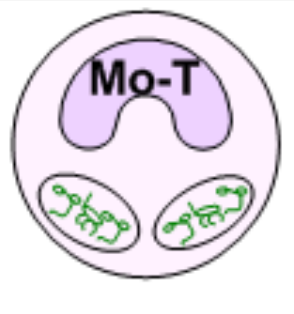 | Mo-T -tolerized monocytes | 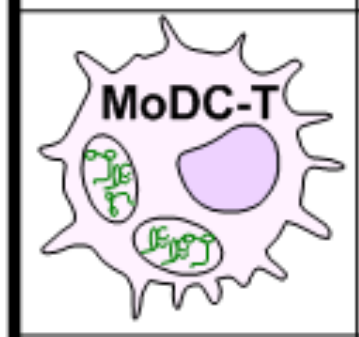 | MoDC-T - dendritic cells, derived from Mo-T |
| 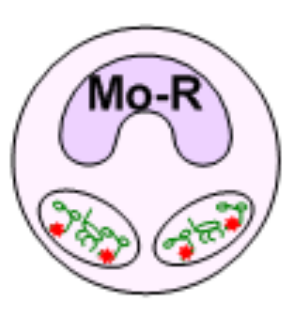 | Mo-R – reprogrammed (tolerized) and repleted by PG-Y monocytes | 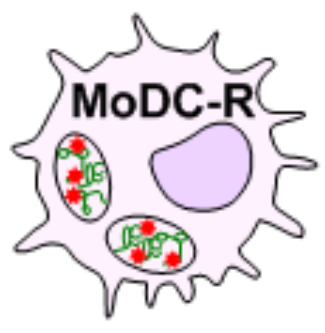 | MoDC-R - dendritic cells, derived from Mo-R |
| 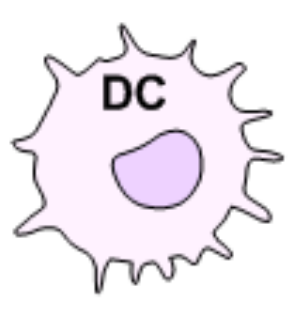 | DC – dendritic cells | 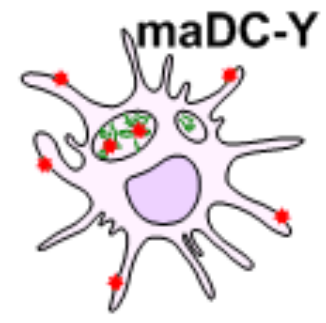 | maDC-Y = mature dendritic cells, presenting Y-antigen |
| 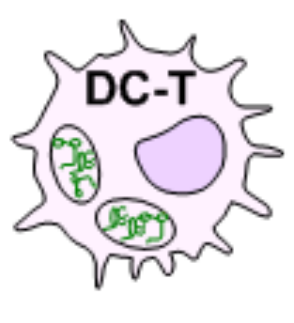 | DC-T – tolerized dendritic cells | 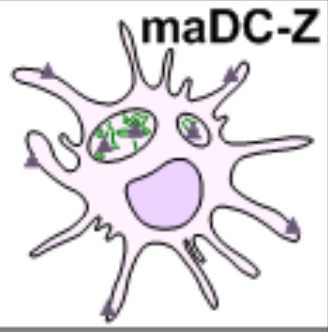 | maDC-Z = mature dendritic cells, presenting Z-antigen |
| 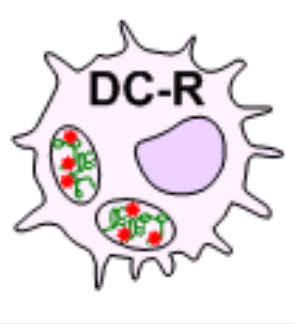 | DC-R – reprogrammed (tolerized) and repleted by PG-Y dendritic cells | 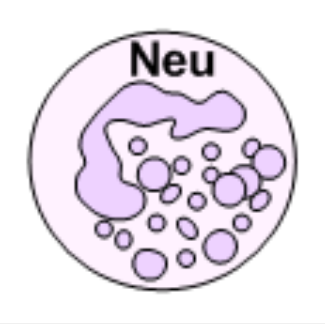 | Neu - neutrophils |
| 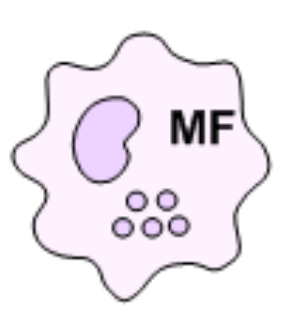 | MF - macrophages, derived from Mo | 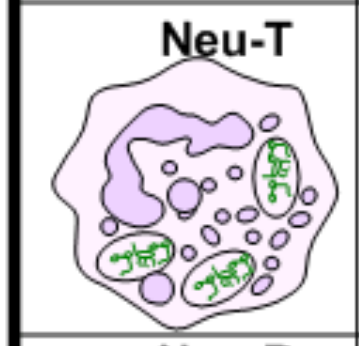 | Neu-T – tolerized Neu |
| 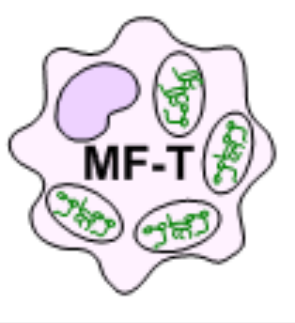 | MF-T - macrophages, derived from Mo-T | 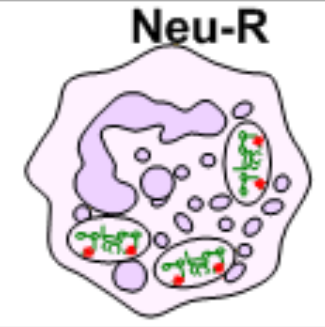 | Neu-R – reprogrammed (tolerized) and repleted by PG-Y neutrophils |
| 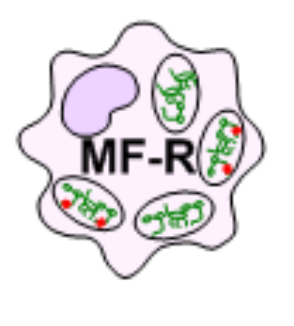 | MF-R - macrophages, derived from Mo-R | 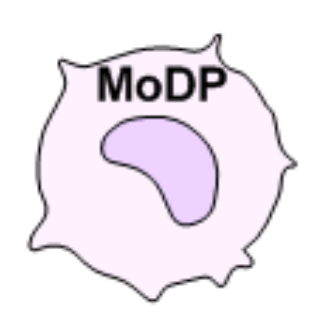 | MoDP - CD34+ cells - precursors of monocytes and immature dendritic cells of bone marrow |
| 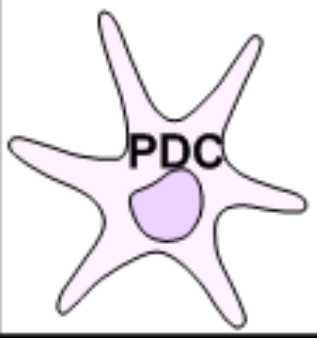 | PDC – plasmacytoid dendritic cells | 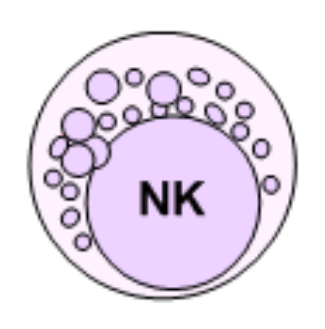 | NK – natural killers |

**Fig. 5. Immune cells (symbols).**

See also Appendix 1.

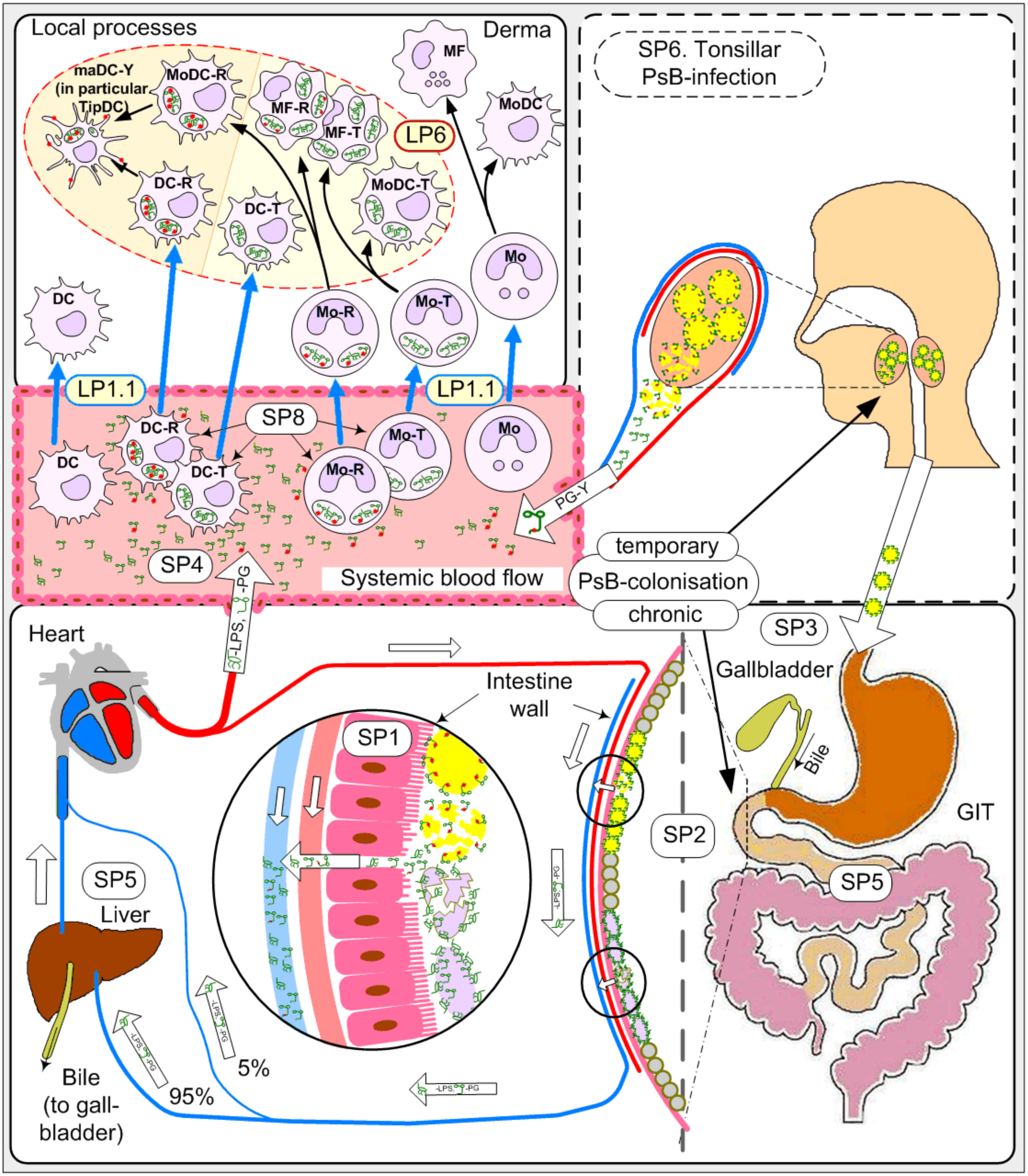

**Fig. 6. SPP (main subprocesses) and some local processes. Illustration.**

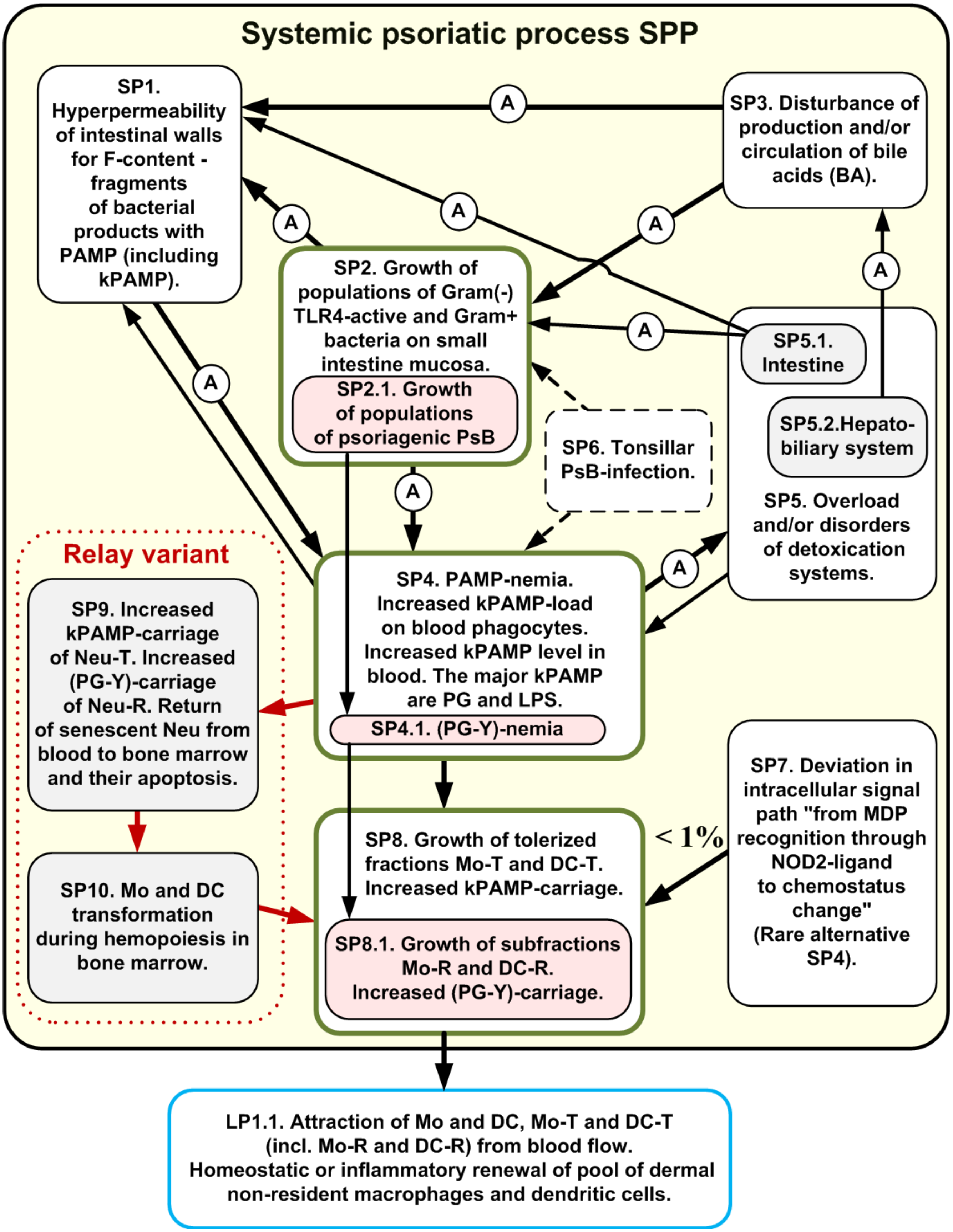

**Fig. 7. SPP and local subprocess LP1.1. The scheme of dependencies.**

Letters A - vicious cycle. Subprocesses SP2, SP4 and SP8 (green contour) make SPP-basis (Fig. 8). Relay variant (SP9, SP10) - see Appendix 5.

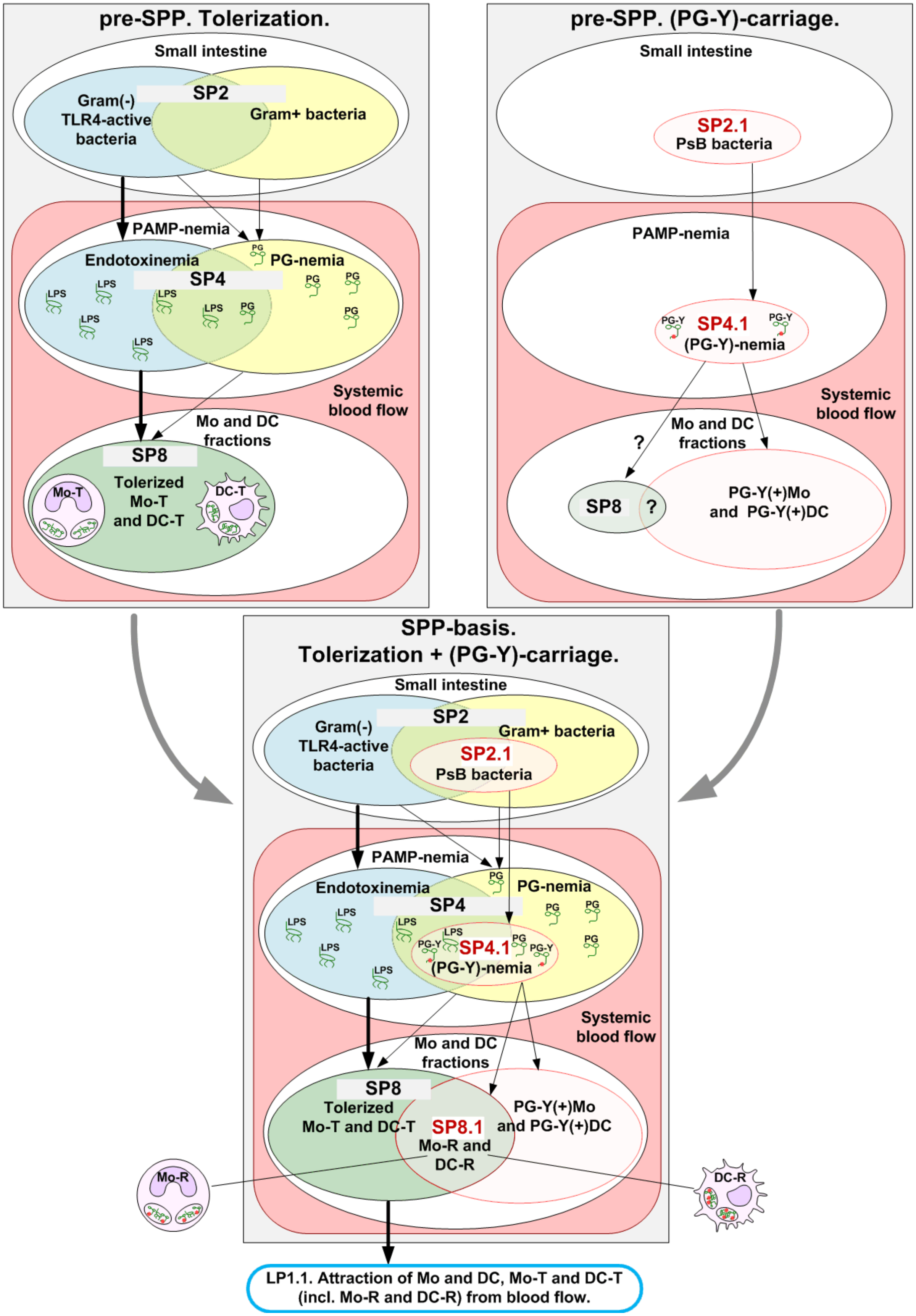

**Fig. 8. SPP-basis. Obligatory subprocesses SP2, SP4 and SP8.**

Two components of SPP-basis: tolerization of phagocytes and (PG-Y)-carriage of phagocytes.

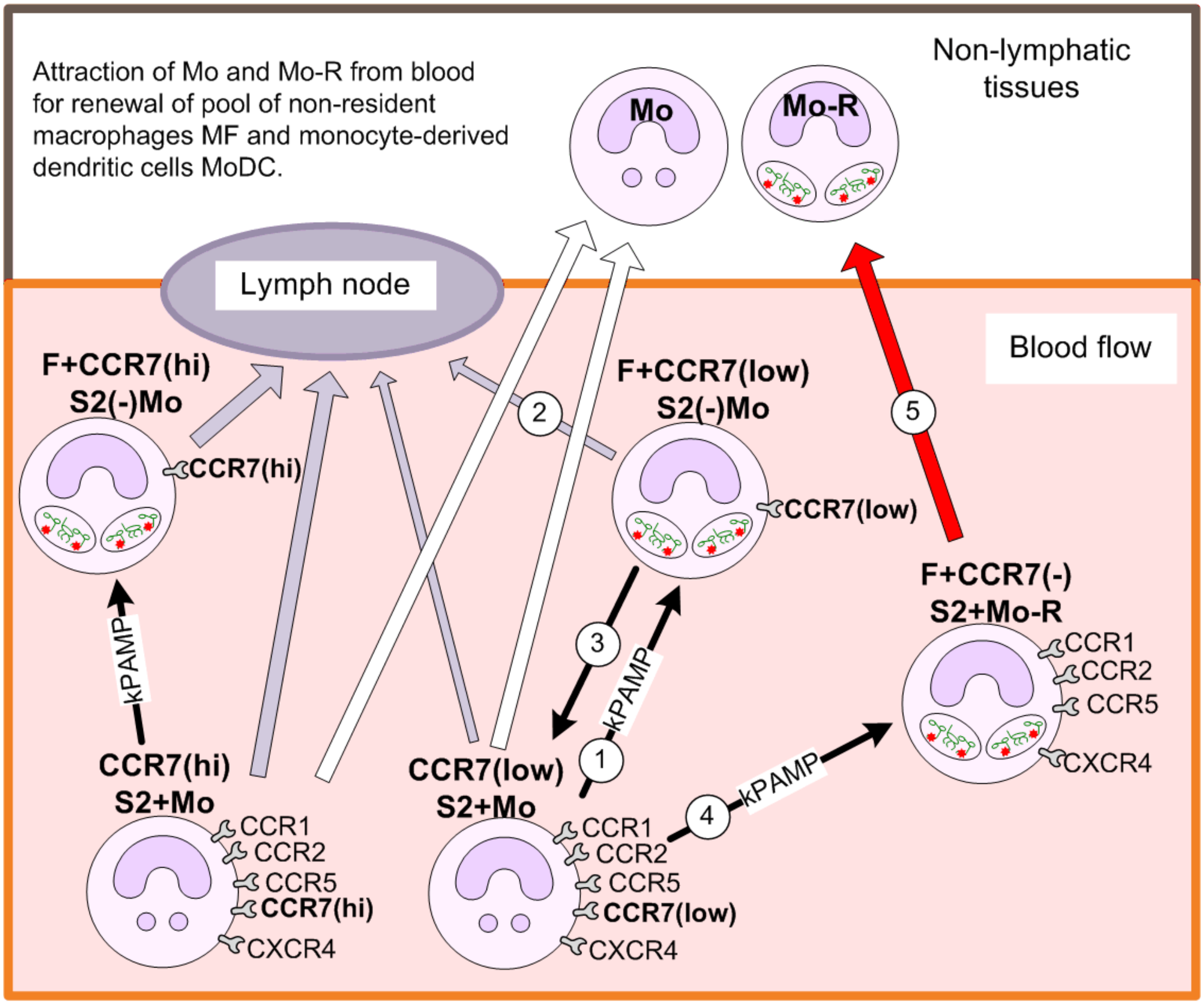

**Fig. 9. CD14+CD16+Mo-R (= F+CCR7(-)S2+Mo-R) formation.**

S2+ - nonactive (or tolerized) chemostatus, S2(-) – active chemostatus of CD14+CD16+Mo. There are main chemokine receptors. Black arrows - transformation, other - traffic.

At LP1.1 non-lymphatic tissues are derma.

Increasing of chronic kPAMP-load and (PG-Y)-load levels is likely to enlarge part of CD14+CD16+Mo-R up to 5-10% among all CD14+CD16+Mo.

It is supposed that in F-content contains PG-Y. In the absence of PG-Y it is possible to replace designation Mo-R on Mo-T.

Subprocess SP8.

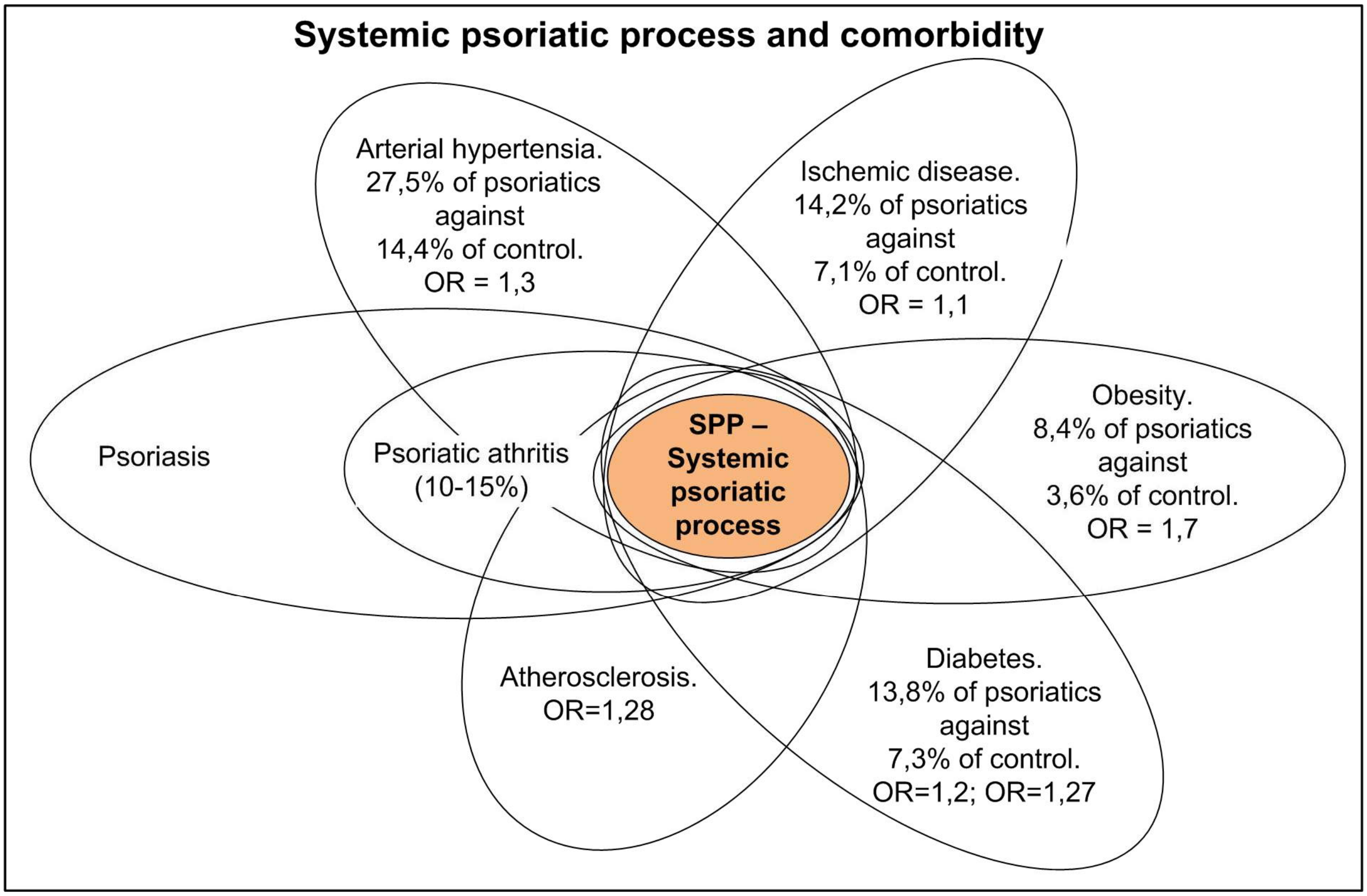

**Fig. 10. Results of two large-scale statistical studies:**
(Shapiro 2007) - more than 48000 psoriatics; (Cohen 2008) - more than 16000 psoriatics.

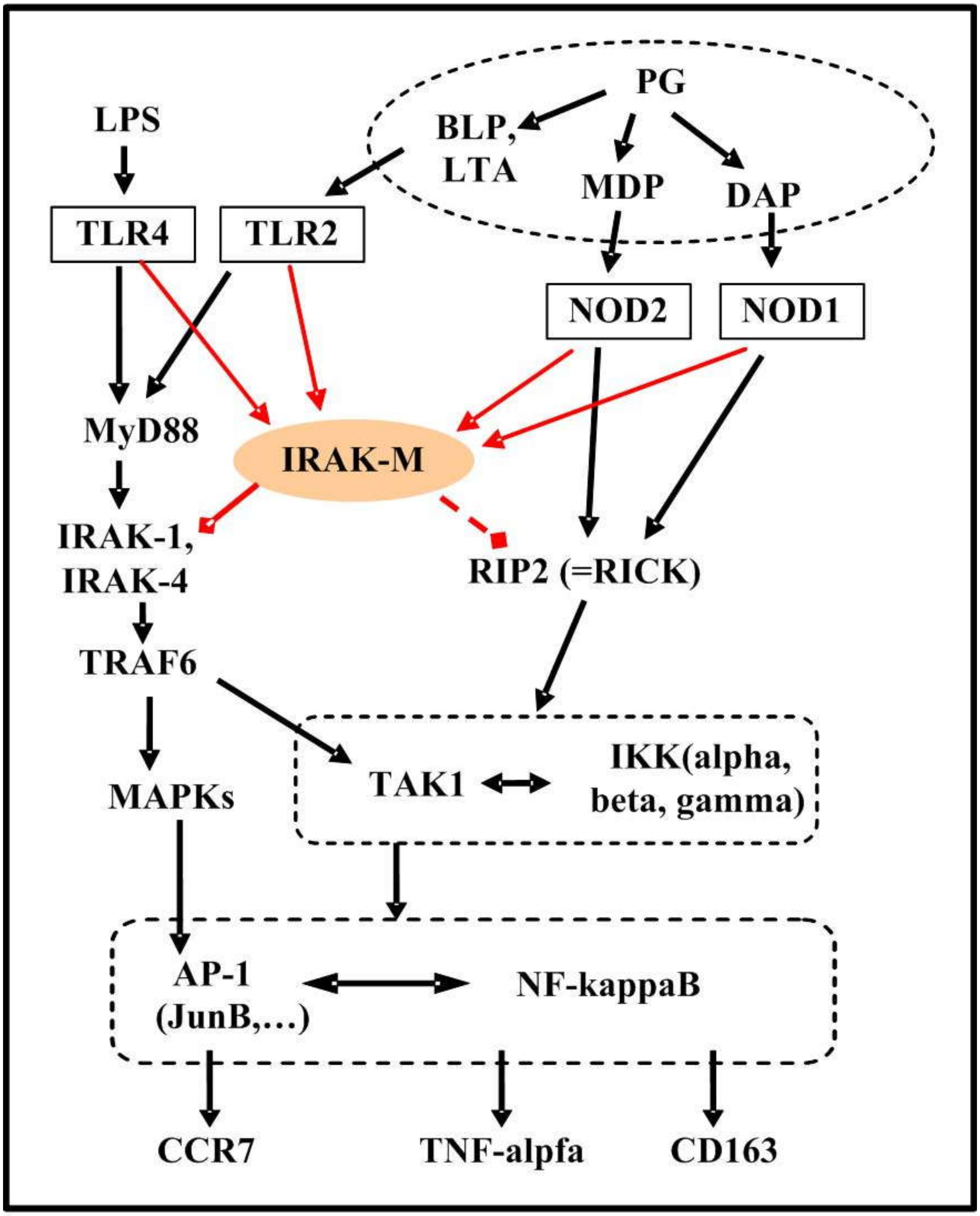

**Fig. 11. Fragment of signal paths from kPAMP-load to CCR7 and CD163 expression and TNF-alpha secretion.**

Red arrows - production increasing, red rhombuses – blocking, red dot-dash - presumable influence.

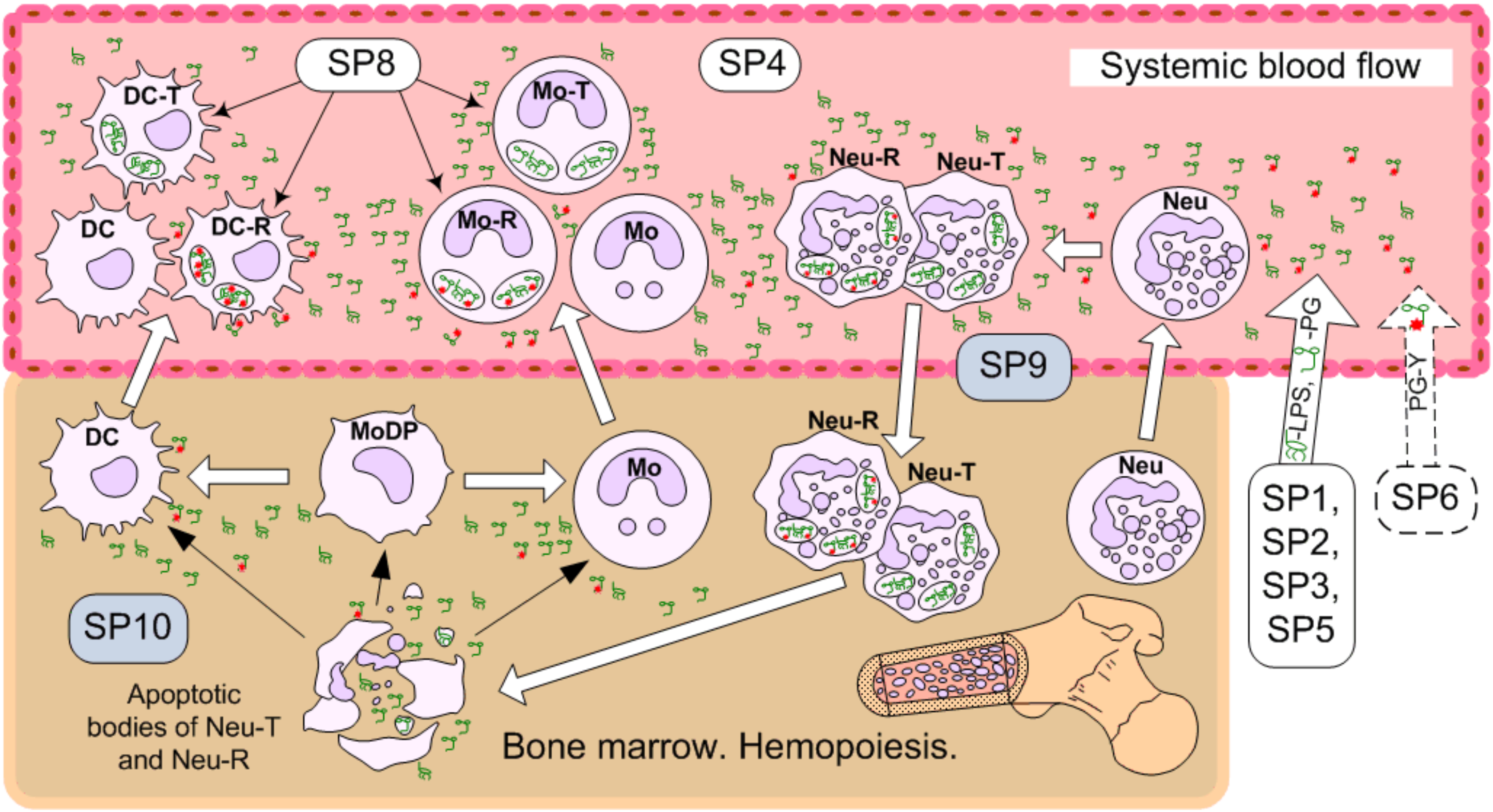

**Fig. 12. Bone marrow transformation Mo and DC with of Neu participation.**

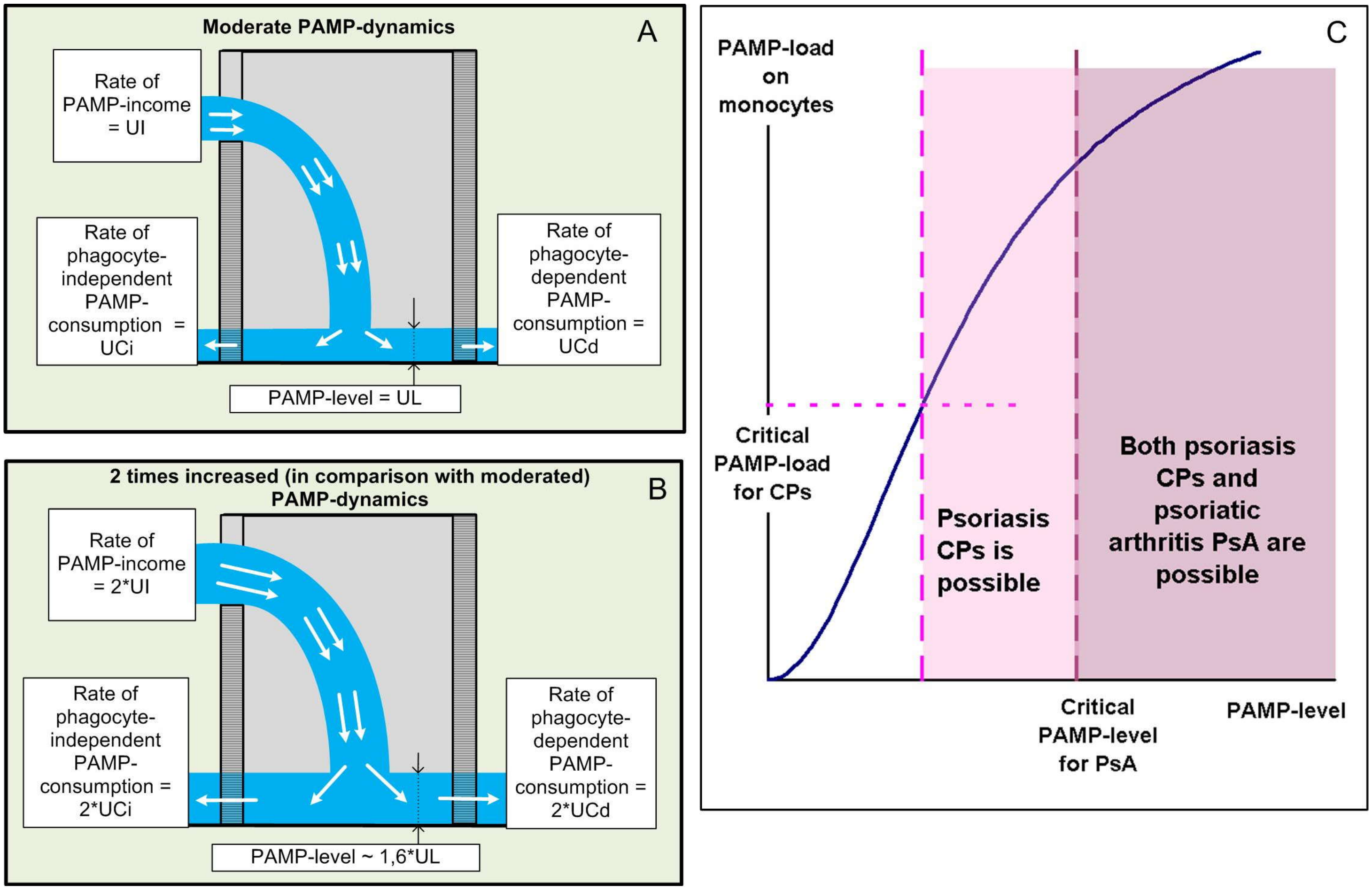

**Fig. 13. PAMP-income and PAMP-consumption.**

Blood flow = container with non-hermetic walls (phagocyte-independent PAMP-consumption is made through left wall, phagocyte-dependent PAMP-consumption is made through right wall). Rates of PAMP-income and RAMP-consumption correspond to length and number of white arrows. They are 2 times larger on scheme B than on scheme A, however PAMP-level raised less than in 2 times (calculation by formula for such container). (C) Critical PAMP-load for CPs possibility is reached earlier than critical PAMP-level for PsA possibility.

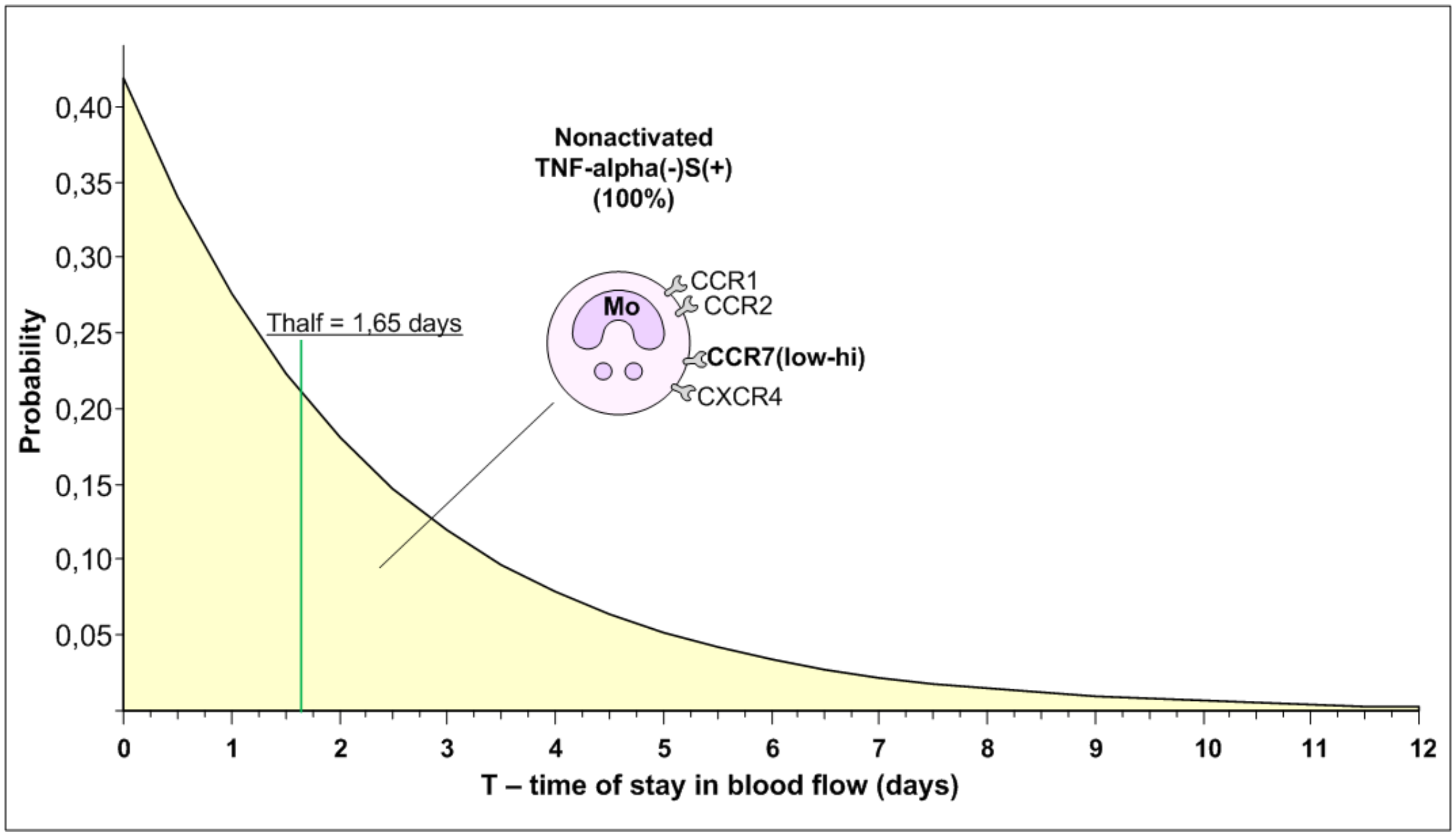

**Fig. 14. Exponential distribution of time of stay of monocytes in the systemic blood flow in norm.**
The chronic PAMP-load is absent. Appendix 10.

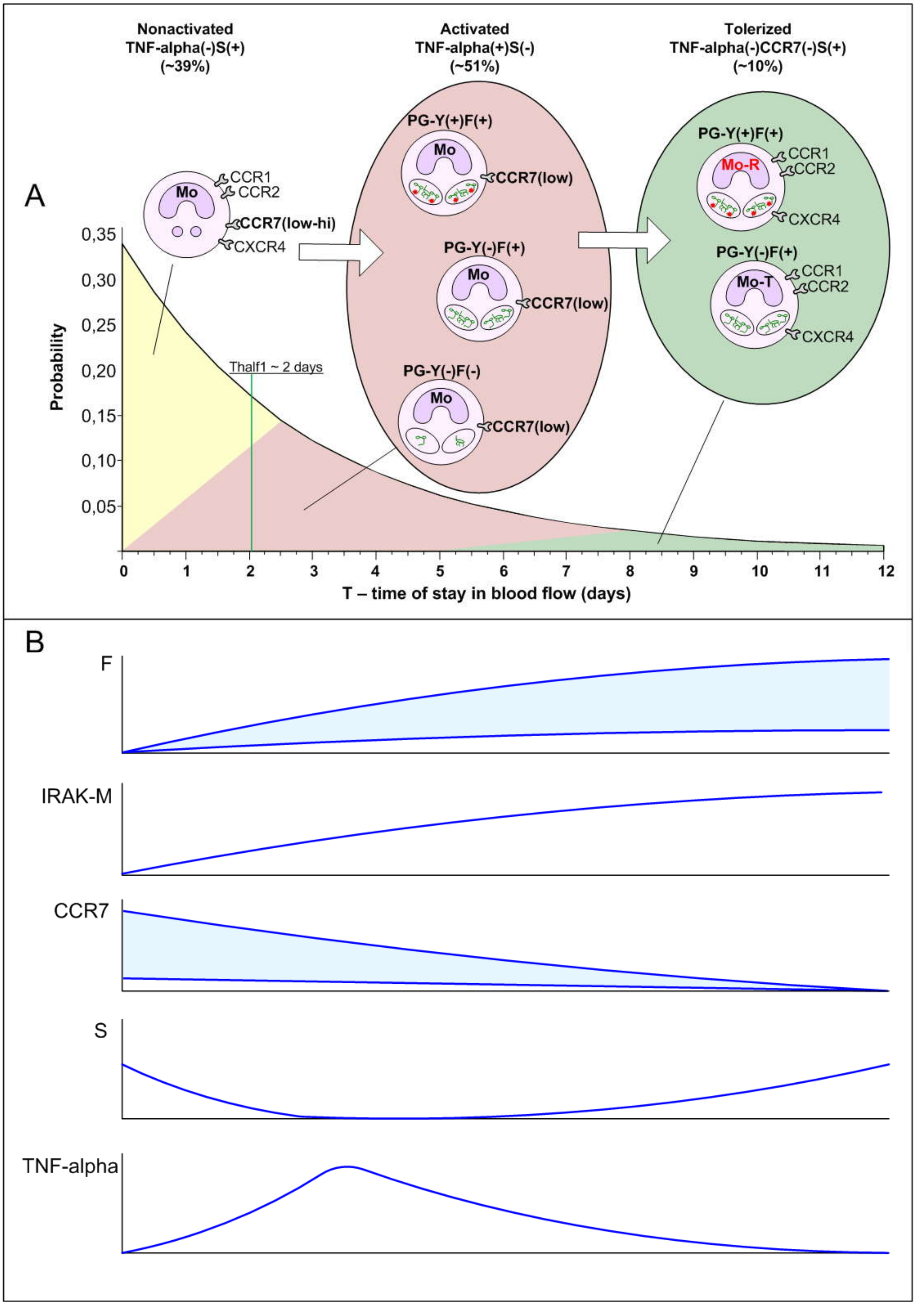

**Fig. 15. Fractionation of blood monocytes without bone marrow activation and possible graphs at SPP.** Fractionation at SPP (A). Possible graphs of expression and secretion of proteins, chemokine receptors and cytokines at blood monocytes depending on time of stay in the blood flow (B). Appendix 10.

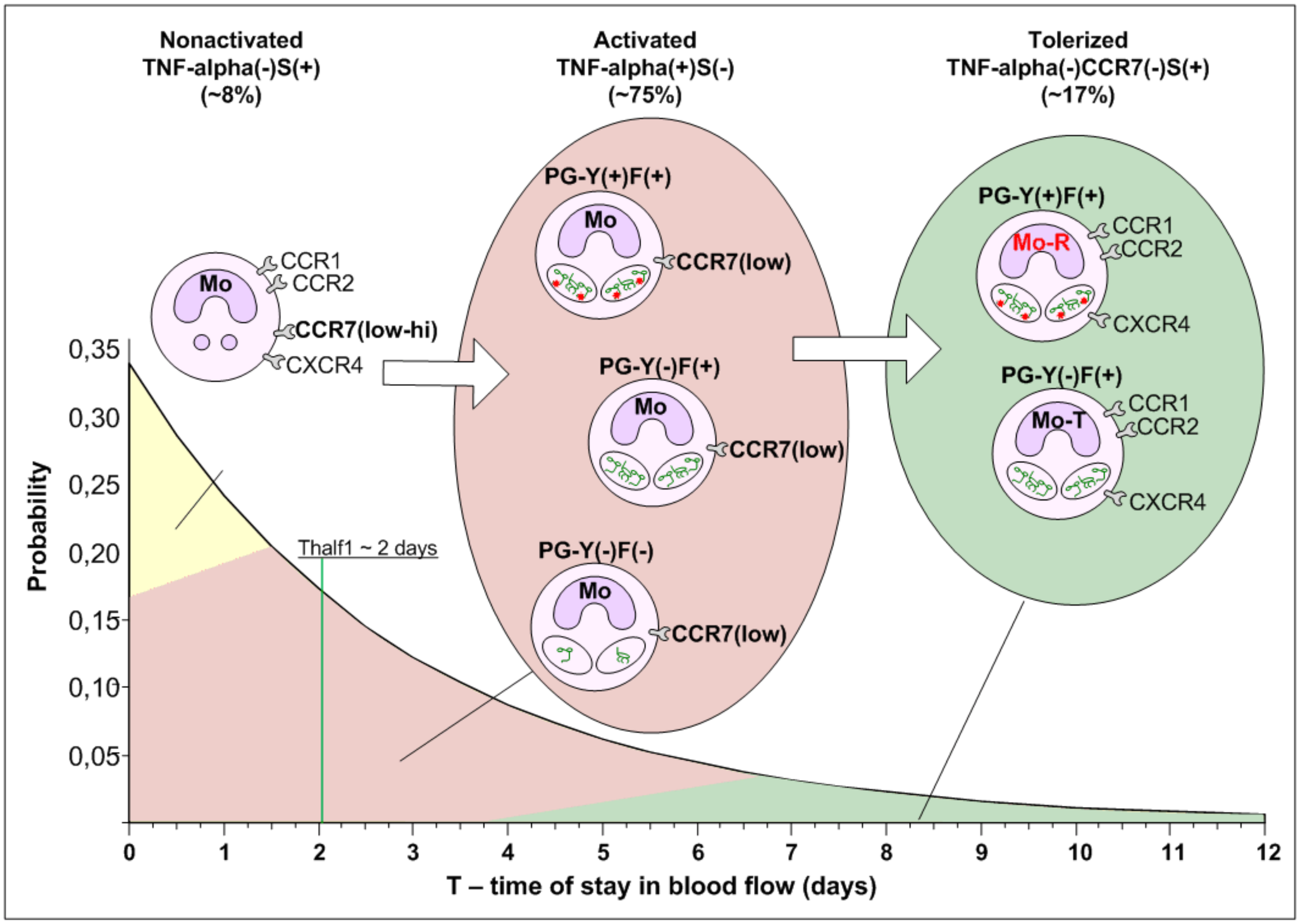

**Fig. 16. Fractionation of blood monocytes at SPP in case of bone marrow activation.**
Thalf1=2,04 days. Appendix 10.

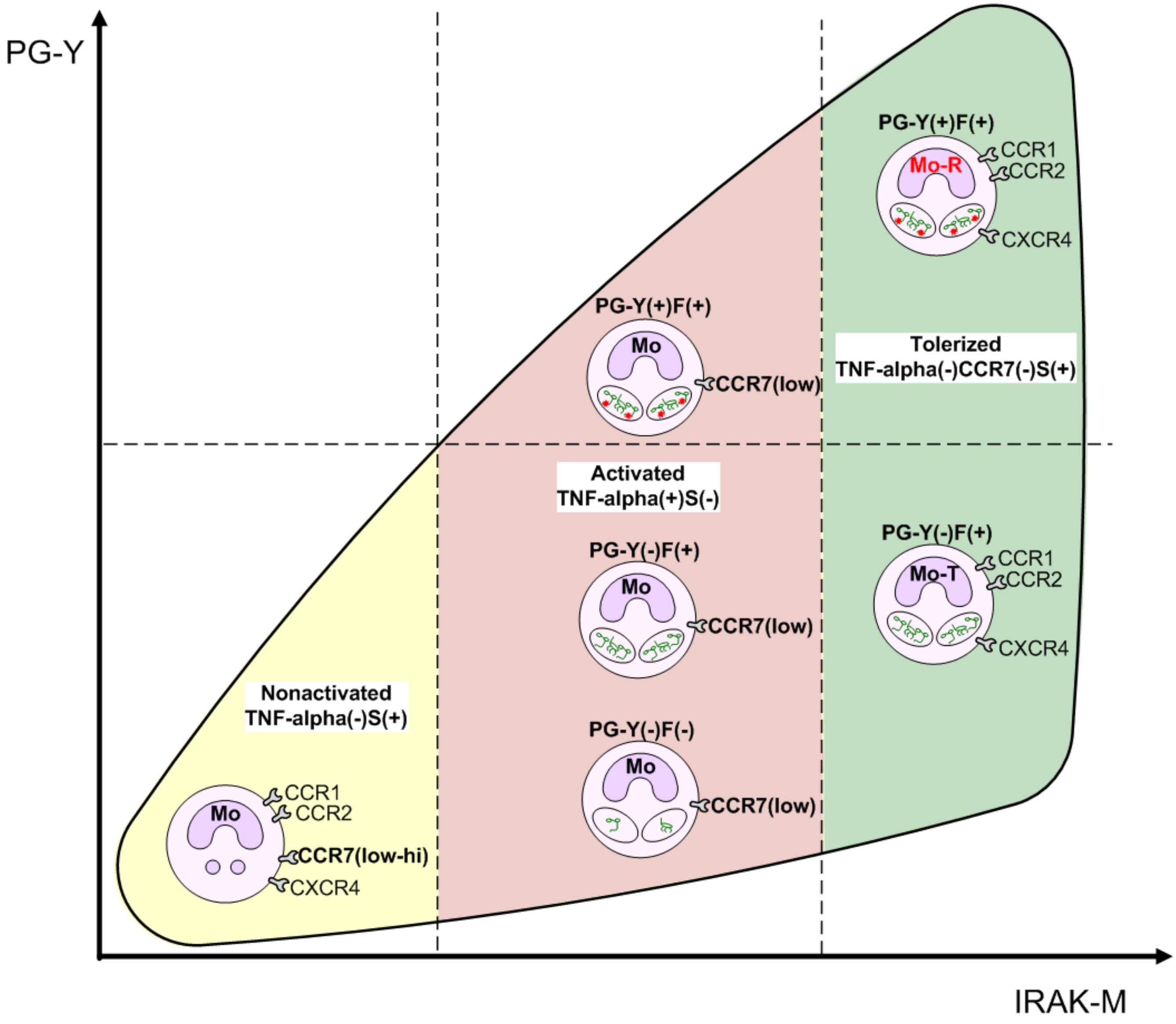

**Fig. 17. Expected (PG-Y, IRAK-M)-distribution of blood monocytes at SPP.**
Appendix 10.